\documentclass[prb,twocolumn,floatfix,10pt,aps]{revtex4-1}
\pdfoutput=1

\usepackage{amsfonts}
\usepackage{amsmath}
\usepackage{amsthm}
\usepackage{amscd}
\usepackage{eucal}
\usepackage{graphicx}
\usepackage{booktabs}
\usepackage{microtype}
\usepackage{hyperref}
\usepackage{times}

\usepackage{tikz}
\usepackage{pgfplots}
\usepackage{mathptmx}

\pgfplotsset{width=0.52\linewidth,height=0.16\textheight}

\let\a=\alpha \let\be=\beta \let\g=\gamma \let\de=\delta
 \let\z=\zeta \let\h=\eta 
\let\eps=\epsilon
   \let\m=\mu
  \let\p=\pi \let\r=\rho 
\let\om=\omega 
\let\ph=\varphi \let\Ph=\phi  
  
 \let\G=\Gamma \let\D=\Delta

\let\qd=\quad  

\def\epp{\, .}
\def\epc{\, ,}

\def\tst#1{{\textstyle #1}}

\theoremstyle{plain}

\newtheorem*{corollary*}{Corollary}

\newtheorem*{conjecture*}{Conjecture}

\theoremstyle{definition}

\newtheorem*{remark}{Remark}

\def\2{\frac{1}{2}} \def\4{\frac{1}{4}}

\def\6{\partial}

\def\+{\dagger}

\def\<{\langle} \def\>{\rangle}

\let\then=\Rightarrow

\def\CH{{\cal H}} 
\def\CO{{\cal O}}

\def\i{{\rm i}}

\def\rd{{\rm d}}
\def\re{{\rm e}}

\DeclareMathOperator{\sh}{sh}
\DeclareMathOperator{\ch}{ch}
\DeclareMathOperator{\tgh}{th}

\DeclareMathOperator{\sech}{sech}
\DeclareMathOperator{\cth}{cth}

\DeclareMathOperator{\tr}{tr}

\DeclareMathOperator{\dn}{dn}

\DeclareMathOperator{\ad}{ad}
\DeclareMathOperator{\id}{id}

\def\hv{\mathbf{h}}

\def\Sv{\mathbf{S}}

\def\fb{\mathfrak{b}}
\def\fbq{\overline{\mathfrak{b}}}

\renewcommand{\tilde}{\widetilde}

\newcommand{\Rarrow}[2][5mm]{
   \xrightarrow{\makebox[#1][c]{#2}}
}

\begin{document}
\title{On the absorption of microwaves by the one-dimensional spin-1/2
Heisenberg-Ising magnet}
\author{Michael Brockmann}
\author{Frank G\"ohmann}
\author{Michael Karbach}
\author{Andreas Kl\"umper}
\affiliation{Fachbereich C -- Physik, Bergische Universit\"at Wuppertal, 42097
Wuppertal, Germany}
\author{Alexander Wei{\ss}e}
\affiliation{Max-Planck-Institut f\"ur Mathematik, P.O. Box 7280,
53072 Bonn, Germany}
\begin{abstract}
We analyze the absorption of microwaves by the Heisenberg-Ising chain 
combining exact calculations, based on the integrability of the model, with
numerical calculations. Within linear response theory the absorbed
intensity is determined by the imaginary part of the dynamical susceptibility.
The moments of the normalized intensity can be used to define the shift
of the resonance frequency induced by the interactions and the line width
independently of the shape of the spectral line. These moments can be calculated
exactly as functions of temperature and strength of an external magnetic field,
as long as the field is directed along the symmetry axis of the chain.
This allows us to discuss the line width and the resonance shift for a
given magnetic field in the full range of possible anisotropy parameters.
For the interpretation of these data we need a qualitative knowledge of the
line shape which we obtain from fully numerical calculations
for finite chains. Exact analytical results on the line shape are out
of reach of current theories. From our numerical work we could extract,
however, an empirical parameter-free model of the line shape at high
temperatures which is extremely accurate over a wide range of anisotropy
parameters and is exact at the free fermion point and at the isotropic
point. Another prediction of the line shape can be made in the zero-temperature and zero magnetic field limit, where the sufficiently
anisotropic model shows strong absorption. For anisotropy parameters
in the massive phase we derive the exact two-spinon contribution to the
spectral line. From the intensity sum rule it can be estimated that
this contribution accounts for more than 80\% of the spectral weight if
the anisotropy parameter is moderately above its value at the isotropic point. 
\end{abstract}

\maketitle

\section{Introduction}
Short-range antiferromagnetic exchange interactions are the predominant
electronic interactions in Mott insulators. They are modeled by the
Heisenberg-Ising Hamiltonian
\begin{equation} \label{ham}
     H = J \sum_{\<i j\>} \bigl( s_i^x s_j^x + s_i^y s_j^y + (1 + \de)
                               s_i^z s_j^z \bigr) \epc
\end{equation}
where the sum is over nearest-neighbor sites, and $J$ measures the strength
of the exchange interaction. The operators $s_i^\a$ are local spin-$\2$ operators,
and the parameter $\de$ takes account of a possible exchange anisotropy and may
include the effects of dipolar interactions as well.

A sensitive experimental probe of magnetic interactions in solids is the
absorption of microwaves, typically in ESR-experiments. In the simplest
experimental setup a circularly polarized wave travels along the direction
of a homogeneous magnetic field. The field is slowly changed and one or
more absorption lines are observed with increasing field, whose precise
location and width depends on the temperature.

For experimentally accessible strengths of the incident microwaves linear
response theory \cite{KuTo54} provides a satisfactory theoretical frame for
the calculation of the absorbed intensity. Then the key quantity to
be calculated for a given Hamiltonian is the (imaginary part of) the
dynamical susceptibility. It is the Fourier transform of a certain dynamical
spin-spin correlation function (see below). Since such a quantity cannot
be calculated for an interacting many-body system as the antiferromagnetic
Heisenberg-Ising model \eqref{ham}, various kinds of approximations have been
tried in the past. Most of these approximations break down, when many-body
correlation effects are strong, in particular in one- and two-dimensional
systems with strong exchange interactions.

In this work we shall concentrate on the one-dimensional spin-$\2$ case
which is not covered by the more traditional approaches. \cite{KuTo54,NaTa72}
It is relevant for the description of quasi one-dimensional compounds
\cite{Ajiro03,KBL10} with strong exchange interactions. This case has been
successfully studied by field theoretical methods \cite{OsAf99,*OsAf02} which
are, however, restricted to small temperatures and to a limited range of
magnetic fields. They also seem to have built in certain {\it a priori}
assumptions about the line shapes. In one dimension purely numerical approaches
\cite{OgMi03,MYO99,*ECM10} are efficient as well. They are unbiased, yet the
extrapolation of the data to the thermodynamic limit of large chains may
require additional justification and support from analytical calculations.

The aim of the present work is to establish a number of exact results for the
microwave absorption of the one-dimensional spin-$\2$ Heisenberg-Ising
magnet, with the homogeneous magnetic field along the magnetic symmetry axis of 
the chain, and to interpret these results in the light of numerical calculations.
In turns the quality and validity of the numerical calculations can be
estimated from the analytical results.

Our work is motivated by the recent progress in calculating static
short-range correlation functions of the integrable spin-$\2$ Heisenberg-Ising
chain at finite temperatures and magnetic fields. This makes it possible
to extend a remarkable result for the resonance shift in one-dimensional
antiferromagnetic chains, which was obtained by Maeda et al.\cite{MSO05}
and which utilizes the exact nearest-neighbor correlation functions of the
isotropic spin-$\2$ Heisenberg chain. We present an alternative framework
for the derivation of the resonance shift which, in the limit of small anisotropy,
reproduces the result of Ref.~\onlinecite{MSO05}. In our approach the
anisotropy is treated non-perturbatively. It allows us, moreover, to derive an
exact formula for the line width at fixed magnetic field.

We utilize the fact that the absorbed intensity $I(\om, h)$ is a positive
function, whose integral over $\om$ exists. The field-dependent moments of the
corresponding normalized intensity turn out to be static short-range correlation
functions which can be calculated directly for the infinite chain. The first
moment is the average absorption frequency. In case that there is a single
pronounced absorption line it gives a measure for the shift of the resonance
compared to the paramagnetic absorption frequency $\om = h$ (in the units
used in this work). This measure is completely independent of the actual shape of
the spectral line. Similarly, the second moment provides a shape-independent
measure of the line width. The idea of using moments was introduced by van
Vleck \cite{VanVleck48} even before the linear response theory was created.
Here we combine it with the finite-temperature linear response theory. The
results of van Vleck are then recovered in the infinite-temperature limit.

Alternatively we may normalize the intensity by its integral over $h$. In this
case the moments cannot be expressed by finite-range correlation functions.
Still, these frequency-dependent moments can be expanded into an infinite
series of field-dependent moments\cite{BGKKW11a} which is a useful starting
point for approximations such as the high-temperature expansion or an
expansion for small anisotropy. Interestingly, the line width defined in
terms of the frequency-dependent moments shows a temperature behavior
rather different from that determined by the field-dependent moments.

When there is more than a single resonance, the interpretation of the moments
is less intuitive. In case of two peaks, for instance, the first moment
would be something like the average location of the two peaks. For this
reason it is desirable to have some knowledge about the full absorption
spectrum (`the line shape'). Hence, we complemented our exact calculation
of the moments with numerical calculations of the dynamical susceptibility
on finite lattices up to 32 sites. The combination of both, the exact
calculation of the moments and the numerics, allows us to propose a model for
the line shape in the high-temperature limit which has no free parameters.
The actual parameters of the corresponding distribution function (a normal-inverse
Gaussian) are determined from the first four exactly calculated moments.

An unbiased but approximate calculation of the line shape is possible 
in the massive ground state phase of the model at vanishing magnetic field.
For an isotropic system there is no absorption without an external field.
For sufficiently high anisotropy, however, the absorption becomes large.
In the massive phase the matrix elements of the local spin operators between
the ground states and excited states (`form factors') are exactly known
\cite{JiMi95} and generally non-vanishing in the thermodynamic limit.
They are classified as $2n$-spinon contributions according to the (even)
number of elementary excitations involved. Here we calculate the two-spinon
contribution to the absorbed intensity exactly. From the intensity sum rule
we infer that for anisotropies moderately above the isotropic point the
two-spin contribution is dominant and amounts to more than 80\% of the
absorbed intensity. For growing anisotropy it rapidly approaches 100\%.
But as opposed to the situation with the dynamic structure factor
for which the relative contribution of the two-spinon excitation is still
dominant in the isotropic limit,\cite{BCK96,BKM98,CMP08} it drops off rapidly for
the dynamical susceptibility.

\section{The method of moments}
\label{sec:methmom}
For any spin system with Hamiltonian $H$ linear response theory relates the
intensity absorbed from a circularly polarized electro-magnetic wave, whose wave
length is large compared to the distance between the spins, to the (imaginary
part of the) dynamical susceptibility \cite{KuTo54}
\begin{equation} \label{defchi}
     \chi_{+-}'' (\om, h) = \frac{1}{2L} \int_{- \infty}^\infty \rd t \:
        \re^{\i \om t} \bigl\< [S^+ (t), S^-] \bigr\>_T \epp
\end{equation}
Here $S^\pm = S^x \pm \i S^y$, and the $S^\a = \sum_{j=1}^L s_j^\a$, $\a = x, y, z$,
are the components of the total spin. $L$ is the number of lattice sites,
$\< \cdot \>_T$ stands for the canonical average at temperature $T$
calculated by means of the statistical operator $\r = \re^{- (H - h S^z)/T}/%
\tr\{ \re^{- (H - h S^z)/T}\}$. Through this average the dynamical susceptibility
depends on $T$ and on an external homogeneous magnetic field $h$ which is
usually applied in ESR experiments. The time evolution of $S^+$ in \eqref{defchi}
must be calculated with $H- h S^z$. The absorbed intensity per spin, normalized
by the intensity of the incident wave and averaged over a half-period $\p/\om$
of the microwave field, is
\begin{equation} \label{int}
     I (\om, h) = \frac\om2 \chi_{+-}'' (\om, h) \epp
\end{equation}
In order to keep this paper self-contained we included a derivation of
\eqref{defchi} and of \eqref{int} in App.~\ref{app:linres}.

In this work we shall exclusively concentrate on the one-dimensional
version of the Heisenberg-Ising (or XXZ) Hamiltonian \eqref{ham}. This
Hamiltonian is in the class of integrable Hamiltonians, but so far this
does not mean that dynamical correlation function such as the expectation
value $\bigl\< [S^+ (t), S^-] \bigr\>_T$ in \eqref{defchi} could be
calculated analytically. We are only aware of three very special cases
where this is possible. These are the `free Fermion case' $\de = - 1$
at $T \rightarrow \infty$, the isotropic limit $\de \rightarrow 0$ and the Ising
limit $\de \rightarrow \infty$. We shall discuss these cases below. In
the general case so far only the short time behavior of $\bigl<[S^+ (t), S^-]
\bigr\>_{T \rightarrow \infty}$ can be accessed by directly calculating
the commutators involved in the time evolution up to a certain power.
We generated this series up to the order $t^{38}$ (cf.~App.~\ref{app:shorttime}).
Still, the results for the short-time behavior alone are not helpful
for calculating the right hand side of~\eqref{defchi}.

Interestingly enough, as we have shown,\cite{BGKKW11a} some more
elementary spectral characteristics, such as the position of the
resonance or the line width, are easier to calculate. They may be
expressed in terms of certain static correlation functions that determine
the moments of a normalized intensity function in one variable. Since
$I (\om, h)$ is a function of the frequency $\om$ and of the magnetic
field $h$ we may normalize it by dividing either by the integral over $\om$
or by the integral over $h$.

In the first case we interpret the resulting normalized function
as a distribution function of frequencies which depends parametrically
on the magnetic field. Then its moments $m_n$ are field dependent.
This corresponds to an experimental situation where the field is
kept fixed and the frequency is varied. We shall see that, from a
theoretical perspective, this case is comparatively simple, since the
moments depend only on static correlation functions of finite
range. The lowest moments which determine the position of the resonance
and its width can be expressed by correlation functions ranging over
up to four lattice sites, which can be calculated exactly.\cite{BDGKSW08,TGK10a}

In the second case, when the intensity is normalized as a function
of the magnetic field, the corresponding moments $M_n$ depend on the
frequency. This corresponds to the standard ESR setup in which the magnetic
field is slowly swept at fixed frequency. As we shall see below this
case is more sophisticated for a theoretical analysis, since static
correlation functions for arbitrary distances are already involved
in the calculation of the lowest moments. Still, the $M_n$ can be
expanded into an infinite series in terms of the moments $m_n$ and
their derivatives, which may serve as a starting point for systematic
approximations.

\subsection{Field-dependent moments}
We temporarily assume that our chain is large but finite. Then the
spectrum is bounded and the integrals
\begin{equation}
     I_n = \int_{- \infty}^\infty \rd \om \: \om^n I (\om, h)
\end{equation}
exist for all non-negative integers $n$. Since $I (\om, h)$ is non-negative
everywhere and since $I_0 > 0$, we may interpret $I (\om, h)/I_0$ as a
probability distribution with moments $I_n$. As we shall see, in our case it is
convenient to express the $I_n$ in terms of another closely related sequence
of integrals
\begin{equation} \label{defm}
	m_n (T, h) = J^{-n} \int_{- \infty}^\infty
	                    \frac{\rd \om}{2 \p} (\om - h)^n \chi_{+-}'' (\om, h)
\end{equation}
which, by slight abuse of language, we shall call (shifted) moments as well.
The existence of the integrals is obvious for every finite chain.

Now by definition the shift of the resonance for fixed $h$ is the deviation
of the average frequency from the paramagnetic resonance at $\om = h$,
\begin{equation} \label{rshiftmom}
     \de \om = \frac{I_1}{I_0} - h = J \frac{J m_2 + h m_1}{J m_1 + h m_0} \epp
\end{equation}
A measure for the line width is the mean square deviation from the
average frequency
\begin{equation} \label{widthmom}
     \D \om^2 = \frac{I_2}{I_0} - \frac{I_1^2}{I_0^2} =
        J^2 \frac{J m_3 + h m_2}{J m_1 + h m_0} - \de \om^2 \epp
\end{equation}

We see that, in order to calculate the resonance shift and the line width, we need
to know the first four shifted moments $m_0$, $m_1$, $m_2$, $m_3$ of the dynamic
susceptibility $\chi_{+-}''$.

In the following we shall employ the notation $\ad_X \cdot = [X, \cdot]$ for
the adjoint action of an operator $X$. Then $S^+ (t) = \re^{- \i ht}
\re^{\i t \ad_{H}} S^+$, since $[H,S^z] = 0$ and $[S^z, S^+] = S^+$, and
it follows with \eqref{defchi} and \eqref{defm} that
\begin{equation} \label{allmoments}
     m_n = \frac1{2L} \bigl\< [S^+, \ad_{H/J}^n S^-] \bigr\>_T \epp
\end{equation}

The latter formula shows that the moments $m_n$ are static correlation
functions whose complexity grows with growing $n$. The first few of them
can be easily calculated. In particular,
\begin{equation} \label{m0}
     m_0 = \frac1{2L} \bigl\< [S^+, S^-] \bigr\>_T
         = \frac1{L} \bigl\< S^z \bigr\>_T
\end{equation}%
which is the magnetization per lattice site. For the subsequent moments,
which do not have such an immediate and simple interpretation, we obtain
\begin{subequations}
\label{m13}
\begin{align} \label{m1}
     m_1 & = \de \< s_1^+ s_2^- - 2 s_1^z s_2^z \>_T \epc \\
     m_2 & = \2 \de^2 \< s_1^z + 4 s_1^z s_2^z s_3^z - 4 s_1^z s_2^+ s_3^- \>_T
               \epc \\
     m_3 & = \4 \de^2 \bigl\< 2 s_1^+ s_2^+ s_3^- s_4^-
                          + 4 s_1^+ s_2^- s_3^+ s_4^-
                          - 2 s_1^+ s_2^- s_3^- s_4^+ \notag \\ & \mspace{27.mu}
                          - 8 s_1^z s_2^z s_3^+ s_4^-
                          - 4 s_1^z s_2^+ s_3^z s_4^-
                          + 8 s_1^z s_2^+ s_3^- s_4^z
			  - 4 s_1^+ s_2^- \notag \\ & \mspace{27.mu}
			  - s_1^+ s_3^- + 8 s_1^z s_2^z s_3^z s_4^z
			  + 2 s_1^z s_3^z - 4 s_1^z s_2^z \notag \\ & \mspace{27.mu}
			  + \de ( 8 s_1^z s_2^+ s_3^- s_4^z + 2 s_1^+ s_2^-
			  - 8 s_1^z s_2^z ) \bigr\>_T \epp
\end{align}
\end{subequations}
These moments are certain combinations of static short-range correlation
functions which implies, in particular, that they all exist in the
thermodynamic limit $L \rightarrow \infty$. Hence, we may relax our restriction
that we are dealing with a finite chain at this point. An interesting
conclusion which can be drawn from the existence of the moments is that
the field-dependent line shape cannot be Lorentzian, as is sometimes assumed
in the literature, since the second moment of a Lorentzian does not exist.
In fact, in our numerical data for finite chains we see an exponential
decay away from the resonance (see below). Note that for finite magnetic
field $m_0$ is of order 1, $m_1$ is of order $\de$, $m_2$ is of order
$\de^2$, but all higher moments are of order $\de^2$ as well. This is
clear from \eqref{allmoments} and will be relevant below.

The formulae \eqref{m13} are appealing from a theoretical perspective,
since, due to recent progress in the theory of integrable systems, static
short-range correlation functions of the Heisenberg-Ising chain can be
calculated exactly. It has been shown that all static correlation
functions of the one-dimensional Heisenberg-Ising model are polynomials
in the derivatives of three functions\cite{JMS08} $\ph$, $\om$, and $\om'$
which, as is common in integrable models, can be expressed in terms of
the solutions of certain numerically well behaved linear and non-linear
integral equations.\cite{BoGo09} We provide the definition of these
functions in the critical case ($- 1 < \de < 0$) in App.~\ref{app:omphi}.
For the massive case ($\de > 0$) the definitions are similar and can
be found in Ref.~\onlinecite{TGK10a}.

Although, in principle, all static correlation functions of the
Heisenberg-Ising chain in the thermodynamic limit are known exactly,
their explicit calculation works out only at short distances. At larger
distances the amount of computer algebra involved in the calculations
grows excessively. In Refs.~\onlinecite{BGKS07},~\onlinecite{BDGKSW08}, and~\onlinecite{TGK10a} we obtained
all correlation functions ranging over at most four lattice sites.
This is just enough to calculate the moments $m_0, m_1, m_2, m_3$.
For the simpler case of the isotropic model in vanishing magnetic
field we obtained the correlation functions ranging over up to
seven lattice sites.\cite{SABGKTT11}

\begin{figure}[!ht]
\begin{center}
\includegraphics[width=.45\textwidth]{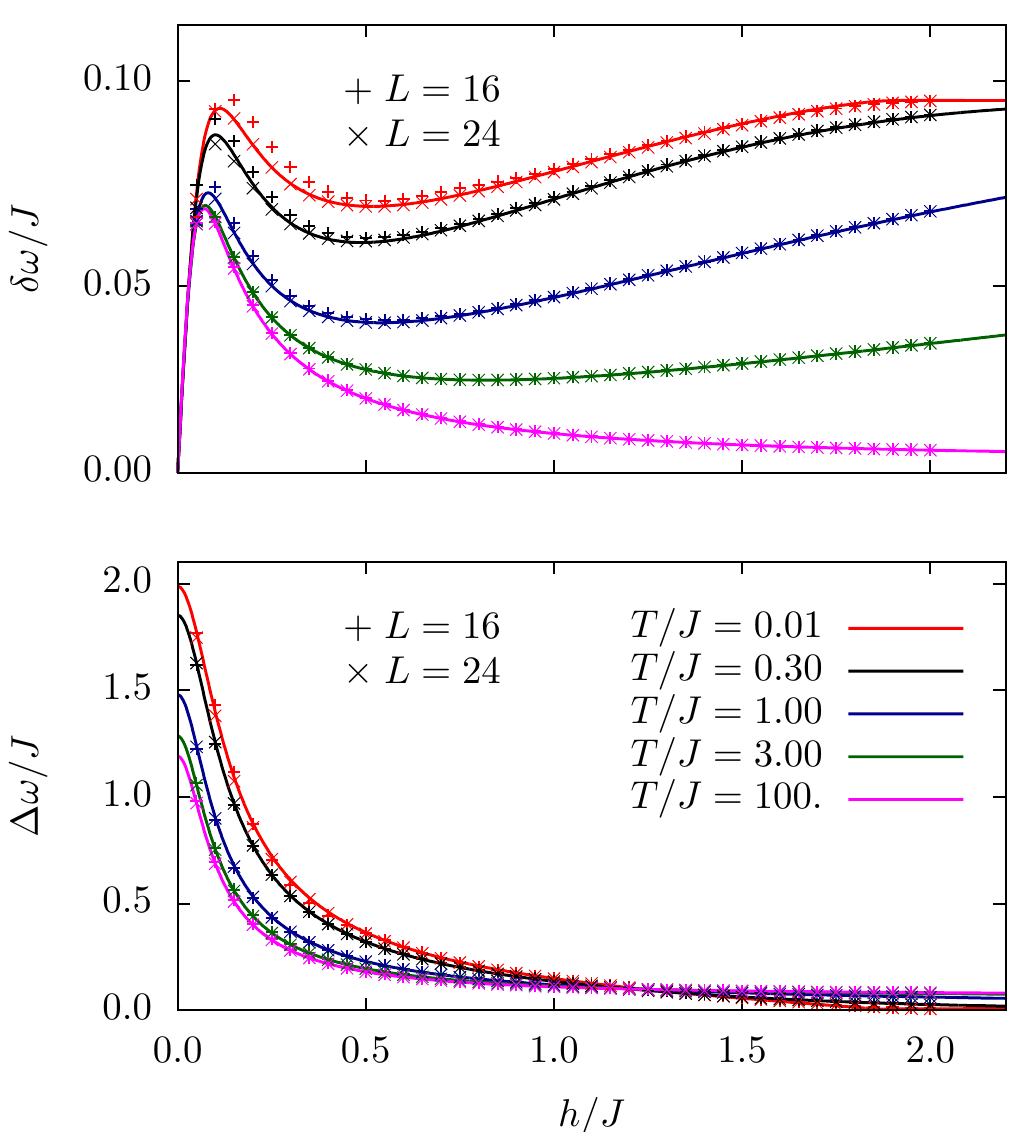}
\caption{\label{fig:shwcritofh} Resonance shift $\de \om/J$ and line width
$\D \om/J$ in the critical regime at $\de = - 0.1$ as function of the
magnetic field. Crosses from fully numerical calculation for finite
chain Hamiltonians of 16 and 24 sites.}
\end{center}
\end{figure}

\begin{figure}[!ht]
\begin{center}
\includegraphics[width=.45\textwidth]{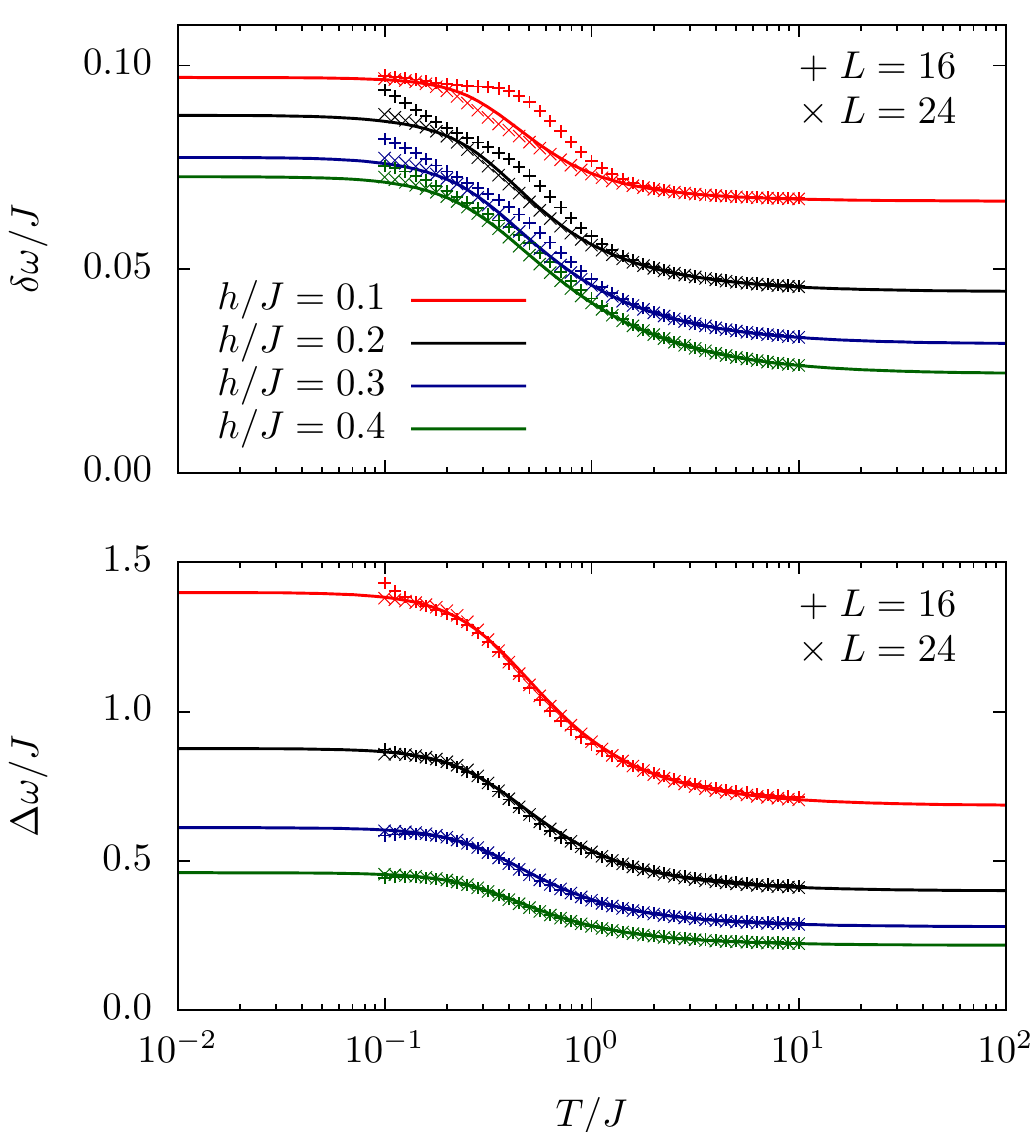}
\caption{\label{fig:shwcritoft} Resonance shift $\de \om/J$ and line
width $\D \om/J$ in the critical regime at $\de = - 0.1$ as function of the
temperature. Crosses from fully numerical calculation for finite chain
Hamiltonians of 16 and 24 sites.}
\end{center}
\end{figure}

\begin{figure}[!ht]
\begin{center}
\includegraphics[width=.45\textwidth]{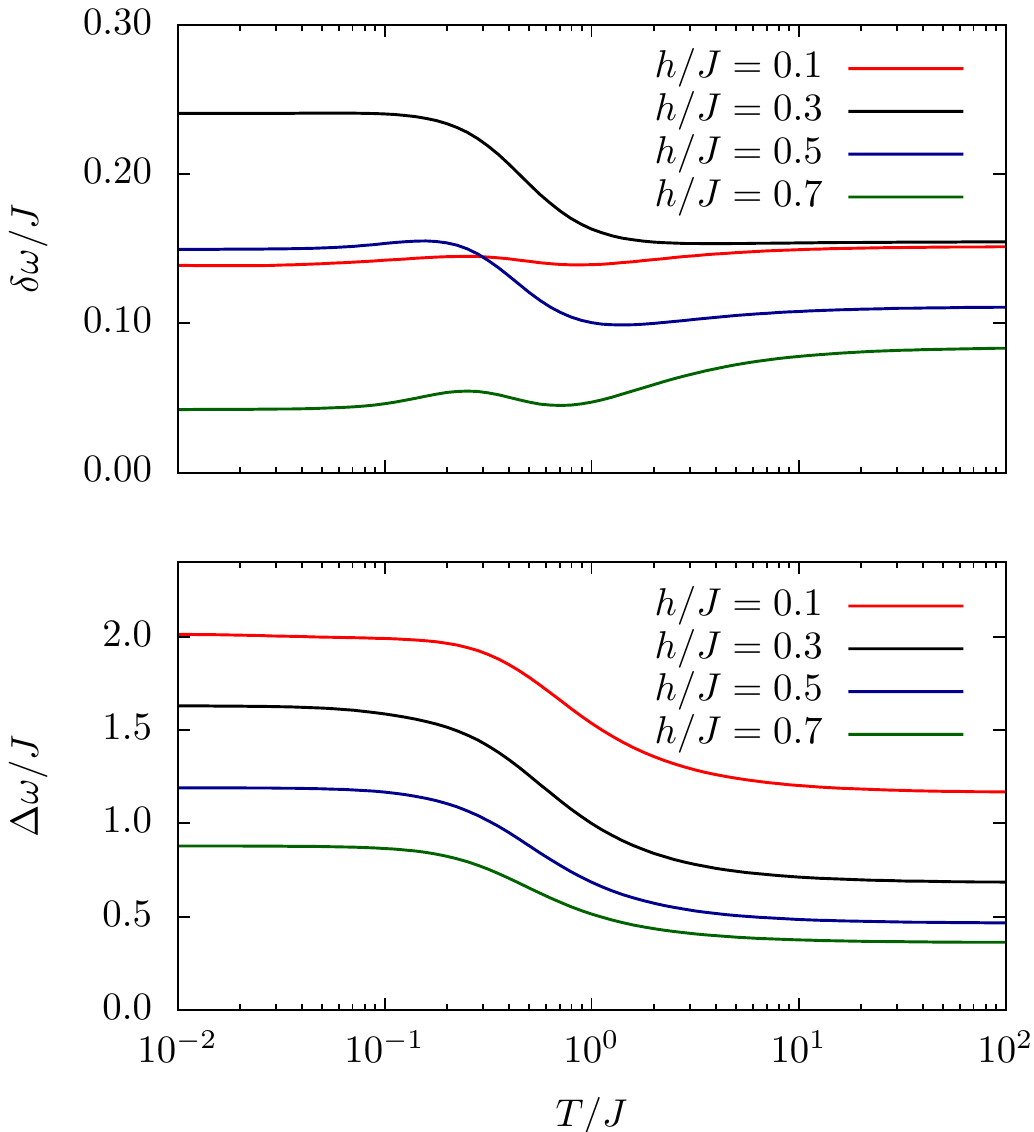}
\caption{\label{fig:shwmassoft} Resonance shift $\de \om/J$ and line
width $\D \om/J$ in the massive regime at $\de =  0.25$ as function
of the temperature.}
\end{center}
\end{figure}

When considering the Heisenberg-Ising Hamiltonian as an integrable
model it is customary to parameterize all functions by a deformation
parameter $q$ in terms of which the anisotropy is $\de = (q - 1)^2/2q$. 
Employing the shorthand notations $\ph_{(n)} = \6_x^n \ph (x)|_{x=0}$,
$f_{(m,n)} = \6_x^m \6_y^n f(x, y)|_{x=y=0}$, for $f = \om, \om'$,
we obtain\cite{BGKKW11a}
\begin{widetext}
\begin{align} \label{ms3}
     m_0 & = - \2 \ph_{(0)} \epc \notag \\
     m_1 & = \frac{(q - 1)^2 (q^2 + 4q + 1) \om_{(0,1)}'}{16q^2}
             - \frac{(q^3 - 1) \om_{(0,0)}}{4 q (q + 1)} \epc \notag \\
     m_2 & = \frac{(q - 1)^2}{256 q^4}
             \bigl[ 4 q (q + 1)(q^3 - 1) (\om_{(0,2)} \ph_{(0)}
	          - 2 \om_{(1,1)} \ph_{(0)} - \om_{(0,0)} \ph_{(2)}) \notag \\
            & \mspace{190.mu}
                  + (q^2 - 1)^2 (q^2 + 4q + 1)
	            (\om_{(1,2)}' \ph_{(0)} + \om_{(0,1)}' \ph_{(2)})
	          - 16 q^2 (q - 1)^2 \ph_{(0)} \bigr] \epc \notag \\
     m_3 & = \frac{(q - 1)^4}{98304 q^8 (q^4 - 1)(q^6 - 1)}
	     \bigl[
             16 q^2 (q^2 - 1)^3 (q^4 - 1)(q^6 - 1)(q^2 + 4q + 1)
	     (\om_{(0,2)} \om_{(0,1)}' + \om_{(0,0)} \om_{(1,2)}') \notag \\ &
	   + 64 q^2 (q^2 - 1)^4 (2 q^{10} - q^9 + 4 q^8 - 4 q^7 - 12 q^6
	             - 14 q^5 - 12 q^4 - 4 q^3 + 4 q^2 - q + 2)
		     \om_{(0,0)} \om_{(1,1)} \notag \displaybreak[0] \\ &
           - 16 q^2 (q^2 - 1)^2 (q^4 - 1)(q^6 - 1)(3 q^4 + 14q^2 + 3) \om_{(0,3)}'
           + 8 q^2 (q^2 - 1) (q^4 - 1)^2 (q^6 - 1) (q + 1)^2
	           (8 \om_{(1,2)}' - \om_{(2,3)}') \notag \\ &
           + 192 q^4 (q^2 - 1)^2 (q^4 - 1) (q^6 + 18 q^4 + 8 q^3 + 18 q^2 + 1)
	              \om_{(0,2)} \notag \displaybreak[0] \\ &
           + 64 q^2 (q^2 - 1)^2 (q^4 - 1) (q + 1)^2 (2 q^8 - 5 q^7 + 26 q^6 - 49 q^5
	             + 28 q^4 - 49 q^3 + 26 q^2 - 5 q + 2) \om_{(1,1)} \notag \\ &
           - 16 q^2 (q^2 - 1)^2 (q^4 - 1) (q + 1)^2 (q^8 - q^7 + q^6 + q^5 + 2 q^4
	             + q^3 + q^2 -  q + 1) (2 \om_{(1,3)} - 3 \om_{(2,2)}) \notag
		     \displaybreak[0] \\ &
           + 64 q^2 (q^4 - 1) (q^6 - 1) (3 q^8 + 2 q^6 + 24 q^5 - 130 q^4 + 24 q^3
	             + 2 q^2 + 3) \om_{(0,1)}' \notag \displaybreak[0] \\ &
           + (q^4 - 1) (q^6 - 1) (q + 1)^2 (q^{10} - 2 q^9 + 25 q^8 + 16 q^7 + 118 q^6
	             + 164 q^5 \notag \\ & \mspace{126.mu} 
		     + 118 q^4 + 16 q^3 + 25 q^2 - 2 q + 1)
	     (\om_{(0,3)}' \om_{(1,2)}' + \om_{(0,1)}' \om_{(2,3)}') \notag \\ &
           - 1536 q^5 (q^4 - 1) (4 q^8 - 9 q^7 - 2 q^6 - 6 q^5 + 8 q^4 - 6 q^3
	             - 2 q^2 - 9 q + 4) \om_{(0,0)}
		     \displaybreak[0] \notag \\ &
           + 4 q^2 (q^6 - 1) (q + 1)^2 (q^2 + 1) (5 q^8 - 2 q^7 + 32 q^6 + 50 q^5
	             + 70 q^4 + 50 q^3 + 32 q^2 - 2 q + 5) \notag \\ & \mspace{54.mu}
	     (2 \om_{(1,3)} \om_{(0,1)}' - 3 \om_{(2, 2)} \om_{(0,1)}'
	      + \om_{(0,2)} \om_{(0,3)}' - 2 \om_{(1,1)} \om_{(0,3)}'
	      - 3 \om_{(0,2)} \om_{(1,2)}' - \om_{(0,0)} \om_{(2,3)}')
	      \notag \displaybreak[0] \\ &
           - 16 q^2 (q + 1)^2 (q^{16} - q^{15} + 8 q^{14} + 9 q^{13} + 47 q^{12}
	             + 45 q^{11} + 96 q^{10} + 91 q^9 + 128 q^8 + 91 q^7 + 96 q^6
		     \notag \\ & \mspace{72.mu}
		     + 45 q^5 + 47 q^4 + 9 q^3 + 8 q^2 - q + 1)
	     (3 \om_{(0,2)}^2 - 6 \om_{(1, 1)} \om_{(0,2)}
	      + 2 \om_{(0,0)} \om_{(1,3)} - 3 \om_{(0,0)} \om_{(2,2)}) \bigr] \epp
\end{align}
\end{widetext}
These are the moments in the thermodynamic limit. We can calculate
them with high numerical accuracy over the whole range of the
phase diagram, for all temperatures and magnetic fields as well as for
arbitrary anisotropy $\de$. In Figs.~\ref{fig:shwcritofh}-\ref{fig:shwmassoft}
we show two examples not too far away from the isotropic point, namely
$\de = - 0.1$ in the critical phase and $\de = 0.25$ in the massive phase.
Values not too far away from the isotropic point are most relevant for
real materials. In both cases we observe an increase of the resonance
shift $\de \om$ and a broadening of the spectral lines, measured as an
increase of $\D \om$, for decreasing temperatures.
\enlargethispage{1ex}

This is interesting as it seems to contradict experimental results \cite{KBL10}
which claim a narrowing. This discrepancy is due to the different measures
for the line width here and in the experimental literature. The mean square
deviation used in Figs.~\ref{fig:shwcritofh}-\ref{fig:shwmassoft} is a
customary measure for the width of wave functions in quantum mechanics.
For Gaussians it is of the order of magnitude of an intuitive line width
drawn by eye, but for distributions which have long and shallow tails this
is no longer the case. Hence, $\D \om$ may strongly deviate from a typical
measure of the line width used in the interpretation of experimental data
as e.g.~the distance between the inflection points right and left to the
maximum of the intensity (`peak-to-peak width'). This discrepancy was
already noted by van Vleck.\cite{VanVleck48} One advantage of the mean
square deviation from the resonance frequency as a measure of the line width
is that it is defined independently of the line shape. In principle it should
be no problem to extract it from experimental data. Yet, we expect that in cases,
where the contributions from the tails of the spectral line are important, a
problem might be to resolve these tails from the `background'.

The difference between different measures of the line width becomes
rather clear from our numerical analysis below (compare also Ref.~\onlinecite{ECM10}).
For high temperatures, where we could extract a model for the line
width from our numerical data, the `peak-to-peak width' is much smaller
than $\D \om$. This can be attributed to the shallow tails of the absorbed
intensity. In experiments these tails may be misinterpreted as stemming
from couplings of the spin chain to other degrees of freedom and may
lead to an overestimation of the background. On the other hand, tails
are expected to have less influence on the resonance shift. As long as
they are not too asymmetric the shift $\de \om$ of the average of the
absorbed intensity should agree with the shift of its maximum, which
is the common measure in experiments.

For other values of the anisotropy parameters the resonance shift and
the line width in the critical regime for, $- 1 \le \de < 0$, show a
qualitatively similar behavior as in Figs.~\ref{fig:shwcritofh} and
\ref{fig:shwcritoft}. Both, $\de \om$ and $\D \om$, increase with
decreasing temperature at fixed magnetic field. Note that in the
massive regime, exemplified with Fig.~\ref{fig:shwmassoft}, the
resonance shift may behave non-monotonically as a function of temperature.

\subsection{Frequency-dependent moments}
In order to obtain the resonance shift and the line width as defined
in the previous section experimentally one would have to measure the
microwave absorption at fixed Zeeman field $h$ for various values of
the frequency $\om$ and then calculate the required averages as
integrals over $\om$. In current ESR experiments different data sets
are recorded. The microwave frequency $\om$ is kept fixed and
the absorbed intensity $I (\om, h) = \om \chi_{+-}'' (\om, h)/2$ is
determined as a function of $h$. This intensity function can be
normalized by dividing by its $h$-integral, and the corresponding
frequency-depending moments define the resonance shift and line width
in `$h$-direction'. Away from the isotropic point ($\de = 0$), where
$\chi_{+-}'' (\om, h)$ is symmetric and the absorption line is
extremely narrow, these may be rather unrelated to their field-dependent counterparts of the previous section.

In analogy with \eqref{defm} we define the frequency-dependent moments
\begin{equation} \label{defM}
	M_n (T, \om) = J^{-n} \int_{- \infty}^\infty
	               \frac{\rd h}{2 \p} (h - \om)^n \chi_{+-}'' (\om, h) \epp
\end{equation}
These can be expressed in terms of the $m_n$ and their derivatives. Denoting
the $k$th derivative with respect to the second argument by a superscript $(k)$
we obtain the representation
\begin{equation} \label{Mom}
     M_n (T, \om) = (- 1)^n
        \sum_{k=0}^\infty \frac{(- J)^k}{k!} m_{k + n}^{(k)} (T, \om) \epc
\end{equation}
for the frequency-dependent moments. This representation involves static
correlation functions for arbitrarily large distances. For this reason the
$M_n$ cannot be calculated by our exact method above. Yet, in certain cases
finitely many terms of the series are sufficient for a good approximation.

We first of all express the resonance shift $\de h = \<h\> - \om$ and the mean
square deviation from the center of the absorption peak $\D h^2 = \<h^2\> - \<h\>^2$
in terms of the $M_n$,
\begin{equation} \label{shiftwidthh}
     \frac{\de h}{J} = \frac{M_1}{M_0} \epc \qd
     \frac{\D h^2}{J^2} = \frac{M_2}{M_0} - \frac{M_1^2}{M_0^2} \epp
\end{equation}
There are at least two cases, where these formulae simplify and finitely many
of the $m_n$ are enough to determine a systematic approximation to $\de h$ and
$\D h$.

The equation for the resonance shift simplifies for small anisotropy, $|\de| \ll 1$.
Since $M_0 = m_0 + {\cal O} (\de)$, $M_1 = - m_1 + {\cal O} (\de^2)$ and, generically,
$m_1$ itself is of order $\de$ (see \eqref{m13}, \eqref{Mom}) we obtain to linear
order in~$\de$
\begin{equation} \label{hshiftapp}
     \frac{\de h}{J} = - \frac{m_1}{m_0} \epp
\end{equation}
In previous work \cite{NaTa72,MSO05} the same equation was obtained by a
more intuitive reasoning. It leads to results which compare rather well
with experiments.\cite{MSO05} However, some care is necessary with the
interpretation of \eqref{hshiftapp}. Since $m_1/\de$ vanishes at $\de = h = 0$,
it follows that $m_1 = \de (a h + b \de + \dots)$ with some coefficients
$a, b$, whence $h/J$ must be large compared to $\de$ for \eqref{hshiftapp}
to be applicable.

Recall that the higher moments $m_n$, $n \ge 2$, are of order $\de^2$. Hence,
for the line width there is no simplification for small anisotropy, like in
\eqref{hshiftapp}. But there is another measurable quantity which does allow
for a systematic small-$\de$ expansion to first order, namely the integrated
intensity $\p \om M_0$, since
\begin{equation}
     M_0 = m_0 - J m_1' \epp
\end{equation}

For the resonance shift it follows to linear order in $\de$ from
\eqref{rshiftmom} and \eqref{hshiftapp} that $\de h (T, \om) =
- \de \om (T, h)|_{h = \om}$. For the line width there is no such simple
relation between $\D \om$ and $\D h$, not even for small $\de$.

The representation \eqref{Mom} is a series in ascending powers of $J/T$ (with
still temperature-dependent coefficients). This can be used to evaluate 
\eqref{shiftwidthh} asymptotically for high temperatures. It turns out that
the leading terms in the $J/T$ expansion of $m_1$ and $J m_2'$ cancel each
other ($\de h \sim \frac{h}{2T} \de \rightarrow 0$ in the high-temperature
limit $T \gg J$) and
\begin{multline} \label{highwidth}
     \frac{\D h}{J} = \frac{|\de|}{\sqrt{2}} \biggl(1 + \frac{(1 + \de) J}{4T}\\
                      -\frac{(6\de^2+10\de+9)J^2+4\om^2}{32T^2} + \dots \biggr) \epc
\end{multline}
where $\om$ is the microwave frequency. This formula provides a simple means
to directly measure the anisotropy parameter $\de$. For $T \rightarrow \infty$
it turns into Eq.~(10) of Ref.~\onlinecite{VanVleck48} upon a proper
identification of parameters.

\begin{figure}[!ht]
\begin{center}
\includegraphics[width=.45\textwidth]{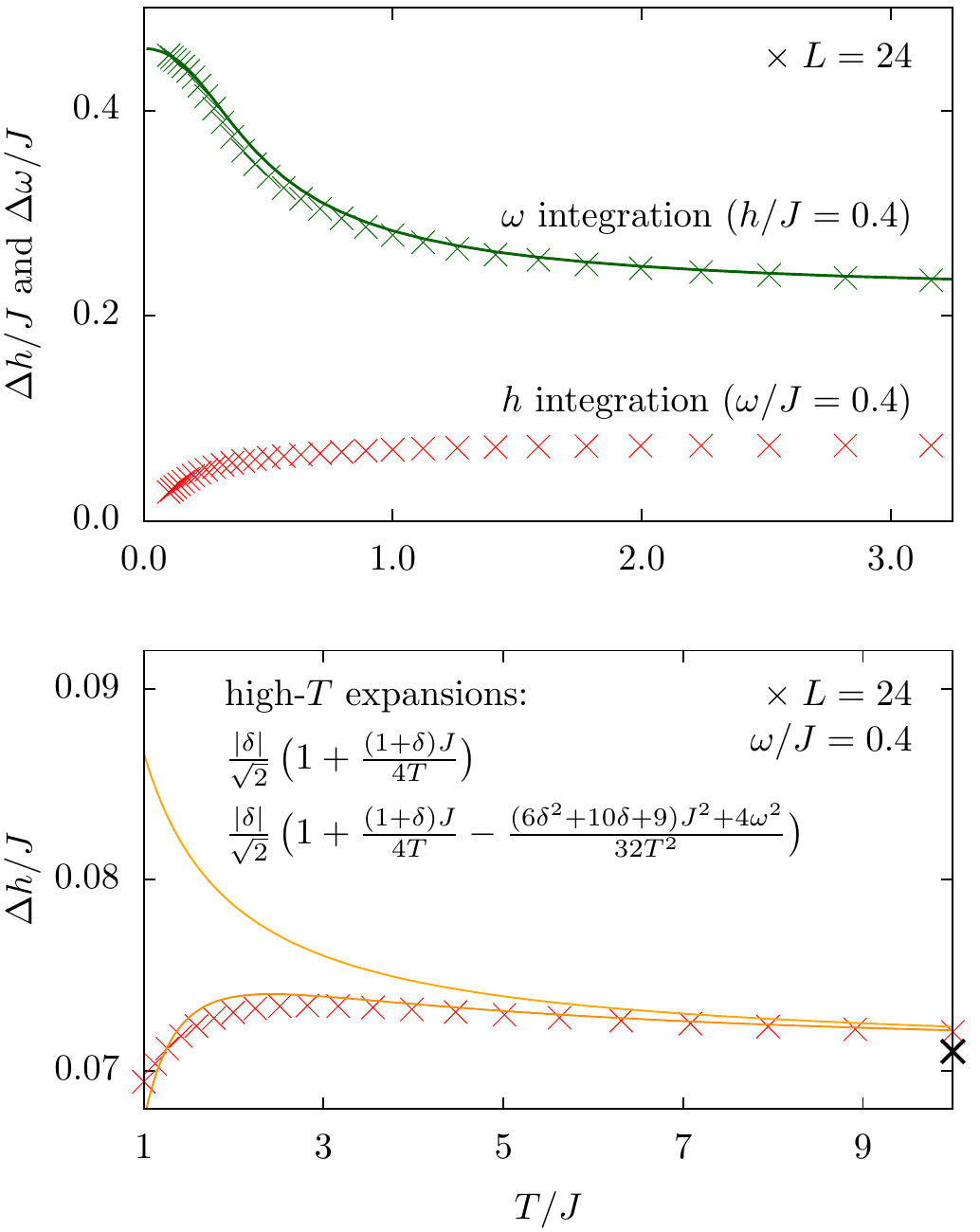}
\caption{\label{fig:omwidthcritoft} Line widths $\D \om/J$ (green) and
$\D h/J$ (red) in the critical regime at $\de = - 0.1$ as functions of the
temperature. Data from a fully numerical calculation for a finite chain Hamiltonian 
of 24 sites. The black cross in the lower panel marks the infinite-temperature
limit $|\de|/\sqrt{2}$. The solid orange lines are high-temperature
expansions of $\Delta h/J$ according to \eqref{highwidth} up to first and
second order in $J/T$.}
\end{center}
\end{figure}

In this case as well the momentum-based line width $\D h$ is initially slightly
increasing from its infinite-temperature limit $|\de|/\sqrt{2}$ when the
temperature is reduced. But our numerical data (see crosses in 
Fig.~\ref{fig:omwidthcritoft}) and the second order term of the high-temperature
expansion show that $\D h$ behaves non-monotonically and decreases again
for temperatures lower than~$J$. For small temperatures it seems to approach
zero linearly. The latter type of behavior is in accordance with field
theoretical predictions\cite{OsAf02} and experimental results.\cite{Ajiro03,KBL10}
We would like to stress, however, that there is no contradiction between
the broadening shown by the upper curve in the first panel of Fig.~\ref{fig:%
omwidthcritoft} and the narrowing shown by the lower curve. In fact, both
curves were obtained from the same numerical data set for the dynamical
susceptibility. The upper curve was calculated with \eqref{widthmom}, whereas
the lower curve was calculated by means of \eqref{shiftwidthh}. What is
important is that the upper curve can be compared with our exact results
(solid line in the upper panel). The good agreement of the crosses with
the exact curve creates confidence in our numerical data. It shows that they
are reliable when used in integrations. This is a non-trivial statement, since
our numerical data happen to be noisy and finite-size affected at low temperatures
(see Sec.~\ref{subsec:numfinite}).

We would like to point out that the resonance shift $\de \om/J$ or $\de h/J$
and the line width $\D \om/J$ or $\D h/J$ defined in terms of moments show
a simple scaling behavior. They depend on the exchange interaction only through
the ratios $T/J$ and $h/J$. In this sense the curves in Figs.~%
\ref{fig:shwcritofh}-\ref{fig:omwidthcritoft} are universal.

The method of moments is not only useful for the integrable Heisenberg-Ising
chain. It may be applied to non-integrable spin chains and to two- and
three-dimensional models as well. The field-dependent moments and the
corresponding shifts and widths may be accurately calculated by approximate
methods, since they are determined by static short-range correlation functions.
For a discussion of the numerical calculation of the moments in one dimension
see below.
In any case, the frequency-dependent moments are harder to obtain,
since they require the calculation of an infinite number of static correlation
functions.

\subsection{Integrated intensity}\label{sec:IntegInten}
An important quantity in experiments of electron spin resonance is the
integrated intensity.\cite{TyGu06} As in the definition of the moments
we may either integrate over the frequency or over the magnetic field.
For fixed magnetic field $h$ our definition \eqref{defm} of the moments
$m_0$ and $m_1$ implies that
\begin{equation}\label{eq:IntegInten}
     \int_{-\infty}^{\infty}d\om\, \frac{\om}{2}\chi_{+-}''(\om,h)
        = \pi\left(J m_1 + h m_0\right)\epp
\end{equation}

As explained in the previous section, in usual ESR experiments the absorbed
intensity is measured as function of $h$ for fixed microwave frequency $\om$.
By definition the corresponding integrated intensity is
\begin{equation}\label{def:IntegInten_h}
      I_{+-}^{\rm (int)} (\om)
         = \int_{-\infty}^{\infty}dh\,\frac{\om}{2}\chi_{+-}''(\om,h)
	 = \pi\om\, M_0(T,\om)\epp
\end{equation}
In the paramagnetic regime, where $h$ is large compared to $J$, the integrated
intensity is proportional to the magnetization $m(T,\om)$.
The high-temperature expansion of the frequency-dependent moment $M_0$,
\begin{equation}
      M_0(T,\om)=\frac{h}{4T} + \frac{J h}{8T^2}(\de-1) + \dots \epc
\end{equation}
following from \eqref{Mom}, provides another means to determine the anisotropy
$\de$ from experimental data.

\section{Exact line shapes}
As we have seen the method of moments allows us, at least in the field-%
dependent case, to obtain exact characterizations of the resonance
shift and the line width for arbitrary temperatures, magnetic fields
and anisotropy parameters for the infinite chain. Unfortunately, it
does not teach us much about the actual line shapes. Even the most
elementary question, how many peaks the line comprises, remains generally
unanswered.

In the remainder of this work we shall try to draw at least
a qualitative picture of how the lines are shaped by considering
all available limiting cases, where exact results are known, and by
complementing these with numerical data. In this section we review
those limiting cases where exact results are known. In the following
section we describe our numerical calculations. Finally, in
Sec.~\ref{sec:two-spinon}, we shall consider the line shapes for $\de > 0$
and $T = h = 0$ in two-spinon approximation.

\subsection{Heisenberg limit}\label{heisenberglimit}
The simplest case where we know the line shape exactly is the isotropic
case $\de = 0$.\cite{OsAf99,*OsAf02} It may be called the Heisenberg
limit of the Heisenberg-Ising chain. In this case $H$ is still a complicated
many body Hamiltonian, but $S^+$ commutes with $H$ and the time evolution
of $S^+$ is driven by $S^z$ alone (see App.~\ref{app:linres}). Thus,
$S^+ (t) = \re^{- \i ht} S^+$, and
\begin{equation} \label{intxxx}
     I (\om, h) = \p \de(\om - h) h\, m(T, h) \epp
\end{equation}
This means that there is a single sharp peak, and the absorbed intensity
is proportional to the magnetic energy $h\, m(T,h)$ per lattice site.
This case includes the familiar paramagnetic resonance (Zeeman effect)
for which the magnetization is known explicitly, namely $m(T,h) =
\2 \tgh \bigl( \frac{h}{2T} \bigr)$ for $J = 0$. In the general case
the magnetization must be calculated from solutions of linear and
non-linear integral equations.\cite{Kluemper93} In our context we infer
from \eqref{m0} and \eqref{ms3} that $m(T,h) = m_0 = - \2 \ph_{(0)}$,
i.e.~the lowest moment $m_0$ alone characterizes the line shape.

\subsection{Ising limit}\label{isinglimit}
The only other limiting case in which the line shape is known for all
temperatures and magnetic fields is the Ising limit.\cite{ShAd81}
For the Ising limit we replace the Heisenberg-Ising Hamiltonian $H \rightarrow
H/\de$ and send $\de \rightarrow \infty$. Then
\begin{equation} \label{hising}
     H \rightarrow H_I = J \sum_{j=1}^L s_{j-1}^z s_j^z \epp
\end{equation}
Due to \eqref{allmoments} this implies that the moments are replaced as
$m_n \rightarrow \de^{- n} m_n$ for $\de \rightarrow \infty$.

In the Ising limit the time evolution of $S^+$ in \eqref{defchi} can be calculated
explicitly. As we show in App.~\ref{app:isinglim} this leads to the formula
\begin{multline} \label{chiising2}
     \frac{1}{2 \p} \chi_{+-}'' (\om, h) = \2 (m_2 - m_1) \de(\om - h + J) \\
        \mspace{-3.mu}
        + (m_0 - m_2) \de(\om - h) + \2 (m_2 + m_1) \de(\om - h - J)
\end{multline}
for the dynamical susceptibility.

Using this formula we can calculate the moments $m_n$ by means of 
\eqref{defm} and verify its consistency. Integrating over $\om$ we obtain
indeed $m_0$. The integrals for the higher moments yield
\begin{equation}
     m_n =
        \begin{cases}
	   m_1 & \text{if $n \in {\mathbb N}$ is odd} \\
	   m_2 & \text{if $n \in {\mathbb N}$ is even} \epc
        \end{cases}
\end{equation}
i.e. there are no `new moments' for $n > 2$. The three independent moments
$m_0$, $m_1$ and $m_2$ correspond to the three $\de$-peaks in the dynamical
susceptibility. They can be calculated by means of the $2 \times 2$
transfer matrix\cite{Babook} of the Ising chain. Explicit expressions are
shown in App.~\ref{app:isinglim}.

The Ising limit is important for us, since it is easy to interpret and since
it provides a physical picture for the massive phase. The eigenstates of
the Ising chain Hamiltonian are tensor products of local $s^z$ eigenstates.
We infer from the spectral representation, App.~\ref{app:spectralrep}, that
transitions can occur only between states which differ by a single flipped
spin. Thus, the following transitions in which the chain absorbs the energy
$\D E$ are possible:
\begin{align*}
     & \cdots \uparrow \downarrow \uparrow \cdots 
       \rightarrow \cdots \uparrow \uparrow \uparrow \cdots, \qd \D E = J - h \\
     & \cdots \downarrow \uparrow \downarrow \cdots 
       \rightarrow \cdots \downarrow \downarrow \downarrow \cdots, \qd \D E = J + h \\
     & \cdots \uparrow \uparrow \downarrow \cdots 
       \rightarrow \cdots \uparrow \downarrow \downarrow \cdots, \qd \D E = h
\end{align*}
The first two correspond to the creation of a pair of domain walls (or spinons)
in one of the N\'eel ground states. The third one is impossible in the ground
states, whence the coefficient $m_0 - m_2$ in front of the corresponding term
in \eqref{chiising2} must vanish at zero temperature.

At finite $\de > 0$ the operator $S^-$ in the spectral representation
\eqref{chispec} does no longer induce transitions between eigenstates. The
three $\de$-peaks in \eqref{chiising2} broaden which reflects the onset
of interactions between the spinons. Still, we believe that three peaks are
characteristic of the massive phase, $\de > 0$, at least at not too large
magnetic fields. This is in accordance with our numerical data and with previous
numerical work.\cite{OgMi03,MYO99,*ECM10} The relative height of the three
peaks is still approximately well described by the relative prefactors
of the $\delta$-peaks in \eqref{chiising2}. At $T = 0$ in the two-spinon
approximation (see below) the central peak vanishes due to the same intuitive
argument as given above.

\begin{figure}[!ht]
\begin{center}
\includegraphics[width=.45\textwidth]{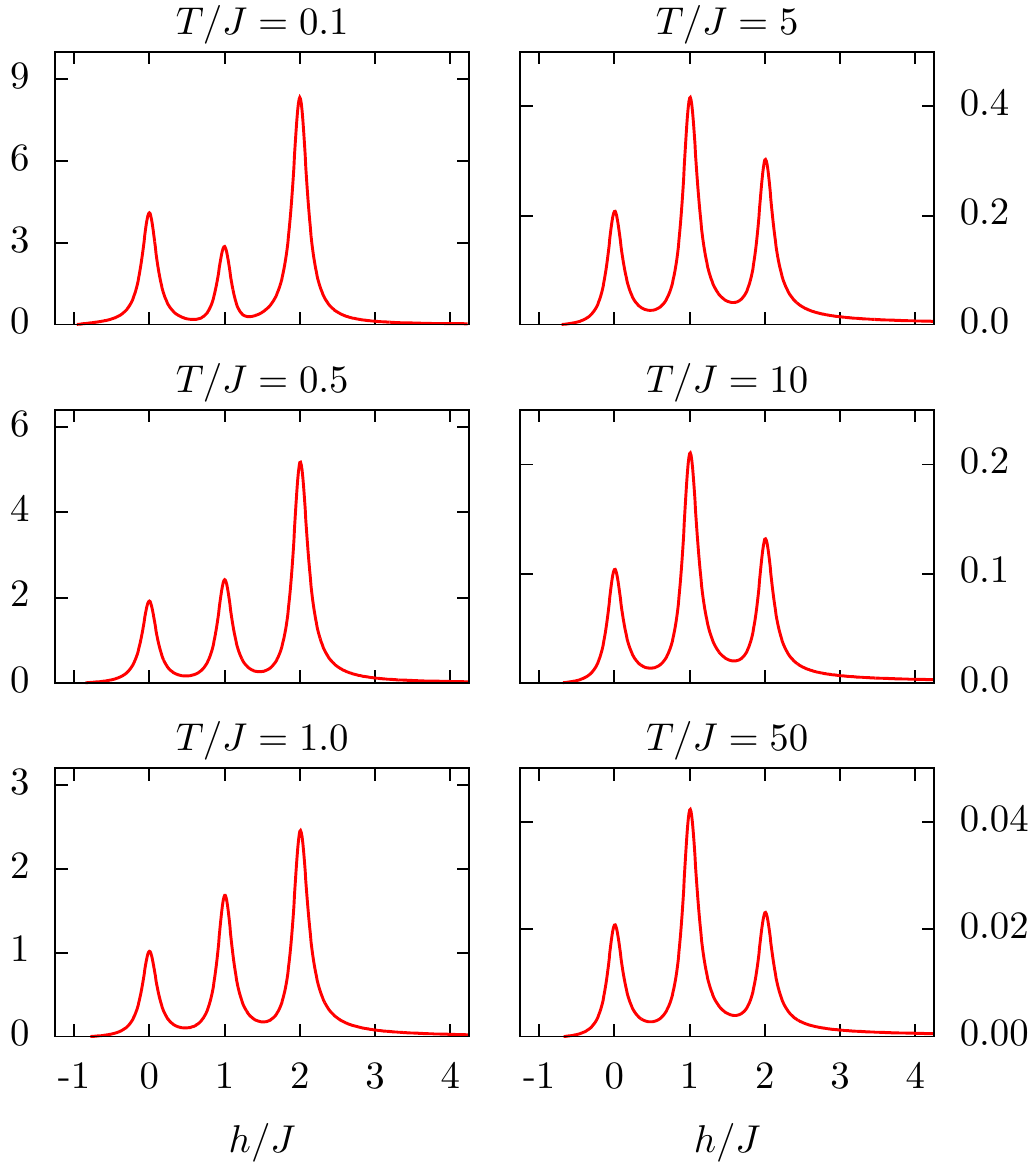}
\caption{\label{fig:Ising_chis_of_om_Lorentz} Dynamical susceptibility
$J\chi_{+-}''$ from \eqref{chiising2} as function of $h/J$ for $\om=J$.
The $\de$-peaks are convolved with a Lorentzian \eqref{Lorentz} with parameter
$\epsilon=0.12\,J$.}
\end{center}
\end{figure}

Figure~\ref{fig:Ising_chis_of_om_Lorentz} shows the dynamical susceptibility
\eqref{chiising2} in the Ising limit as a function of $h/J$. We visualize
the $\delta$-peaks by convolving them with a Lorentzian of the form
\begin{equation}\label{Lorentz}
     \mathcal{L}(h)=\frac{J}{\pi}\frac{\epsilon}{h^2+\epsilon^2}\epp
\end{equation}
While in the low-temperature limit ($T/J=0.1$) the right peak at $h =
\om + J$ is the highest, in the intermediate-tem\-pera\-ture regime ($T/J
= 0.5,\, 1.0,\, 5$) the relative height of the central peak at $h = \omega$
increases rapidly with increasing temperatures. For even higher temperatures
($T/J = 5,\, 10,\, 50$) the relative height of the left peak at $\om=h-J$
increases as well and reaches values comparable to those of the right peak.
The relative heights of the central peak and the right peak are compatible
with the numerical data for $\de = 1$ shown in Fig.~8 of
Ref.~\onlinecite{OgMi03}. The absolute value of the height of the central
peak differs approximately by a factor of $8$, because the authors of
Ref.~\onlinecite{OgMi03} plot the function $\chi_{xx}''= (\chi_{+-}''
+ \chi_{-+}'')/4$ instead of $J\chi_{+-}''$ and scale it with a factor of
$1 + \delta = 2$. Concerning the location of the three peaks as well
as the relative and the absolute heights of the central peak and the right
peak, the curves in Fig.~8 of Ref.~\onlinecite{OgMi03} can be qualitatively
explained by the exact result \eqref{chiising2} for the dynamical susceptibility
in the Ising limit.

If we use \eqref{chiising2} and the formulae \eqref{IsingExplExp} for the
moments $m_1$ and $m_2$, we can explicitly perform the infinite-temperature
limit and obtain the function $\phi(\om-h)$ defined in the next section in
\eqref{hightchi} and \eqref{defphi}. We find that its $\de$-peaks are weighted
by $1/2$ for the central peak and by $1/4$ for the two side-peaks. 

%

\subsection{High-temperature limit}\label{XXmodel}
In the infinite-temperature limit the dynamical susceptibility $\chi_{+-}''$
vanishes identically. This follows, for instance, from Eq.~\eqref{chialt}.
From the same equation and from the sum rule \eqref{sumalt} we obtain the
leading high-temperature contribution to $\chi_{+-}''$,
\begin{equation} \label{hightchi}
     \chi_{+-}'' (\om, h)
        = \frac{\om \p}{2T} \Ph (\om - h) + \mathcal{O} (T^{-2}) \epc
\end{equation}
where
\begin{equation} \label{defphi}
     \Ph (\om) = \frac{2^{-L}}{\p L} \int_{- \infty}^\infty \rd t \;
        \re^{\i \om t} \tr \bigl\{ (\re^{\i t \ad_H} S^+) S^- \bigr\} \epp
\end{equation}
The function $\Ph$ is even, non-negative and normalized. Hence, it may be
interpreted again as a distribution function. 

The function $\Ph$ is simpler as compared to $\chi_{+-}''$. Still, in general,
we are unable to calculate it exactly. In App.~\ref{app:shorttime} we comment on
the small-$t$ expansion of the integrand, which we have calculated up
to the order $t^{38}$, and draw some conclusions. In the free Fermion case
$\de = - 1$ it is possible to calculate it to all orders. The terms sum up
to a Gaussian,\cite{BrJa76} and
\begin{equation}
     \Ph (\om) = \frac{\re^{- (\om/J)^2}}{J \sqrt \p} \epp 
\end{equation}

From this explicit result we can calculate the field- and frequency-dependent
line widths of the previous section,
\begin{equation}
     \frac{\D \om^2}{J^2} = \frac{3/2 + 2(h/J)^4}{\bigl(1 + 2(h/J)^2\bigr)^2} \epc \qd 
     \frac{\D h}{J} = \frac{1}{\sqrt{2}} \epp
\end{equation}
The second equation is in accordance with the high-\-tem\-pera\-ture result
\eqref{highwidth}. Our explicit example clearly shows that $\chi_{+-}''$ is
asymmetric in $h$ and $\om$ and that the two line widths $\D \om$ and $\D h$
are rather different quantities.

\begin{figure}[!ht]
\begin{center}
\includegraphics[width=.45\textwidth]{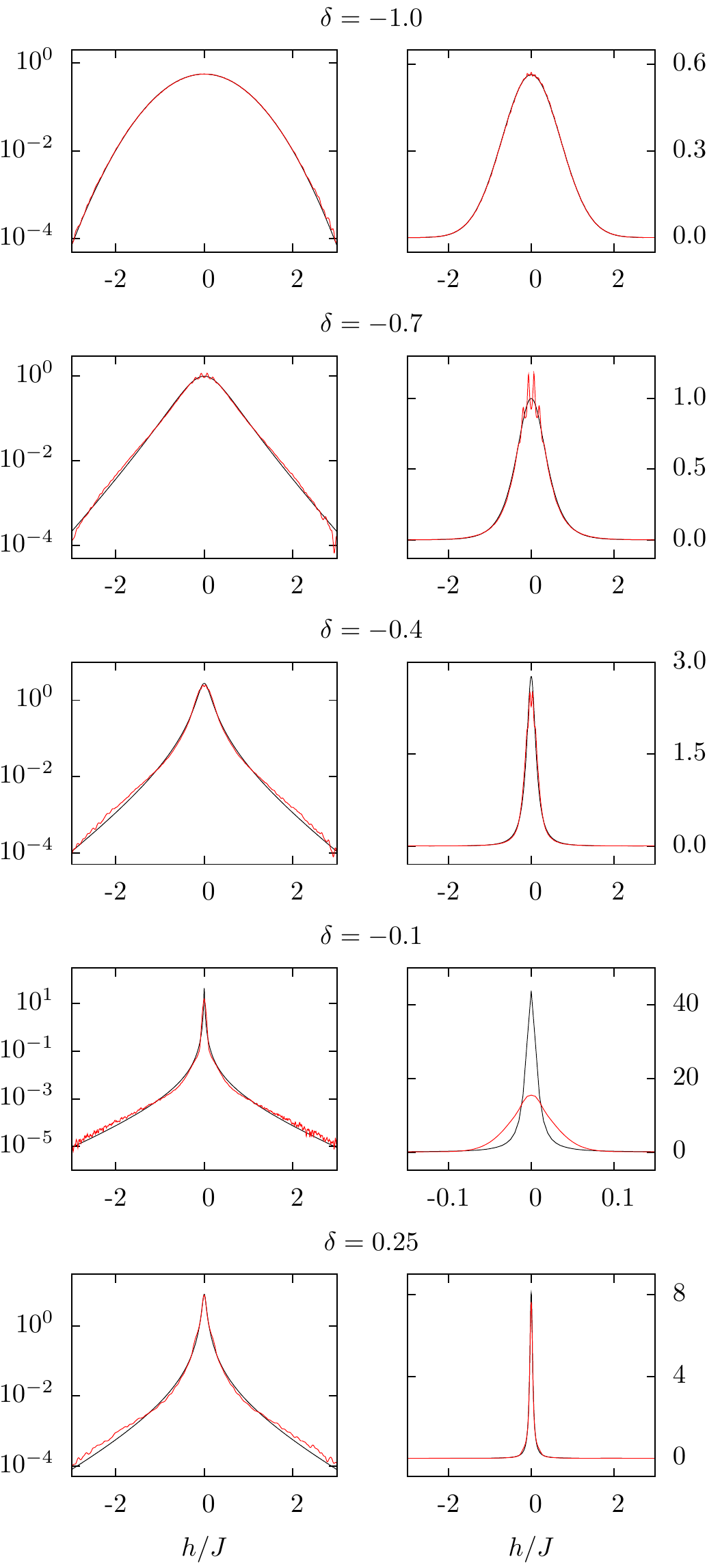}
\caption{\label{fig:highline} Normal-inverse Gaussian (black lines) as a model
for the high-temperature line shape, comparison with numerical data (red
lines). Parameters of the normal-inverse Gaussian as calculated in \eqref{nigpams}.
All panels show $J\chi_{+-}''(\om,h)$ as a function of $h/J$, left panels
logarithmic scale, right panels linear scale. Note the different scale on the x-axis in
the right panel for $\de = - 0.1$. In general the numerical data were obtained
for $T/J = 100$, $\om/J = 0.4$, $L = 16$, and $M = 1024$ (see below). For $\de =
- 0.1$ the chain length was increased to $L = 20$ and the resolution to $M = 4096$.}
\end{center}
\end{figure}

We learn from the above discussion that $2T \chi_{+-}'' (\om, h)/\p \om$ is a
`good function'. It converges to a normalized function which depends only on
the difference $\om - h$ for $T \rightarrow \infty$. In the free Fermion
case $\de = - 1$ and in the isotropic case this function has a single peak
at $\om = h$. From our numerical data (see Fig.~\ref{fig:highline}) we see
that this seems to be true for all values of $\de$ between $- 1$ and $0$ and
even for small positive $\de$. The Gaussian decay for large $\om$ seems to be
peculiar of the free Fermion point. At all values of $\de$ which are larger
than $-1$ our logarithmic plots in Fig.~\ref{fig:highline} indicate an
exponential decay.

When looking for a simple model for such type of line shape we found the
so-called `normal-inverse Gaussian',
\begin{equation}\label{def:NIG}
     \mathcal{N} (x |\a, \be)
        = \frac{\a \be \re^{\a \be} K_1 (\a \sqrt{x^2 + \be^2})}
	       {\p \sqrt{x^2 + \be^2}} \epc \qd \a, \be > 0 \epc
\end{equation}
where $K_1$ is a modified Bessel function. It becomes a Gaussian in the limit
$\a \rightarrow \infty$, a Lorentzian for $\a \rightarrow 0$  and a
$\delta$-function for $\be \rightarrow 0$. Its moments can be easily calculated
from its characteristic function
\begin{equation}
     \widetilde{\mathcal{N}} (k|\a, \be)
        = \int_{- \infty}^\infty \mspace{-9.mu} 
	       \rd \om \; \re^{\i k x} \mathcal{N} (x |\a, \be)
	= \re^{\be(\a - \sqrt{k^2 + \a^2})} \epp
\end{equation}
For instance,
\begin{equation}
     \<x^2\>_\mathcal{N} = \frac{\be}{\a} \epc \qd
     \<x^4\>_\mathcal{N} = 3 \biggl( \frac{\be^2}{\a^2} + \frac{\be}{\a^3} \biggr)
        \epp
\end{equation}%

This can be compared with the dimensionless moments of the distribution function
$\Ph$, which follow from \eqref{defm} and \eqref{hightchi},
\begin{equation}
     \frac{\<\om^{2n}\>_\Ph}{J^{2n}}
        = \lim_{T \rightarrow \infty} \frac{4T}{J} m_{2n-1} (T, h) \epp
\end{equation}
Here the right hand side can be easily calculated. If we demand that
\begin{subequations}
\begin{align}\label{momsNIG}
     \<x^2\>_\mathcal{N} & = \frac{\<\om^2\>_\Ph}{J^2} = \frac{\de^2}{2} \epc \\
     \<x^4\>_\mathcal{N} & = \frac{\<\om^4\>_\Ph}{J^4}
        = \frac{\de^2}{2} \biggl( \frac32 + \de + \de^2 \biggr) \epc
\end{align}
\end{subequations}
we obtain
\begin{equation} \label{nigpams}
     \a = \sqrt{\frac{6}{(1 + \de)(3 - \de)}} \epc \qd
     \be = \frac{\de^2}{2} \sqrt{\frac{6}{(1 + \de)(3 - \de)}} \epp
\end{equation}
Figure~\ref{fig:highline} compares the normal-inverse Gaussian with
parameters \eqref{nigpams} with our numerical high-temperature line shapes.

For $\delta = - 1$ the model line shape is exact. This is no longer
the case for $\delta > - 1$ which can be seen by comparing the sixth normalized
moment of $\Ph$ with the sixth moment of $\mathcal{N}$. Still, we find it remarkable
how well the model line shape fits our numerical data. Especially the exponential
tails visible in the left panels of Fig.~\ref{fig:highline} have not been fitted
to the numerical data. The good agreement comes out automatically. On the
other hand, the deviation of the center of the peak in the right panel
for $\de = - 0.1$ in Fig.~\ref{fig:highline} does not seem to be due to a
resolution problem of our numerical calculation. We rather attribute it
to a slight mismatch of the normal inverse Gaussian at small anisotropy.

If we fix the parameters $\alpha$ and $\beta$ of $\mathcal{N}(x|\alpha,\beta)$
according to \eqref{nigpams}, we see from \eqref{momsNIG} that the width,
calculated by its second moment, behaves as $\sqrt{\<x^2\>_\mathcal{N}}
\sim |\delta|$. On the other hand, it is easy to calculate the peak-to-peak
width of the normal-inverse Gaussian, which is the distance of its two inflection
points. Setting the second derivative $\partial_x^2\mathcal{N}(x|\alpha,\beta)$
to zero and using the differential equation defining $K_1$,
\begin{equation}
     y^2K_1''(y)=(y^2+1)K_1(y)-y K_1'(y)\epc \quad y=\alpha \sqrt{x^2+\beta^2}\epc
\end{equation}
we see that the positive inflection point $y_{0}$ is the solution of the equation
\begin{equation}
  \frac{1}{y}+\frac{y(y^2-(\alpha\beta)^2)}{3y^2-4(\alpha\beta)^2}
     =\partial_y\ln{\left(K_1(y)\right)}\epp
\end{equation}
In principle this algebraic equation can be solved numerically, but for our purposes
the following argument leading to an estimation for the inflection point is sufficient.
Since $\partial_y\ln{(K_1(y))} < -1$ for all $y\geq 0$ and the left-hand side
becomes large and positive for $y\to 0,\infty$, the solution $y_{0}$ has to be
located close to the left of the pole. Hence, we obtain
\begin{equation}
     y_{0}\lesssim \sqrt{\frac{4}{3}}\alpha\beta
        \quad \Rightarrow \quad x_{0} \lesssim \frac{\beta}{\sqrt{3}}\epp
\end{equation}
An upper limit of the peak-to-peak width is therefore given by
\begin{equation}\label{peak-to-peak_width}
     \frac{\Delta_{\text{pp}}x}{J}
        = 2x_0 \lesssim \delta^2\sqrt{\frac{2}{(1+\delta)(3+\delta)}} \sim \delta^2 \epp
\end{equation}
Accordingly, the normal-inverse Gaussian is an example of a distribution for
which the width calculated by its second moment ($\sim |\de|$) and the
peak-to-peak width ($\sim \de^2$) behave asymptotically differently for small
anisotropies $\de$ and can therefore differ strongly in value. Fitting experimental
data at high temperatures to a normal-inverse Gaussian, this offers a way to determine
$\de$ independently of the exchange integral $J$ from the ratio of the two line widths,
\begin{equation}\label{ratio_width}
     \frac{\Delta_{\text{pp}}x}{\sqrt{\<x^2\>_\mathcal{N}} }
        \approx \frac{2|\delta|}{\sqrt{(1+\delta)(3+\delta)}}\epp
\end{equation}

If we estimate the peak-to-peak width of Fig.~6 of Ref.~\onlinecite{Exp_LiCuVO_01}
to $\Delta_{\text{pp}}h \approx 2\,\text{kOe}\; \widehat{=}\; 0.27\, K$, we can
solve \eqref{peak-to-peak_width} numerically ($J=22\,K$) and obtain $\delta
\approx -0.12$. This value is compatible with the prediction of Maeda et
al.\cite{MSO05} ($\delta=-0.15$) obtained from a fit for the resonance shift
based on the data of Ref.~\onlinecite{Exp_LiCuVO_01}. The authors of
Ref.~\onlinecite{Exp_LiCuVO_02} obtain for the same material as in
Ref.~\onlinecite{Exp_LiCuVO_01} ($\text{LiCuVO}_4$) the exchange integral $J=30\, K$.
From Fig.~4 therein we can read off $\Delta_{\text{pp}}h=1.5\, \text{kOe}$, and
obtain together with Eq.~\eqref{peak-to-peak_width} the anisotropy $\delta\approx -0.088$
which is compatible with the value $J_{zz}/J \approx -2\, K / 30\,K \approx -0.067$
of Ref.~\onlinecite{Exp_LiCuVO_02}.

\section{Numerical line shapes}
\label{sec:numerics}
The numerical approach we are using for the calculation of the
dynamical susceptibility $\chi_{+-}''(\om, h)$, Eq.~\eqref{defchi},
has been described in detail elsewhere.~\cite{We04,*WWAF06,*WF08b} We
will therefore give only a short outline of the method and discuss a
few special tricks beneficial for the present project.

Starting point of the numerics for finite chains is the spectral
representation (see also App.~\ref{app:spectralrep}),
\begin{align}
  \chi_{+-}'' (\om, h) & = \frac{\p}{L Z} \sum_{m, n}
  \bigl( \re^{- E_n/T} - \re^{- E_m/T} \bigr) \nonumber{}\\ 
  \label{chispec:alex}
  & \qquad \times |\<m|S^-|n\>|^2 \de(\om - E_m + E_n)\\
  & = \frac{\pi}{L Z} \int_{-\infty}^{\infty} \rd y \:
  s(y+\omega,y)\,\bigl( \re^{- y/T} - \re^{- (y+\omega)/T} \bigr)\,,
  \nonumber{}
\end{align}
which can be written as an integral over thermal weighting factors
and the temperature-independent function
\begin{equation}
  s(x,y) = \sum_{m, n} |\<m|S^-|n\>|^2 \de(x - E_m) \de(y - E_n)\,.
\end{equation}

At first sight the calculation of this function seems to require
knowledge of all eigenvectors and eigenvalues of the Heisenberg-Ising
Hamiltonian on a finite lattice. However, it is significantly more
efficient to rescale all energies $E\to \tilde E=a E + b$, such that
$\tilde E\in[-1,1]$, and to expand $s(\tilde x,\tilde y)$ in terms of
Chebyshev polynomials of the first kind $T_i$,
\begin{equation}\label{chebsers}
  s(\tilde x,\tilde y) = \sum_{i,j=0}^{M-1} 
  \frac{\mu_{ij} g_i g_j (2-\de_{i0})(2-\de_{j0}) 
    T_i(\tilde x)T_j(\tilde y)}{\pi^2\sqrt{(1-\tilde x^2)(1-\tilde y^2)}}\,.
\end{equation}
The problem then reduces to the calculation of the expansion coefficients
$\mu_{ij}$ which are given by traces,
\begin{equation}
  \mu_{ij} 
  = \iint_{-1}^{1} \rd x \rd y \: s(x,y)\,T_i(x) T_j(y)
  = \tr [S^+ T_i(\tilde H) S^- T_j(\tilde H)]\,.
\end{equation}
Instead of summing over the whole Hilbert space, these traces are well
approximated by averages over a few random states. The action of
$T_k(\tilde H)$ on an arbitrary state can be quickly evaluated with
the recursion relations of the Chebyshev polynomials. Taking into account
symmetries of the Hamiltonian ($S^z$-conservation and translation), it
is thus feasible to calculate the $\mu_{ij}$ for systems of up
to $L=32$ lattice sites and expansion orders up to $M=4096$ on average
hardware.

Once we have a complete set of expansion coefficients $\mu_{ij}$ for a
given lattice size $L$, anisotropy $\delta$ and all $S^z$ sectors, we obtain
$s(x,y)$ from Eq.~\eqref{chebsers} using fast Fourier methods. In
Eq.~\eqref{chebsers}, the damping factors $g_k$ cure the Gibbs
oscillations inherent to truncated Chebyshev (and Fourier) expansions,
and ensure good convergence properties (see
Ref.~\onlinecite{We04,*WWAF06,*WF08b} for details). Given $s(x,y)$ we
can calculate $\chi_{+-}'' (\om, h)$ for all temperatures $T$,
frequencies $\omega$ and magnetic fields $h$ via straightforward
numerical integration. Changing any of these three parameters does
\emph{not} require a new Chebyshev expansion, which is the most
time-consuming part of the simulations.

A little more care is required for low temperatures. Here the
Boltzmann factors put most of the weight on very few states at the
lower edge of the spectrum. The sums in Eq.~\eqref{chispec:alex}
should then be split up into contributions from these low-energy
states and from the rest of the spectrum, and the low-energy
eigenstates should be calculated exactly with Lanczos
recursion.~\cite{La50}  For the data in Figs.~\ref{fig:shwcritofh}
and~\ref{fig:shwcritoft} we separated two states per $S^z$ and
momentum sector from the rest of the spectrum. This procedure does not
increase the overall computation time, but the book keeping is
slightly more elaborate.

Another trick improves the precision of the numeric estimates of the
moments $m_n(T,h)$ and $M_n(T,\omega)$ of $\chi_{+-}''(\om, h)$,
Eqs.~\eqref{defm} and~\eqref{defM}, and of the resonance shifts and
line widths. Since the expansion coefficients $\mu_{ij}$ are based on
averages over random vectors, they are subject to a low level of noise
which is carried over to $\chi_{+-}''(\om, h)$. Even though the error
is hardly visible in $\chi_{+-}''(\om, h)$ itself, it is amplified when
$\chi_{+-}''(\om, h)$ is multiplied by powers of $(\omega-h)$ in the
course of the moment integration. Large values of $(\omega-h)$ taken
to the power $2$ or $3$ then induce noticeable errors in $m_n(T,h)$ or
$M_n(T,\omega)$. This can be avoided by doing the multiplication with
$(\omega-h)$ in the space of Chebyshev moments. Consider a
one-dimensional Chebyshev expansion of a function
$f:[-1,1]\to\mathbb{R}$, where the expansion coefficients are given by
\begin{equation}
  \mu_i = \int_{-1}^{1} \rd x \: f(x)\, T_i(x)\,.
\end{equation}
Then, the expansion coefficients of $x f(x)$ are
\begin{align}
    \tilde\mu_i & = \int_{-1}^{1} \rd x \: x f(x)\, T_i(x)
    = \int_{-1}^{1} \rd x \: f(x)\, T_1(x) T_i(x) \notag \\
    & = \frac{1}{2}\int_{-1}^{1} \rd x \: f(x)\,(T_{i+1}(x) + T_{i-1})
    = \frac{1}{2} (\mu_{i+1} + \mu_{i-1})\,.
\end{align}
Hence, multiplication of the expanded function with the independent
variable corresponds to taking a kind of mean value in the space of
expansion coefficients. The application of this procedure to the
two-dimensional expansion required for $(\omega-h)^n \chi_{+-}''(\om,
h)$ leads to a cancellation of noise and to much better estimates of
the moments $m_n(T,h)$, $M_n(T,\omega)$, and of the resonance shifts
and widths.

\section{Two-spinon line shapes}\label{sec:two-spinon}
Spectral representations such as \eqref{chispec}, \eqref{chispec0} have been
used in the past as a starting point for the approximate calculation of
dynamical correlation functions. For the ground state case, Eq.~\eqref{chispec0},
approximate calculations can be built upon the partial summation of matrix
elements of local operators between the ground state and excited states,
so-called form factors, since these are sometimes known exactly. Such
type of procedure is efficient if sub-classes of form factors can be identified
which contribute dominantly to the considered correlation function.

For $- 1 < \D := 1 + \de \le 1$ the Heisenberg-Ising chain is at a critical
point in the ground state. In this case all form factors vanish algebraically
in the thermodynamic limit.\cite{AKMW06,KKMST09b,KKMST11a} Thus, the summation
of form factors and the thermodynamic limit do not commute. In this case
good results for the dynamic structure factor were obtained from a summation
of form factors for finite chains,\cite{BKM02a,BKM03,SST04,CaMa05} which
requires a considerable amount of numerical calculation, though. More recently,
an exact summation of the leading contribution to the large distance
asymptotics of two-point functions was obtained in Ref.~\onlinecite{KKMST11b}.

In the massive phase at $\D > 1$ the situation is mathematically less
involved. Certain classes of multi-spinon form factors stay finite in the
thermodynamic limit.\cite{JiMi95} In calculations of the dynamic structure
factor,\cite{BKM98} it turned out that the two-spinon contribution is
always dominant at $T = h = 0$. Here we use the results of Refs.~%
\onlinecite{BCK96,BKM98,CMP08} to discuss the two-spinon contribution to
the dynamical susceptibility for $\D > 1$. As we shall see, the dynamical
susceptibility is dominated by the two-spinon states only if $\D$ is large
enough. For $\Delta > 3/2$ the two-spinon contribution $I^{(2)}(\om)$ to
the absorbed intensity amounts to the main part of the total intensity
$I(\om,h=0)$, but it is marginal in the isotropic limit $\Delta \to 1$.
In the Ising limit $\Delta \to \infty$ both intensities are identical,
$I^{(2)}(\om) = I(\om,h=0)$, and the result of Sec.~\ref{isinglimit} is
reproduced.

A comparison of the two-spinon line shapes with numerical line shapes
calculated for finite chains shows the high quality of our numerical data.

\subsection{Line shape}\label{sec:two-spinon line shape}
The dynamical susceptibility $\chi_{+-}''$ decomposes into two terms,
one for positive and the other one for negative frequencies. In order
to calculate these terms separately we define the function
\begin{equation}
 \chi(\om)=\frac{1}{2L} \int_{-\infty}^\infty \rd t \:
        \re^{\i \om t} \bigl\<S^+ (t)S^-\bigr\>_T \epp
\end{equation}
Using the invariance of the Hamiltonian \eqref{ham} under spin flip we obtain
\begin{equation} 
     \chi_{+-}'' (\om, h=0) = \chi(\om)-\chi(-\om)\epc
\end{equation}
where $\chi(\om)$ vanishes for $\om<0$ and is non-negative for $\om\geq 0$.

The space of excited states decomposes into scattering states of an even 
number of 2n spinons.\cite{JKM73} The precise mathematical structure of the 
space of states of the infinite XXZ chain in the massive phase was identified in 
Refs.~\onlinecite{JMMN92,JiMi95}. In order to utilize the results of
Ref.~\onlinecite{JiMi95} we have to adapt our conventions. The Hamiltonian
\begin{equation}\label{hamJM}
     H_{JM}=-J \sum_{j= -\infty}^\infty \bigl( s_{j-1}^x s_j^x + s_{j-1}^y s_j^y
            - \Delta\, s_{j-1}^z s_j^z \bigr)
\end{equation}
used in Ref.~\onlinecite{JiMi95} is related to our Hamiltonian \eqref{ham}
by a unitary transformation $s_j^{x,y} \mapsto (-)^j s_j^{x,y}$
and $s_j^{z} \mapsto s_j^{z}$. Under this transformation the dynamical
susceptibility turns into
\begin{equation}\label{chi2spinon}
     \chi(\om)= \frac{1}{4}\sum_{\substack{ k=-\infty \\ j=0,1}}^{\infty}
                \int_{-\infty}^\infty \rd t \:
		\re^{\i \om t} {_{j}}\< 0|(-)^k s_k^+ (t)s_0^-|0\>_{j} \epc
\end{equation}
where the time evolution in $s_k^+ (t)$ has to be evaluated by means of the
Hamiltonian $H_{JM}$ instead of $H$, and where the states $|0\rangle_{0,1}$
are the two degenerate ground states of $H_{JM}$. Here we have used $S^\pm
= \sum_k s_k^\pm$, the invariance of the Hamiltonian under translations
as well as the unitary transformation defined above. 

Now we insert the resolution of the identity into multi-spinon states,\cite{JiMi95}
\begin{multline}\label{idJM}
     id_\mathcal{F}=\sum\limits_{j=0,1}\sum\limits_{n\geq 0}
                    \sum\limits_{\eps_n,\ldots,\eps_1}\frac{1}{n!}
		    \oint\frac{d\xi_n}{2\pi \i \xi_n}\dots
		    \oint\frac{d\xi_1}{2\pi \i \xi_1}\times\\
        \times|\xi_n\ldots\xi_1\>_{\eps_n\ldots\eps_1,j}\,
	{_{j,\eps_1\ldots\eps_n}}\<\xi_1\ldots\xi_n| \epc
\end{multline}
in between the spin operators in \eqref{chi2spinon}
and consider only the term with $n=2$ which is the two-spinon contribution.
The subindices $j$ refer to the two ground-state sectors and the $\eps_\ell$
are spin indices labeling the scattering states of even numbers of spinons.
In this language the two-spinon contribution is
\begin{multline}\label{chi2spinonid}
     \chi^{(2)}(\om) =  \frac{1}{8}\sum_{\substack{ j,j'=0,1 \\ \eps_1,\eps_2=\pm}}
        \sum\limits_k\int_{-\infty}^{+\infty} \rd t \:\re^{\i \om t}
	\oint\prod_{i=1}^{2}\frac{d\xi_{i}}{2\pi\i \xi_{i}}\\
        {_{j'}}\< 0|(-)^{k}s^{+}_{k}(t)|\xi_2,\xi_{1}\>%
	_{\eps_2\eps_1,j}\, {_{j,\eps_1\eps_2}}\< \xi_1,\xi_{2}|s_0^-|0\>_{j'}\epp
\end{multline}

Very similar expressions were evaluated elsewhere.\cite{BKM98,CMP08}
The time evolution of ${_{j'}}\< 0|(-)^{k}s^{+}_{k}(t)|\xi_2,\xi_{1}\>$
as well as the remaining form factors in \eqref{chi2spinonid} were
obtained in Ref.~\onlinecite{JiMi95}. Inserting those results and calculating
the remaining sums and integrals we end up with
\begin{subequations}\label{chi2spinonend}
\begin{align}\label{chi2spinonend_onlychi}
     \chi^{(2)}(\om) &=\frac{k'}{4I}
        \frac{\Theta(1-\hat{\om})\Theta(\hat{\om}-k')}
	     {\hat{\om}\sqrt{1-\hat{\om}^2}\sqrt{\hat{\om}^2-{k'}^2}}
	     \frac{\vartheta_A^2(\theta)}{\vartheta_n^2(\theta)}\epp
\end{align}

Here the factors $\Theta$ in the numerator denote unit-step functions. The
variables $\theta$ and $\om$ are related as
\begin{equation}\label{hatomega}
     \hat{\om}=\frac{\om}{2I}
              ={\rm dn}\left(\frac{2K}{\pi}\theta\right),
	       \quad 0\leq\theta\leq\frac{\pi}{2}\epc
\end{equation}
where $\dn$ is a Jacobi elliptic function and where
\begin{equation}\label{Idef}
     I=\frac{JK}{\pi}\sh{\left(\frac{\pi K'}{K}\right)}\epp
\end{equation}
The anisotropy parameter $-(1+\de)=(q+q^{-1})/2$ is related to the nome
\begin{equation}
    q = -\exp(-\pi K^\prime/K), 
\end{equation}
and thus determines the moduli $k$, $k' = \sqrt{1-k^2}$ of the elliptic
integrals $K = K(k)$, $K' = K(k')$.
The remaining theta functions in \eqref{chi2spinonend} are standard
and defined by
\begin{align}
  \label{thetan}
  \vartheta_n(\theta) &=
     \frac{\vartheta_4(\theta,p)}{\vartheta_4(0,p)}, \quad p:=-q \epc\\
  \label{thetaA}
  \vartheta_A^2(\theta) &=
     \frac{\gamma(\xi^4)\gamma(\xi^{-4})}{\gamma(q^{-2})\gamma(q^{-2})},
     \quad \xi=-\i \re^{\i\theta} \epc\\
  \label{gamma}
     \gamma(u) &=
     \frac{(q^4u,q^4,q^4)(u^{-1},q^4,q^4)}{(q^6u,q^4,q^4)(q^2 u^{-1},q^4,q^4)} \epc\\
  \label{xyz}
     (x,y,z) &= \prod\limits_{m,n=0}^{\infty}(1-xy^mz^n)\epp
\end{align}
\end{subequations}

\begin{figure}[!ht]
\begin{center}
\includegraphics[width=.45\textwidth]{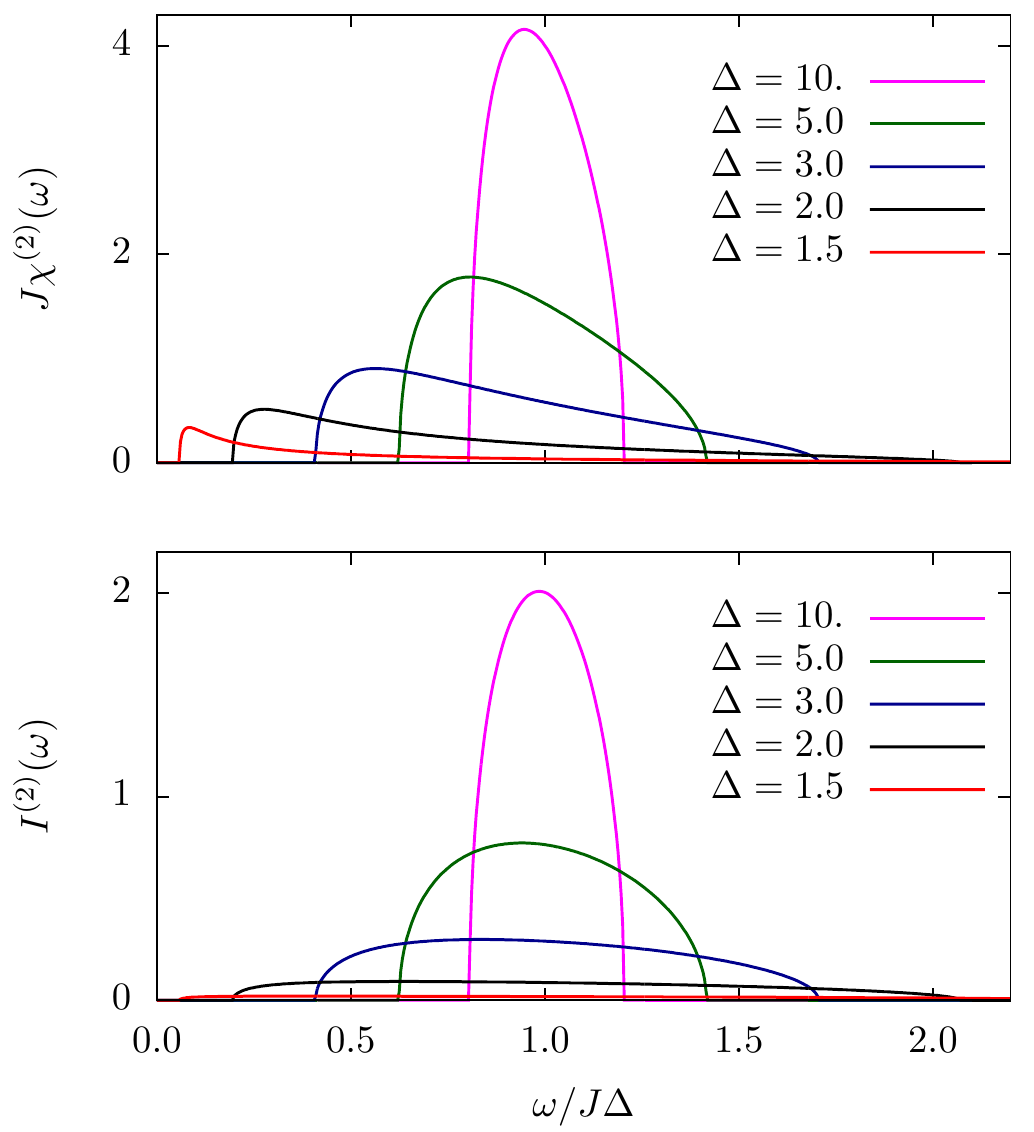}
\caption{\label{fig:chi2spinon} Two-spinon contribution $J\chi^{(2)}$ to
the dynamical susceptibility $J\chi_{+-}''$ and corresponding intensity
$I^{(2)}$ as functions of $\om/J \D$ for different anisotropy parameters $\Delta>1$
at $T=h=0$.}
\end{center}
\end{figure}

The line shape of the function $\chi^{(2)}$ is shown in Fig.~\ref{fig:chi2spinon}
for several values of $\Delta$. One can observe that the broadened peak is
very asymmetric for small $\Delta$. For increasing $\Delta$ the peak becomes
narrower and more symmetric. The Ising limit $\Delta \to \infty$ is analyzed
in the next section. 

\subsection{Ising limit in the two-spinon case}\label{isinglimit2spinon}
As the Hamiltonian \eqref{hamJM} diverges for $\Delta\to\infty$ we rescale all
energies by the factor $\Delta$. Replacing, in particular, $J$ by $J/\Delta$
in \eqref{hamJM}, we obtain the Ising Hamiltonian \eqref{hising} in the limit
$\Delta\to\infty$.

In the expression \eqref{chi2spinonend} for $\chi^{(2)}$, the rescaling only pertains 
to the definition \eqref{Idef} of $I$, where $J$ must be
replaced by $J/\Delta$. All other relations \eqref{thetan}-\eqref{xyz} and
particularly \eqref{hatomega} remain unaffected, and the Ising limit can be easily
performed. Since $\Delta=(p+p^{-1})/2$, we conclude that $p \to 0$ and consequently
$k\to 0$, $k' \to 1$, $K \to \pi/2$ as well as $K'\to -\frac{1}{2}\ln{(p)}\to\infty$.
The rescaled $I$ becomes
\begin{equation}
     I=\frac{J K}{\pi\Delta}\sh{\left(\pi K'/K\right)} \to
        -\frac{J}{2}\frac{p-p^{-1}}{p+p^{-1}} \to \frac{J}{2}\epp
\end{equation}
On the one hand the product of the two unit-step functions in the numerator of
\eqref{chi2spinonend} ensures that $\chi^{(2)}(\om)=0$ for all $\om \neq J$,
on the other hand we show that $\int_{-\infty}^{\infty} \rd \om \: \chi^{(2)}(\om)
=\pi/2$ in App.~\ref{m0ising}. Hence, $\chi^{(2)}$ is a $\de$-function with
prefactor $\pi/2$,
\begin{equation}\label{chi2ising2}
  \chi^{(2)}(\om) = \frac{\pi}{2}\de(\om-J)\epc
\end{equation}
which coincides with the result \eqref{chiising2} of Sec.~\ref{isinglimit},
because $m_0=m_2=0$ for $h=0$ and $m_1\to 1/2$ for $h=0$ and $T\to 0$. For the
integrated intensity we easily obtain
\begin{equation}
     \int_{-\infty}^{\infty} \rd \om \: \frac{\om}{2} \chi_{+-}''(\om)
        =\int_0^\infty \rd \om \: \om\chi^{(2)}(\om)=\frac{\pi J}{2}\epc
\end{equation}
which agrees with Eq.~\eqref{eq:IntegInten}
of Sec.~\ref{sec:IntegInten} for $h=0$ and $m_1=1/2$.

\subsection{Heisenberg limit and integrated intensity}\label{twospinonheisenberglimit}
In the Heisenberg (or isotropic) limit $\Delta\to 1$ we have $p\to 1$ and
consequently $k\to 1$, $k'\to 0$, $K'\to\pi/2$ as well as $K\to \infty$.
Fig.~\ref{fig:chi2spinon} shows that the function $\chi^{(2)}$ tends to zero
uniformly. This is consistent with the behavior of $\chi_{+-}''$ in the isotropic
limit, Eq.~\eqref{chixxx}, 
because $m(0,T)=0$ for all temperatures $T$.

In order to obtain a measure for
the relative contribution of $\chi^{(2)}$ to the full susceptibility $\chi_{+-}''$
for all $\Delta$, especially for the isotropic limit $\Delta\to 1$, we compare
the two-spinon contribution of the integrated intensity 
\begin{equation}
     I_{\rm int}^{(2)}(\D)=\int_0^\infty \rd \om \: \om\chi^{(2)}(\om)
\end{equation}
with the total integrated intensity  
\begin{equation}
 I_{\rm int}(\D)=\int_0^\infty \rd \om \: \om\chi(\om) \epp
\end{equation}
We denote their ratio 
\begin{equation}\label{ratio}
     r(\D)=\frac{I_{\rm int}^{(2)}(\D)}{I_{\rm int}(\D)}\epp
\end{equation}

If we substitute $\om$ by $\theta$ by means of \eqref{hatomega}, the numerator
of \eqref{ratio} becomes
\begin{equation}\label{2spinonintint}
     I_{\rm int}^{(2)}(\D)
        =\frac{2k'KI}{\pi}\int_{0}^{\pi/2} \rd \theta \:
	 \frac{\vartheta_A^2(\theta)}{\vartheta_n^2(\theta)}\epp
\end{equation}
The integral on the right hand side can be easily evaluated numerically.
Furthermore, we can express the behavior of $I_{\rm int}^{(2)}$ in the isotropic 
limit analytically in terms of $1-p = \sqrt{\Delta^2-1}-(\Delta-1)$,
\begin{equation}\label{overexpdecay}
      I_{\rm int}^{(2)}(\D) \Rarrow[8mm]{$\scriptstyle \D\to 1$}
         C \re^{-\frac{\pi^2}{2(1-p)}}\left(1+\mathcal{O}(1-p)\right)\epp
\end{equation}
The derivation of this formula and the value of the constant $C$ are shown
in App.~\ref{2spinonIsoLimes}. 

For the denominator of \eqref{ratio} we use a sum rule and obtain
\begin{equation}
     I_{\rm int}(\D)=\frac{2\pi\delta J}{\Delta}
        \left(\<s_1^x s_2^x\>_0 -\<s_1^z s_2^z\>_0 \right)\epp
\end{equation}
Using Eqs.~(28), (34), and (35) of Ref.~\onlinecite{TKS04} the two-point
correlation functions on the right hand side can be expressed by integrals
which again are easy to compute numerically. We obtain 
\begin{multline}
      \<s_1^x s_2^x\>_0 -\<s_1^z s_2^z\>_0 = -\frac{1}{4}
	+ \int_{-\infty}^\infty \frac{\rd x}{\eta\ch{\left(\frac{\pi x}{\eta}\right)}}\\
      \times\frac{3 \sin^2{\hspace{-0.3ex}x}\ch^4{\hspace{-0.3ex}\frac{\eta}{2}}
        + \cos^2{\hspace{-0.3ex}x}\sh^4{\hspace{-0.3ex}\frac{\eta}{2}}
	- \frac{x}{\eta}(\ch^2{\hspace{-0.3ex}\frac{\eta}{2}}
	+ \frac{1}{2})\sin{\hspace{-0.3ex}2x}\sh{\hspace{-0.3ex}\eta}}
	  {4(\sh^2{\hspace{-0.3ex}\frac{\eta}{2}}+\sin^2{\hspace{-0.3ex}x})^2}\epc
\end{multline}
where the parameter $\eta$ is defined by $\D=\ch{\eta}$. An expansion close
to the isotropic point $\eta=0$ yields
\begin{equation}
     I_{\rm int}(\D) \sim \frac{(7-4\ln{2})\pi J}{120}\eta^4\epp
\end{equation}
Accordingly, the total integrated intensity $I_{\rm int}(\D)$ tends to zero
for $\D\to 1$, but nowhere nearly as fast as the two-spinon contribution
$I_{\rm int}^{(2)}(\D)\sim \re^{-\pi^2/2(1-p)}$. We conjecture that, close
to the isotropic point $\D=1$, all higher spinon contributions $\chi^{(2n)}$,
$n\geq 2$, are as marginal as $\chi^{(2)}$, but in such a way that for the
full susceptibility $\chi = \sum_{n\geq 1}\chi^{(2n)}$ still holds. 

\begin{figure}[!ht]
\begin{center}
\includegraphics[width=.47\textwidth]{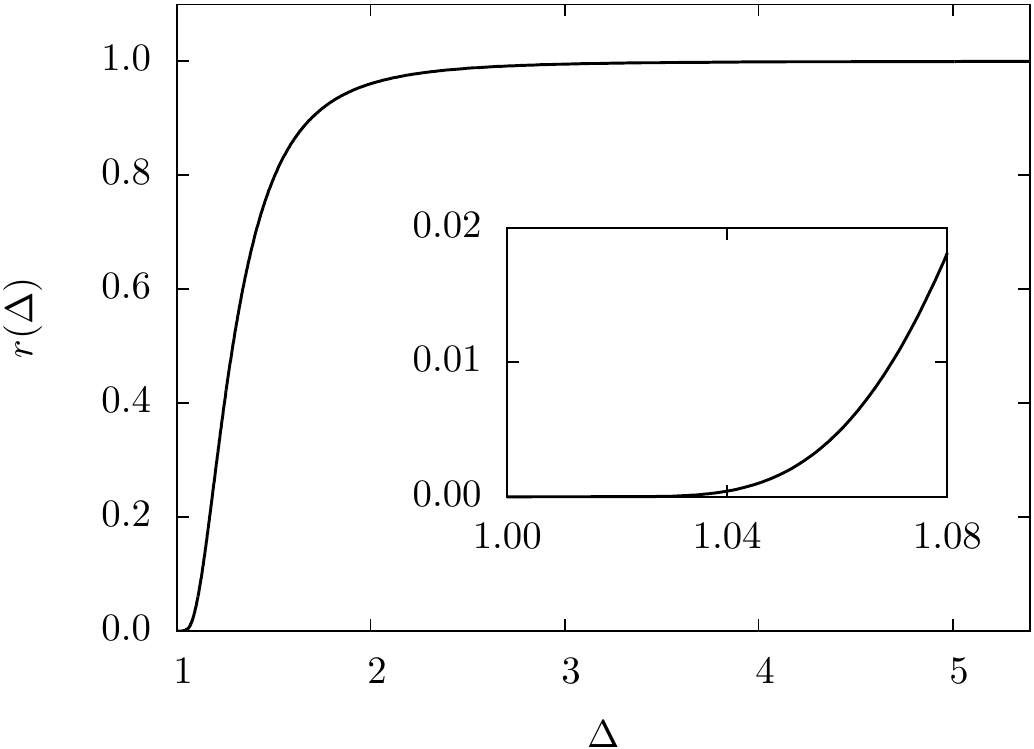}
\caption{\label{fig:ratio} Ratio $r$ as a function of the anisotropy $\D>1$. The
inset shows the behavior of $r$ close to the isotropic point $\D=1$, where it
decays over-exponentially.}
\end{center}
\end{figure}

In Fig.~\ref{fig:ratio} the ratio $r$ is plotted as a function of the
anisotropy $\D>1$. For $\D>3/2$ the two-spinon contribution accounts for
more than 80\% of the integrated intensity, for $\D>2$ for even more
then~96\%. In the limit $\D\to\infty$ it rapidly approaches 100\%.
Additionally, one can observe in the inset of Fig.~\ref{fig:ratio} the
over-exponential decay \eqref{overexpdecay} for $\D\to 1$. 

\subsection{Line width and some moments}
\begin{figure}[!ht]
\begin{center}
\includegraphics[width=.47\textwidth]{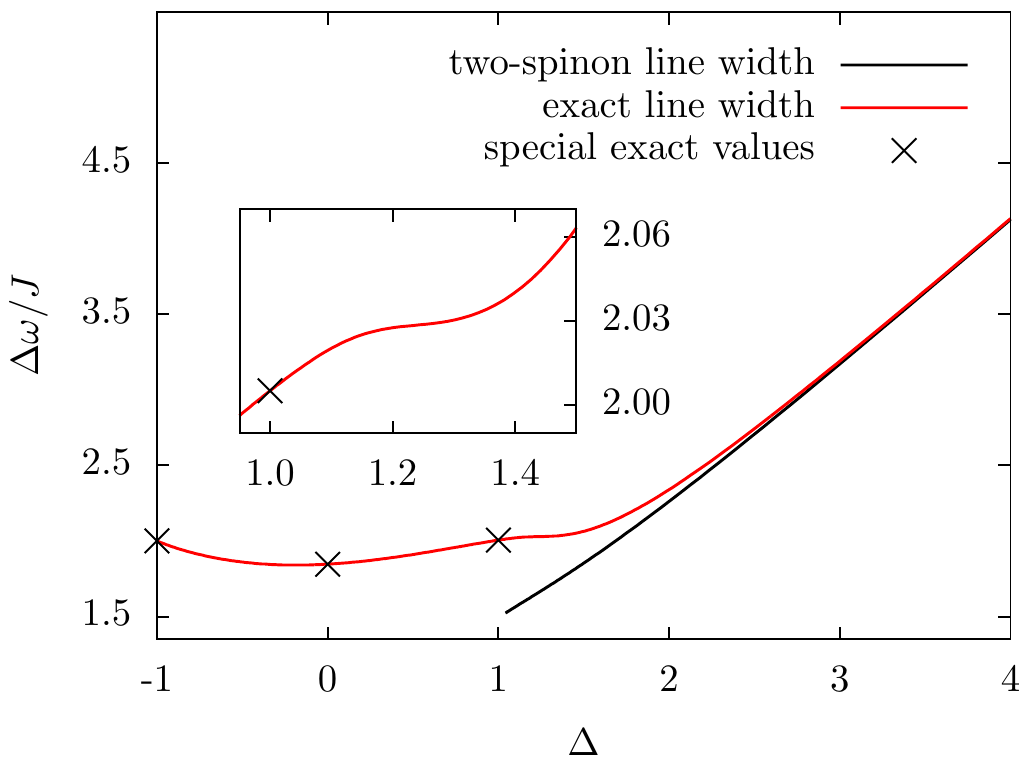}
\caption{\label{fig:two_spinon_line_width} Two-spinon line width $\D\om/J$
compared with the exact line width for $T=h=0$. The black crosses marks the
special values at $\D=1$, $\D=0$, and $\D=-1$.}
\end{center}
\end{figure}

In this section we compare the line width of the two-spinon contribution to
the dynamical susceptibility with the exact line width. For this purpose we
set $T=h=0$ in the expressions for $\ph$, $\om$, and $\om'$ in 
App.~\ref{app:omphi}. Then, all integrals involving functions defined
as solutions of integral equations vanish and $\ph$, $\om$, and $\om'$ are
determined solely by their explicit contributions. The formula for the line
width simplifies to $\Delta\om/J=\sqrt{m_3/m_1}$ which, for $T = h = 0$,
is then expressed in terms of explicit integrals by means of Eq.~\eqref{ms3}.
Equivalently, one may use the ground-state results for short-range correlation
functions of Ref.~\onlinecite{KSTS04}. Fig.~\ref{fig:two_spinon_line_width}
shows the exact line width as a function of the anisotropy for $\Delta>-1$.

For the two-spinon contribution the moments $m_1$ and $m_3$ can be expressed
by the integrals
\begin{equation}
      m_n = \frac{(2I)^n k'K}{\pi}\int_0^{\pi/2}d\theta\,
        \text{dn}^{n-1}\left(\frac{2K\theta}{\pi}\right)
	\frac{\vartheta_A^2(\theta)}{\vartheta_n^2(\theta)}\epc
\end{equation}
which can be evaluated to arbitrary precision. The two-spinon line width is
shown in Fig.~\ref{fig:two_spinon_line_width} as a function of $\D>1$. As
argued in the previous section the two-spinon contribution and the full
susceptibility become identical in the Ising limit $\D\to\infty$. An expansion
in $1/\D$ of both line widths shows that they agree up to the order $1/\D^2$. 

When $q$ is a root of unity the moments $m_1$ and $m_3$ can be
evaluated.\cite{KSTS04} Here, we present the results for $\D=1$, $\D=0$, and
$\D=-1$, respectively, 
\begin{subequations}\label{secialvalues}
\begin{align}
  \label{m1D1}m_1(\D=1) &= \frac{7-4\ln{2}}{15}\epc\\
  \label{m3D1}m_3(\D=1) &= -\frac{2}{5}+\frac{169}{10}\zeta(3)
     -\frac{18}{5}\zeta^2(3)-\frac{65}{4}\zeta(5)\nonumber\\ \displaybreak[0]
  &\qquad+\ln{(2)}\,(-4-10\zeta(3)+20\zeta(5))\epc\\
  \label{m1D0}m_1(\D=0) &= \frac{2(\pi-2)}{\pi^2}\epc\\
  \label{m3D0}m_3(\D=0) &= \frac{64-48\pi-6\pi^2+27\pi^3}{9\pi^4}\epc\\
  \label{m1D-1}m_1(\D=-1) &= 1/2\epc\\
  \label{m3D-1}m_3(\D=-1) &= 2\epp
\end{align}
\end{subequations}
The function $\z$ is Riemann's zeta function. The numerical values
of the line widths at these anisotropies are $\Delta\om/J \approx 2.00518$
for $\Delta=1$, $\Delta\om/J \approx 1.84606$ for $\Delta=0$, and $\Delta\om/J
= 2$ for $\Delta=-1$. Note that the exact line width as a function of $\D$
is continuous and non-zero except at the isotropic points $\D = 1, -1$ where 
it is not defined. However, the line width can be continued continuously at 
these points which yields the curve plotted in Fig.~\ref{fig:two_spinon_line_width}.

\subsection{Comparison with numerical calculations for finite chains}
\label{subsec:numfinite}
We can now compare the two-spinon contribution to the dynamical susceptibility
$\chi^{(2)}$, Eq.~\eqref{chi2spinonend}, with the full susceptibility
obtained numerically by the method described in Sec.~\ref{sec:numerics}.
For this purpose we use numerical data for $\chi_{+-}''$ as a function of $\om$
for chains of lengths $L=24$ and $L=32$ at $h = 0$ and $\D = 2$. We shall
indicate the length by a subscript and briefly write $\chi_L''$. Based on
the discussion of Sec.~\ref{twospinonheisenberglimit} we expect that
the two-spinon contribution amounts to the main part ($\sim$ 96\%) of the full 
susceptibility $\chi_{+-}''$ in the limit $T\to 0$. For larger temperatures
we expect deviations.

\begin{figure}[!ht]
\begin{center}
\includegraphics[width=.43\textwidth]{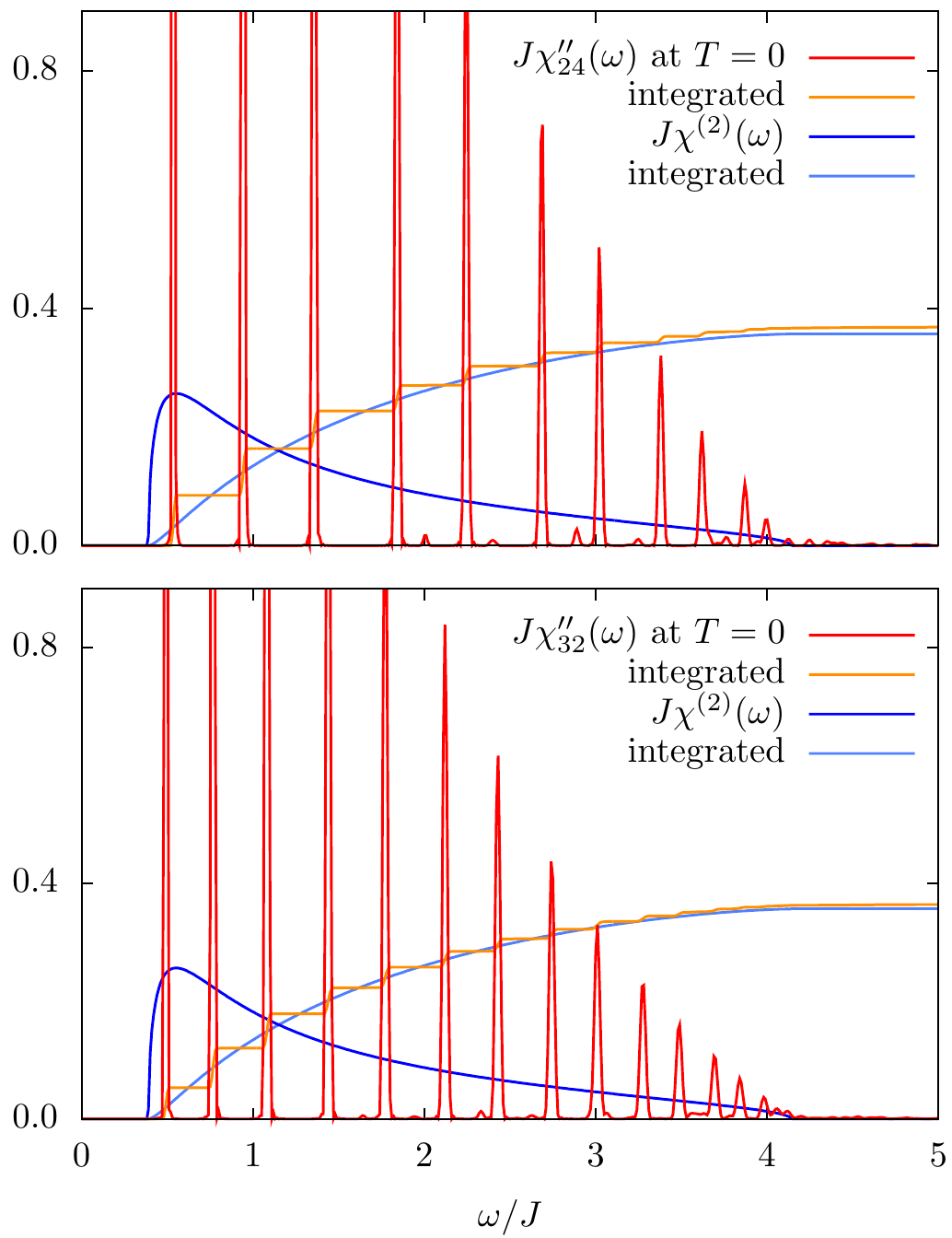}
\caption{\label{fig:two_spinon_vs_L24_L32_T0} Two-spinon contribution $J\chi^{(2)}$
compared with the dynamical susceptibility $J\chi_{L}''$ of finite chains as
functions of $\om/J$ for $\Delta = 2$ at $T=h=0$. $L=24$ in the upper panel,
$L=32$ in the lower panel.}
\end{center}
\end{figure}

We have to comment on the limit $T\to 0$. For finite chains the ground state
with energy $E_0$ is non-degenerate and for $L \mod 4 = 0$ carries momentum
$q = 0$.
Yet, the gap $\Delta E = E_\pi-E_0$ to the low-lying $q=\pi$-state is very
small compared to the gaps to all other states. In the thermodynamic limit
these two states degenerate and stay separated from the rest of the spectrum.
Hence, for a better comparison with the two-spinon contribution, we averaged
over the $q=0$- and $q=\pi$-states in our numerical calculation shown in
Fig.~\ref{fig:two_spinon_vs_L24_L32_T0}. 

\begin{figure}[!ht]
\begin{center}
\includegraphics[width=.43\textwidth]{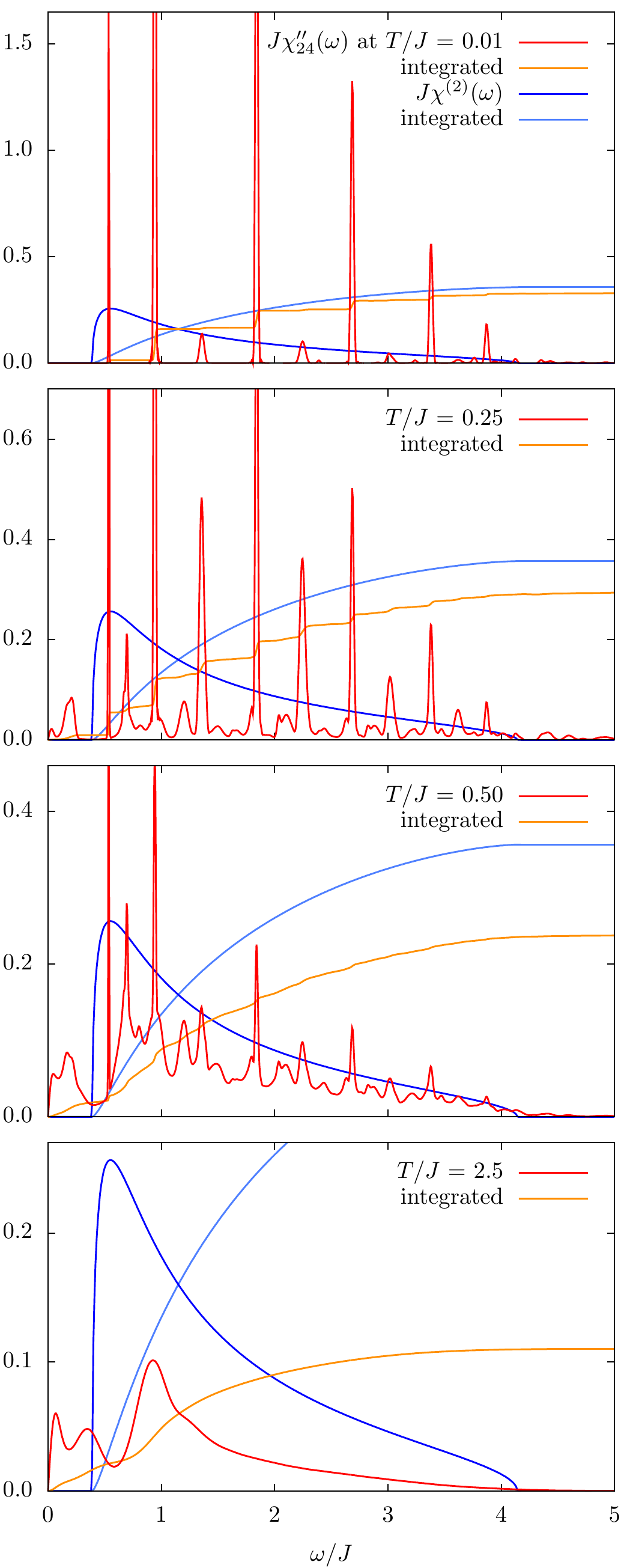}
\caption{\label{fig:two_spinon_vs_L24_diffTs} Two-spinon contribution $J\chi^{(2)}$
(dark-blue line) compared with the dynamical susceptibility $J\chi_{24}''$ (red lines)
as functions of $\om/J$ for $\Delta = 2$, $h=0$ and $L=24$ in all panels. The
temperature is increased from $T/J=0.01$ to $T/J=2.5$. Note the different
scales at the vertical axis.}
\end{center}
\end{figure}

As one can see, at low temperature the dynamical susceptibilities
$\chi_{24}''$ and $\chi_{32}''$ (red lines) consist of a multitude of
narrow peaks. This peak-structure is due to the finiteness of the
chain and the small number of eigenstates that contribute to the $T=0$
response.  The Chebyshev expansion approach, whose resolution is
inversely proportional to the expansion order $M$, can then distinguish
all contributing matrix elements.  With increasing temperature the
Boltzmann factors $\exp(-E_n/T) - \exp(-E_m/T)$, see App.~\ref{app:spectralrep},
suppress fewer states, and eventually the density of the peaks becomes
higher than the numeric resolution. The dynamical susceptibilities then
evolve into smooth curves, as is illustrated in
Fig.~\ref{fig:two_spinon_vs_L24_diffTs} and also by the
high-temperature data in Fig.~\ref{fig:highline}.

Similar behavior occurs for increasing lattice size $L$. Comparing
the two panels of Fig.~\ref{fig:two_spinon_vs_L24_L32_T0} one observes
that the peak-structure of $\chi_L''$ becomes tighter for larger
$L$. Additionally, the heights of the peaks decrease. Although
$\chi_L''$ (red lines) and $\chi^{(2)}$ (dark-blue line) do not look
alike, the integrals of these two functions (orange and light-blue
line) match very well. This indicates that the step function
$\int_0^\om \rd \om' \: \chi_L''(\om')$ (orange line) converges
uniformly to $\int_0^\om \rd \om' \: \chi_{+-}''(\om')$. For the
thermodynamic limit $L\to\infty$ we expect that the peak-structure
smears out and the dynamical susceptibility becomes a smooth curve
akin to the two-spinon contribution $\chi^{(2)}$.

We conclude that at low temperature the numerically calculated line
shape of the dynamical susceptibility shows strong dependence on the
size of the chain and on its finite-size spectrum. By way of contrast,
the moments of the dynamical susceptibility and other integrals over
the whole range of frequencies seem to be well approximated by our
numerical data.

\section{Conclusions}
The Heisenberg-Ising chain considered in this work is a prototypical model
of a quasi one-dimensional anisotropic antiferromagnet. Its collective
spinon excitations are created in pairs. Their spectrum is  a scattering
continuum characteristic of one-dimensional interacting systems. In microwave
absorption it becomes visible in a broadening of resonances away from the
isotropic point.

Within the linear response theory the absorbed intensity is basically equal
to the imaginary part of the dynamical susceptibility multiplied by the
absorption frequency. Although the model is exactly solvable as long as the
magnetic field is directed along the axis of magnetic anisotropy, the calculation of
such type of dynamical correlation functions at finite fields and temperatures
is still beyond the possibilities of contemporary theoretical methods. Due
to recent progress in the calculation of static short-range correlation
functions, however, it became possible to calculate certain global characteristics
of the spectrum: the field-dependent moments of Sec.~\ref{sec:methmom} that
determine the average absorption frequency (the resonance shift $\de \om$)
as well as a field-dependent line width $\D \om$ at arbitrary temperatures
and magnetic fields.

In this work we compared the exact data for the resonance shift $\de \om$
and the line width $\D \om$ with data extracted from a numerical calculation
of the dynamical susceptibility for chains of finite length. We used the
exact data to improve the numerical calculation and to validate the quality
of the numerical data. When looking at the numerical data for the susceptibility
they appear spiky and finite-size dependent. In any case, they look rather
different from the smooth curves obtained for the two-spinon contribution
to the dynamical susceptibility of the infinite chain at $T = h = 0$, which
is quasi-exact at large enough anisotropy. However, and this is an important
part of the moral of our work, the picture changes if we look at integrated
quantities. The integrated susceptibility for $L = 32$ in Fig.~%
\ref{fig:two_spinon_vs_L24_L32_T0} appears already rather similar to the
integrated two-spinon susceptibility. When turning to moments the picture
becomes even better. The numerical finite-chain data for the width $\D \om$
in Fig.~\ref{fig:shwcritoft} (where they are compared with the exact data
for the infinite chain) look as if they are almost free of finite-size
corrections. This gives us confidence that our data for the frequency-dependent
line width $\D h$, calculated from the same numerical data set for the dynamical
susceptibility, are reliable as well.

It was a big surprise for us that $\D h$ shows the opposite temperature
dependence in the critical phase as $\D \om$ (see Fig.~\ref{fig:omwidthcritoft}).
The line width $\D \om$ increases as the temperature decreases, but $\D h$
decreases.  This markedly distinct behavior can be attributed to the asymmetry
of the dynamical susceptibility in $\om$ and $h$. Not only the temperature
dependence of the two measures of the line width defined by the two types of
moments $m_n$ and $M_n$ is different, but also their absolute values. We observe
that $\D h < \D \om$.

In general, the peak-to-peak width, usually measured in experiments\cite{Ajiro03,KBL10}
and decreasing with temperature, cannot be extracted from our numerical data,
since they are not smooth enough at low temperatures. At high temperatures, however,
where we can use the normal-inverse Gaussian as a model for $\chi_{+-}'' (\om, h)/\om$,
we find a peak-to-peak width $\D h_{\text{pp}}$ which is again smaller, $\D h_{\text{pp}}
< \D h < \D \om$, and whose magnitude seems to be compatible with experiments.

The conclusion for theoretical attempts to extract the line width from
approximations to the dynamical susceptibility is that the seemingly simple
and intuitive concept of a line width is rather delicate. The peak-to-peak
width, popular in the analysis of experimental data, is shape dependent and
influenced by {\it a priori} assumptions on the line width. By contrast,
our moment-based line width $\D \om$ is not based on assumptions about the
shape of the spectral lines and can be calculated exactly for the
Heisenberg-Ising chain. It is, moreover, universal with respect to a scaling of
all quantities with the exchange interaction $J$. For these reasons we are
curios if it will be possible in practice to obtain $\D \om$ from experimental
data.

This will depend on how well background and noise can be separated from the
signal. From clean data one could even directly extract the moments $m_1$, $m_2$,
$m_3$, \dots, defined in \eqref{defm}, which would mean to directly measure
certain short-range correlation functions ranging over $2$, $3$, $4$, \dots
lattice sites.

As opposed to the line width the resonance shift is expected to be a more
robust quantity. We expect our results for $\de \om$ to compare rather
directly with experimental data as long as the observed line shapes are not
too much asymmetric. In the latter case the definition \eqref{rshiftmom})
should be taken seriously and should be used to calculate the average absorption
frequency from the experimental data.

From the two-spinon result for the absorbed intensity (see Fig.~\ref{fig:chi2spinon})
it can be seen that the spectral lines at low temperatures can be expected
to be broad and asymmetric. In principle, the amount of asymmetry of the
lines can be expressed in terms of the higher moments $m_4$, $m_5$ of the
dynamical susceptibility. And, in principle, these higher moments can be
calculated exactly at any temperature and magnetic field, which we leave
as project for future research. Another interesting project for the future
may be the calculation of the $T = 0$ dynamical susceptibility in the
critical regime by means of form factors in the finite volume, in analogy with
the work of Refs.~\onlinecite{BKM02a,BKM03,SST04,CaMa05} on the dynamical
structure factor.

\begin{acknowledgments}
The authors would like to thank R. K. Kremer and K. Sakai for stimulating
discussions and are grateful to K. Sakai for his Mathematica notebook with
results for the short-range ground state correlation functions. AK and FG
are indebted to Y. Maeda for a discussion in 2007 which inspired this work.
MB~acknowledges partial financial support by the Volkswagen Foundation. 
\end{acknowledgments}

\appendix

\section{Absorption of energy in quantum spin chains}
\label{app:linres}
In order to keep this work self-contained we include a summary of
the linear response theory of energy absorption and its application
to quantum spin chains.

\subsection{Time evolution of the statistical operator}
We consider a quantum system with Hamiltonian $\CH$ possessing a
discrete spectrum $(E_n)_{n=0}^\infty$ and corresponding eigenstates
$\{|n\>\}_{n=0}^\infty$. At time $t_0$ we adiabatically switch on
a time-dependent perturbation $V(t)$. We are interested in the time
evolution of the system, assuming it was initially, at times $t < t_0$,
in an equilibrium state described by the statistical operator
\begin{equation}
     \r_0 = \frac1Z \sum_{n=0}^\infty \re^{- \frac{E_n}T} |n\>\<n|
\end{equation}
of the canonical ensemble. We denote the temperature by $T$ and the
canonical partition function by $Z$.

Let $U(t)$ the time evolution operator of the perturbed system,
\begin{equation} \label{ut}
     \i \6_t U(t) = \bigl( \CH + V(t) \bigr) U(t) \epc \qd U(t_0) = \id \epp
\end{equation}
Under the influence of the perturbation the state $|n\>$ evolves
into $|n, t\> = U(t) |n\>$, and the statistical operator at time $t$
becomes
\begin{equation} \label{rhot}
     \r (t)  = \frac1Z \sum_{n=0}^\infty \re^{- \frac{E_n}T} |n, t\>\<n, t|
             = U (t) \r_0 U^{-1} (t) \epp
\end{equation}

We define
\begin{subequations}
\begin{align}
     & R(t) = \re^{\i \CH t} (\r(t) - \r_0) \re^{- \i \CH t} \epc \\
     & W(t) = \re^{\i \CH t} V(t) \re^{- \i \CH t} \epp
\end{align}
\end{subequations}
Then
\begin{multline}
	\i \6_t R(t) = \i \6_t \re^{\i \CH t} U (t) \r_0
	                 \bigl( \re^{\i \CH t} U(t) \bigr)^{-1} \\
	             = [ W(t), \re^{\i \CH t} \r(t) \re^{- \i \CH t}]
		     = [ W(t), R(t) + \r_0] \epp
\end{multline}
Since $R(t_0) = 0$ by construction, we obtain
\begin{equation}
     R(t) = - \i \int_{t_0}^t \rd t' \: [W(t'), R(t') + \r_0] \epp
\end{equation}
This Volterra equation is an appropriate starting point for a
perturbation theory. Assuming that $W(t)$ be small we conclude that
\begin{equation}
     R(t) = - \i \int_{t_0}^t \rd t' \: [W(t'), \r_0] + \CO (W^2) \epc
\end{equation}
i.e.~to lowest order in $W$
\begin{equation} \label{rhoborn}
     \r (t) = \r_0 - \i \re^{- \i \CH t} \int_{t_0}^t \rd t' \: [W(t'), \r_0]
              \re^{\i \CH t} \epp
\end{equation}
This is the statistical operator in Born approximation. In the following
$t_0$ will be sent to $ - \infty$.

\subsection{Time evolution of expectation values}
Using \eqref{rhoborn} we can calculate the time evolution of the expectation
value of an operator $A$ due to the perturbation. Writing $A(t) = \re^{\i \CH t}
A \re^{- \i \CH t}$ and using the invariance of the trace under cyclic permutations
we obtain
\begin{align}
  \de \< A\>_T &= \tr \{ (\r (t) - \r_0) A\} \nonumber\\ 
               &= -\i\int_{- \infty}^t \rd t' \: \tr \bigl\{ [W(t'), \r_0] \re^{\i \CH t}
                   A \re^{- \i \CH t} \bigr\} \nonumber\\
               &= - \i \int_{- \infty}^t \rd t' \: \bigl\< [A(t - t'), V(t')]
	           \bigr\>_T \epp
\end{align}
Here $\< \cdot \>_T = \tr \{ \r_0 \cdot \}$ denotes the thermal average.
A typical example of a perturbation, which will be relevant for our
discussion below, is a classical time-dependent field $h^\a (t)$ coupling
linearly to operators $X^\a$,
\begin{align} \label{classpert}
     V(t) & = h^\a (t) X^\a \epc \\ \label{aborn}
     \de \< A\>_T & = - \i \int_{- \infty}^t \rd t' \: \bigl\< [A(t - t'), X^\a ]
	                \bigr\>_T h^\a (t') \epp
\end{align}

\subsection{Absorption of energy}
The absorbed energy per unit time is
\begin{align}
     \frac{d E}{dt} & = \frac{d}{dt} \tr \{ (\r (t) - \r_0) (\CH + V(t)) \}
                        \notag \\[1ex] & =
			- \i \tr \{ [\CH + V(t), \r (t)] (\CH + V(t)) \}
			\notag \\ & \mspace{90.mu}
			+ \tr \{ (\r (t) - \r_0) \dot V(t) \} \notag \\[1ex] & =
			\tr \{ (\r (t) - \r_0) \dot V(t) \} = \de \< \dot V(t) \>_T \epp
\end{align}
Here we used \eqref{ut},~\eqref{rhot} in the second equation and the cyclic
invariance of the trace in the third equation. Assuming that $V(t)$ is of the
form \eqref{classpert} and using \eqref{aborn} we obtain
\begin{equation}
     \frac{d E}{dt} = - \i \int_{- \infty}^t \rd t' \:
                        \bigl\< [X^\a (t - t'), X^\be ] \bigr\>_T \,
			\dot h^\a (t) h^\be (t') \epp
\end{equation}

\subsection{Application to quantum spin chains}
Let us now apply the above formalism to the Hamiltonian of the Heisenberg-Ising spin
chain in a longitudinal static magnetic field of strength $h$,
\begin{equation} \label{hamapp}
     \CH = J \sum_{j=1}^L \bigl( s_{j-1}^x s_j^x + s_{j-1}^y s_j^y + (1 + \de)
                               s_{j-1}^z s_j^z \bigr) - h S^z \epp
\end{equation}

We perturb the spin chain by a circularly polarized electro-magnetic
wave propagating in $z$-direction. We assume that the wave length is 
large compared to the length of the spin chain and idealize this
assumption by setting the wave number $k=0$. Then the magnetic field
component of the wave is
\begin{equation} \label{extfield}
     \hv (t) = A \begin{pmatrix}
	        \cos (\om t) \\ - \sin(\om t) \\ 0
             \end{pmatrix} \epc \qd
	     A > 0 \epp
\end{equation}
It couples to the total spin as
\begin{equation}
     V(t) = h^\a (t) S^\a \epp
\end{equation}
Thus,
\begin{align} \label{dedt}
     & \frac{d E}{dt}
        = - \i \int_{- \infty}^t \rd t' \:
	         \bigl\< [S^\a (t - t'), S^\be ] \bigr\>_T \,
		          \dot h^\a (t) h^\be (t') \notag \\
        & = \frac{A^2 \om}{4} \int_0^\infty \rd t'
	         \bigl\{ \re^{\i \om (2t - t')} \bigl\< [S^+ (t'), S^+] \bigr\>_T
		 \notag \\ & \mspace{27.mu}
		 - \re^{- \i \om (2t - t')} \bigl\< [S^- (t'), S^-] \bigr\>_T
		 + \re^{\i \om t'} \bigl\< [S^+ (t'), S^-] \bigr\>_T
		   \notag \\ & \mspace{172.mu}
		 - \re^{- \i \om t'} \bigl\< [S^- (t'), S^+] \bigr\>_T \bigr\} \epp
\end{align}

The ability to absorb radiation is a material property. Hence, we generally
expect the absorbed energy per unit time to be proportional to the number of
constituents of a physical system and to diverge in the thermodynamic limit.
In order to define a quantity that truly characterizes the material and is
finite in the thermodynamic limit we should therefore normalize by the average
intensity $A^2$ of the incident wave and by the number of lattice sites $L$.
Further averaging the normalized absorption rate over a half-period $\pi/\om$ of the applied
field, we obtain the normalized absorbed intensity
\begin{align}
     I(\om, h) & = \frac{\om}{L A^2 \p} \int_0^{\frac\p\om} \rd t \: \frac{d E}{dt}
	      \notag \\[1ex]
            & = \frac{\om}{4 L} \int_{- \infty}^\infty \rd t \:
	        \re^{\i \om t} \bigl\< [S^+ (t), S^-] \bigr\>_T \epp
\end{align}
Introducing the familiar (imaginary part of) the  dynamical susceptibility per spin,
\begin{equation} \label{defchiapp}
     \chi_{+-}'' (\om, h) = \frac{1}{2L} \int_{- \infty}^\infty \rd t \:
        \re^{\i \om t} \bigl\< [S^+ (t), S^-] \bigr\>_T \epc
\end{equation}
the normalized absorbed intensity reads
\begin{equation}
     I (\om, h) = \frac\om2 \chi_{+-}'' (\om, h) \epc
\end{equation}
which is Eq.~\eqref{int} of the main text.

\subsection{The isotropic chain}
The full dynamical susceptibility can only be calculated in certain special cases.
In order to understand its behavior qualitatively we first of all consider
the isotropic point $\de = 0$ of the Heisenberg-Ising chain. In this case
\begin{equation}
     [\CH, \Sv] = - h [S^z, \Sv] \epc
\end{equation}
and the Heisenberg equation of motion for the total spin $\Sv$ can be
solved,
\begin{align} \label{spmxxxt}
     & \dot S^\pm = \i [\CH, S^\pm]
                  = - \i h [S^z, S^\pm] = \mp \i h S^\pm \epc \notag \\
     & \then S^\pm (t) = \re^{\mp \i ht} S^\pm \epc
\end{align}
and $S^z (t) = S^z$. Hence, the total spin behaves as
\begin{equation} \label{rots}
     \Sv (t) = \begin{pmatrix}
                  \cos(ht) & \sin(ht) & \\
		  - \sin(ht) & \cos(ht) & \\
		  && 1
               \end{pmatrix} \Sv \epp
\end{equation}
It rotates clockwise about the $z$ axis.

On the other hand, inserting \eqref{spmxxxt} into \eqref{defchiapp} we obtain
\begin{align} \label{chixxx}
     \chi_{+-}'' (\om, h) & = \frac{1}{2L} \int_{- \infty}^\infty \rd t \:
                              \re^{\i (\om - h) t} \bigl\< [S^+, S^-] \bigr\>_T
			      \notag \\
                          & = 2\p \de(\om - h) m(T, h) \epp
\end{align}
where $m(T, h) = \<S^z\>_T/L$ is the magnetization per lattice site. The
corresponding normalized absorbed intensity is Eq.~\eqref{intxxx} in the main
body of this article.

Comparing \eqref{extfield}, \eqref{rots} and \eqref{chixxx} we interpret
the absorption of energy as a resonance between the rotating field of the
incident wave and the rotating total spin of the chain, both spinning
clockwise with frequency $\om = h$. If we are off the isotropic point
$\de = 0$ of the Hamiltonian \eqref{hamapp} we may expect that energy is
transferred  from the `coherent motion of the total spin' to `other modes',
causing a damping of the spin precession and hence a shift and a broadening
of the spectral line.

\begin{remark}
In other treatments of the same problem the incident wave was considered
to be linearly polarized, leading to an absorbed intensity
\begin{equation} \label{chixxxxx}
     \frac{\om}2 \chi_{xx}'' (\om, h) =
        \frac{\p h}4 m(T, h) \bigl( \de (\om - h) + \de (\om + h) \bigr) \epp
\end{equation}
This can be understood by taking into account that a linearly polarized
wave can be decomposed into a superposition of two circularly polarized
waves of opposite circular polarization. For this reason the two spectral
lines in \eqref{chixxxxx} are, in fact, one and the same, when either the
right circularly polarized wave has frequency $\om$ or the left circularly
polarized wave has frequency $- \om$.
\end{remark}

\subsection{The Ising chain}
\label{app:isinglim}
For the Hamiltonian $H_I$ of Eq.~\eqref{hising} the time evolution $\re^{\i t \ad_{H_I}}
S^+$ can be calculated explicitly. One easily proves by induction that
\begin{equation}
     \ad_{H_I}^n S^+ = J^n \sum_{j=1}^L (s_{j-1}^z + s_{j+1}^z)^n s_j^+
\end{equation}
for all non-negative integers $n$. Furthermore
\begin{equation}
     (s_{j-1}^z + s_{j+1}^z)^n =
        \begin{cases}
	   s_{j-1}^z + s_{j+1}^z & \text{if $n$ is odd,} \\[1ex]
	   \2 + 2 s_{j-1}^z s_{j+1}^z & \text{if $n$ is even}
        \end{cases}
\end{equation}
for all $n \in {\mathbb N}$. It follows that
\begin{equation} \label{stising}
     \re^{\i t \ad_{H_I}} S^+ = S^+ - A + \frac{A + B}{2} \re^{\i J t}
                                    + \frac{A - B}{2} \re^{- \i J t} \epc
\end{equation}
where
\begin{equation}
     A = \sum_{j=1}^L (\tst{\2} + 2 s_{j-1}^z s_{j+1}^z) s_j^+ \epc \qd
     B = \sum_{j=1}^L (s_{j-1}^z + s_{j+1}^z) s_j^+ \epp
\end{equation}
Inserting \eqref{stising} into the definition of the dynamical susceptibility
\eqref{defchi} in the Ising limit we obtain
\begin{multline} \label{chiising1}
     \chi_{+-}'' (\om, h) = \frac{\p}L \Bigl\{
                            \<[S^+ - A, S^-]\>_T \de (\om - h) \\
                            + \2 \<[A + B, S^-]\>_T \de (\om - h + J)  \\
                            + \2 \<[A - B, S^-]\>_T \de (\om - h - J)  \Bigr\} \epp
\end{multline}
The coefficients in front of the $\de$-functions can be easily expressed in terms
of the moments in the Ising limit,
\begin{subequations}
\label{spinoncoeffs}
\begin{align}
     \frac{1}{2L} \<[A, S^-]\>_T & = m_2 = \2 \<s_1^z + 4 s_1^z s_2^z s_3^z \>_T \epc \\
     \frac{1}{2L} \<[B, S^-]\>_T & = - m_1 = 2 \< s_1^z s_2^z \>_T \epp
\end{align}
\end{subequations}
For these correlation functions explicit expressions in terms of $h$ and $T$
can be obtained by means of the $2 \times 2$ transfer matrix of the Ising chain,%
\cite{Babook}
\begin{subequations}
\label{IsingExplExp}
\begin{align}\label{IsingExplExp_a}
     \< s_1^z \>_T & = \frac{\sh \bigl(\frac{h}{2T}\bigr)}
                          {2 \sqrt{\sh^2 \bigl(\frac{h}{2T}\bigr) + \re^{J/T}}}\epc \\
     \label{IsingExplExp_b}
     \< s_1^z s_n^z \>_T & = \< s_1^z \>_T^2 + \Bigl( \frac14 - \< s_1^z \>_T^2 \Bigr)
                           f^{n - 1} \Bigl( \frac{h}{2T}, \frac{J}{T} \Bigr) \epc
\end{align}
\end{subequations}
where
\begin{equation}
     f(x,y) = \frac{\ch(x) - \sqrt{\sh^2 (x) + \re^y}}
                   {\ch(x) + \sqrt{\sh^2 (x) + \re^y}}
\end{equation}
and
\begin{equation}
     \< s_1^z s_2^z s_3^z \>_T
        = \< s_1^z \>_T \bigl(2\< s_1^z s_2^z \>_T - \< s_1^z s_3^z \>_T \bigr) \epp
\end{equation}

Inserting \eqref{spinoncoeffs} into \eqref{chiising1} we obtain
Eq.~\eqref{chiising2} of the main text.

\subsection{Spectral representation of the dynamical susceptibility}%
\label{app:spectralrep}
The dynamical susceptibility has the spectral representation
\begin{multline} \label{chispec}
     \chi_{+-}'' (\om, h) = \frac{\p}{L Z} \sum_{m, n}
        \bigl( \re^{- \frac{E_n}T} - \re^{- \frac{E_m}T} \bigr) \\ \times
	|\<m|S^-|n\>|^2 \de(\om - E_m + E_n)
\end{multline}
following from \eqref{defchiapp}. Here the $E_n$ are eigenvalues of $\CH$,
i.e.~they include the dependence of the magnetic field .

Let $\om > 0$. Then the only non-zero terms under the sum are those with
$E_m > E_n$ and those are positive. Hence, $\chi_{+-}'' (\om)$ is non-negative
for $\om > 0$. Similarly, $\chi_{+-}'' (\om)$ is non-positive for $\om < 0$. It
follows that $I (\om)$ is non-negative as was expected on physical grounds.
The spectral representation simplifies for $T \rightarrow 0$,
\begin{multline} \label{chispec0}
     \chi_{+-}'' (\om, h) \rightarrow \frac{\p}{L n_g} \sum_{n, g}
	\bigl\{|\<n|S^-|g\>|^2 \de(\om - E_n + E_g) \\ -
	       |\<g|S^-|n\>|^2 \de(\om - E_g + E_n) \bigr\} \epp
\end{multline}
Here we average over the $n_g$ degenerate ground states $|g\>$.
In general, $\chi_{+-}''$ is non-vanishing even for vanishing magnetic field.

\subsection{Alternative representation and a sum rule}
From the spectral representation \eqref{chispec} we infer that
\begin{multline} \label{chialt}
     \chi_{+-}'' (\om, h) \\
        = \frac{\p}{L Z} \bigl( 1 - \re^{- \frac\om T} \bigr)
	   \sum_{m, n} \re^{- \frac{E_n}T} |\<m|S^-|n\>|^2 \de(\om - E_m + E_n) \\
        = \frac{1}{2L} \bigl( 1 - \re^{- \frac\om T} \bigr)
	   \int_{- \infty}^\infty \rd t \: \re^{\i \om t} \bigl\< S^+ (t) S^- \bigr\>_T
	   \epp
\end{multline}
This representation immediately implies the sum rule
\begin{equation} \label{sumalt}
     \int_{- \infty}^\infty \frac{\rd \om}{2 \p}
        \frac{\chi_{+-}'' (\om, h)}{1 - \re^{- \frac\om T}}
	= \frac{1}{2L} \< S^+ S^- \>_T
	= \2 \sum_{j=0}^{L-1} \< s_1^+ s_{j+1}^- \>_T \epp
\end{equation}
Since the correlation functions $\< s_1^+ s_{j+1}^- \>_T$ decay exponentially in
$j$ in the thermodynamic limit, the integral on the left hand side exists for
$L \rightarrow \infty$. If $\chi_{+-}'' (\om, h)$ is analytic in the thermodynamic
limit, it follows that the constant term in the Taylor expansion at $\om = 0$ must
vanish. Then the intensity $I (\om, h) = \om \chi_{+-}'' (\om, h)/2$ has a double
zero at $\om = 0$.  Hence, it must have at least two maxima, one for $\om > 0$ and
another one for $\om < 0$.

\section{Technical details of the two-spinon calculations}

\subsection{Integrated susceptibility in the Ising limit}\label{m0ising}
We calculate the integral of the susceptibility $\chi^{(2)}$ over positive frequencies $\om$ in the Ising limit $\Delta\to\infty$, i.e.~$p\to 0$, 
\begin{equation}
 \int_0^\infty \rd \om \: \chi^{(2)}(\om)
    = \frac{k'K}{\pi}\int_0^{\pi/2} \rd \theta \: \frac{1}{{\rm dn}\left(\frac{2K}{\pi}\theta\right)}\frac{\vartheta_A^2(\theta)}{\vartheta_n^2(\theta)}\epc
\end{equation}
where we have inserted \eqref{chi2spinonend} and substituted $\om$ by $\theta$ via the relation \eqref{hatomega}. Due to $K\to\pi/2$, $k'\to 1$ we obtain ${\rm dn}(2K\theta/\pi)\to{\rm dn}(\theta)\to 1$, $\vartheta_n(\theta)\to 1$ for all $\theta\in(0,\pi/2)$ and thereby
\begin{equation}
 \int_0^\infty \rd \om \: \chi^{(2)}(\om) \to
    \frac{1}{2}\int_0^{\pi/2} \rd \theta \: \vartheta_A^2(\theta) \epp
\end{equation}
The limit $p\to 0$ for the function $\vartheta_A$  is more complicated. Using the definitions \eqref{thetaA}-\eqref{xyz} of $\vartheta_A$
and the relation
\begin{multline}
  \ln\left(\prod_{n_{1},...,n_{m}\geq 0}(1-zp_{1}^{n_{1}}\cdots p_{m}^{n_{m}})\right)
  = \\-\sum_{n=1}^{\infty}\frac{1}{(1-p_{1}^{n})\cdots(1-p_{m}^{n})}\frac{z^{n}}{n}\epc
\end{multline}
we obtain
\begin{equation}
  \frac{\gamma(z)}{\gamma(q^{-2})} = \exp\left(-\sum_{n=1}^{\infty}\frac{\sh^{2}(n(\epsilon-2\i\theta))}{\sh(2n\epsilon)\ch(n\epsilon)}\frac{\re^{n\epsilon}}{n}\right)\epc
\end{equation}
where $z = \re^{2\i\theta}$ and $\epsilon=\pi K'/K $, and therefore
\begin{equation}\label{thetaAend}
  \vartheta_{A}^{2}(\theta) = \exp\left(-\sum_{n=1}^{\infty}\frac{\ch(2n\epsilon)\cos(4n\theta)-1}
    {\sh(2n\epsilon)\ch(n\epsilon)}\frac{\re^{n\epsilon}}{n}\right)\epp 
\end{equation}
We now convert \eqref{thetaAend} into the form
\begin{multline} 
  \ln(\vartheta_{A}(\theta)) = \ln{\left[2\sin(2\theta)\right]} + \sum_{n=1}^{\infty}\frac{1}{n}\frac{p^{2n}}{1-p^{2n}}\sin^2{(2n\theta)} \\
+ \sum_{n=1}^{\infty}\frac{1}{n}\frac{p^{2n}}{1+p^{2n}}\cos^2{(2n\theta)} \\+ 2\sum_{n=1}^{\infty}\frac{1}{n}\frac{p^{2n}}{(1+p^{2n})^2}\cos^2{(2n\theta)}
\end{multline}
and see that $\vartheta_{A}(\theta)\Rarrow[4.3mm]{$\scriptstyle p\to 0$} 2\sin{(2\theta)}$. We finally obtain
\begin{equation}
 \int_0^\infty \rd \om \: \chi^{(2)}(\om) \Rarrow[7mm]{$\scriptstyle \D\to 1$}
    2\int_0^{\pi/2} \rd \theta \: \sin^2(2\theta) = \frac{\pi}{2}\epp
\end{equation}

\subsection{Integrated intensity and Heisenberg limit}\label{2spinonIsoLimes}
We analyze the behavior of the two-spinon contribution of the integrated intensity \eqref{2spinonintint}, 
\begin{equation}\label{2spinonintintappendix}
 I_{\rm int}^{(2)}(\D)=\frac{2k'KI}{\pi}\int_{0}^{\pi/2} \rd \theta \:
    \frac{\vartheta_A^2(\theta)}{\vartheta_n^2(\theta)}\epc
\end{equation}
in the isotropic limit $\Delta\to 1$, i.e.~$p\to 1$. On the one hand, it is known that $\vartheta_n(\theta)\to\ch{(2K/\pi\theta)}\to\infty$,  $k'\to 0$ and $I\to 0$, on the other hand we have $K\to\infty$ and $\vartheta_A(\theta,p)\to\infty$. In order to obtain more explicit results we define the function
\begin{equation}
 f\left(\frac{2K}{\pi}\theta\right)=\lim_{p\to 1}I\vartheta_A^2{(\theta)}\epc
\end{equation}
which can be calculated as
\begin{multline}
 f(x)=\frac{\re^{-12\zeta'(-1)}}{2^{4/3}}x\sh{(x)}\left[\psi\left(\frac{x}{\pi}\right)\right]^{-2\i x/\pi}\\
  \times \exp\left\{\frac{2\i}{\pi}\int_0^x \rd x' \:
  \ln{\left[\psi\left(\frac{x'}{\pi}\right)\right]}\right\}\epc
\end{multline}
where $\psi$ is defined by
\begin{equation}
 \psi(x)=\frac{\G(1-\i x)\G(1/2+\i x)}{\G(1+\i x)\G(1/2-\i x)}\epp
\end{equation}
Hence, we obtain
\begin{multline}
     \kappa :=\lim_{p\to 1}\left(\frac{I_{\rm int}^{(2)}}{k'}\right)
        = \lim_{p\to 1}\left(\frac{2 K I}{\pi}\right)
	  \int_0^{\pi/2} \rd \theta \:
	  \frac{\vartheta_A^2(\theta)}{\vartheta_n^2(\theta)}\\
      = \lim_{p\to 1}\int_0^{K} \rd x \:
      \frac{I\vartheta_A^2{\left(\frac{\pi x}{2K}\right)}}
      {\vartheta_n^2{\left(\frac{\pi x}{2K}\right)}}
      =\int_0^\infty \rd x \: \frac{f(x)}{\ch^2{x}} = 2,6471\ldots
\end{multline}
A straightforward calculation yields
\begin{align}
 \ln{\left( k'\right)} &= 4\ln{\left[(p,p^2)\right]}-4\ln{\left[(-p,p^2)\right]}\nonumber \\ 
  &\Rarrow[5mm]{$\scriptstyle p\to 1$} -\frac{\pi^2}{2(1-p)} + 2\ln{2} + \frac{\pi^2}{4} + \mathcal{O}(1-p)\epp
\end{align}
The asymptotic behavior of $I_{\rm int}^{(2)}$ therefore reads
\begin{equation}\label{overexpdecayappendix}
 I_{\rm int}^{(2)}(\D) \Rarrow[8mm]{$\scriptstyle \D\to 1$} C \re^{-\frac{\pi^2}{2(1-p)}}\left(1+\mathcal{O}(1-p)\right)\epc
\end{equation}
where $C=4\re^{\pi^2/4}\kappa \approx 124,6$.

\section{Short time expansion at infinite temperature}
\label{app:shorttime}
In the limit of infinite temperature the dynamical correlation
functions reduce to traces of the considered time-dependent operators,
and it is tempting to express the time evolution in terms of nested
commutators, evaluate the traces, and obtain the leading terms of a
Taylor series near time $t=0$. For the spin-pair correlation function
$\<s^z_n(t) s^z_0(0)\>$ such series have been calculated to orders as
high as $t^{30}$ already two decades ago.~\cite{BoLe92} Unfortunately,
these series converge slowly and with the accessible number of terms
precise values of the correlation functions can only be obtained for
rather short times $t\lesssim 5$. Nevertheless, we performed such an
expansion for $\tr[S^+(t) S^-]$, which is at the core of the function
$\phi(\omega)$ defined in Eq.~\eqref{defphi}. Using the highly
efficient computer algebra program FORM,~\cite{Ver00} we computed the
series up to the order $t^{38}$. The first few terms read
\begin{align}
    2 \text{Tr} & [S^+(t) S^-] \approx 
     1 - \frac{1}{4}(\Delta-1)^2 t^2 \notag \\
    & + \frac{1}{96}(\Delta-1)^2(3-2\Delta+2\Delta^2)t^4\notag \\
    & - \frac{1}{11520} (\Delta-1)^2 
    (30 - 15\Delta + 50 \Delta^2 - 16 \Delta^3 + 8 \Delta^4) t^6\notag \\
    & + \cdots\epp
\end{align}
Taking the logarithm we find 
\begin{align}
\label{shortt:log}
    \ln& (2 \text{Tr}[S^+(t) S^-]) \approx -\frac{1}{4}(\Delta-1)^2 t^2 
    + \Delta(\Delta-1)^2 \Big[ \frac{4-\Delta}{2^2 4!}t^4 \notag \\
    & + \frac{15-140\Delta+76\Delta^2-8\Delta^3}{2^4 6!} t^6 \notag \\
    & + \frac{56 - 2730 \Delta + 10948 \Delta^2 - 8792 \Delta^3 + 2256 \Delta^4 - 136 \Delta^5}{2^6 8!} t^8 \notag \\
    & + \cdots\Big]\,,
\end{align}
and we can immediately read off the two known results for $\Delta=0$
and $\Delta=1$, $-t^2/4$ and $0$, respectively. From the terms in the
square bracket we were only able to sum the $\Delta$-free
contributions and the terms of highest order in $\Delta$. The
coefficients of the $\Delta$-free terms read $4$, $15$, $56$, $210$,
$792$, $\dots$, and correspond to $\binom{2k}{k-1}$ with
$k=2,3,\dots$ leading to the Bessel function $I_2$. The coefficients of the terms with the highest power of
$\Delta$ for each power of $t$ read $1$, $8$, $136$, $3968$, $176896$,
$\dots$, and constitute the expansion coefficients of $\tan(x)^2$,
\begin{equation}
  \frac{\tan(x)^2}{2} = \frac{1}{2!}x^2 + \frac{8}{4!}x^4 + \frac{136}{6!}x^6
    + \frac{3968}{8!}x^8 + \cdots
\end{equation}
Summing these two sub-series of~\eqref{shortt:log}, we obtain
\begin{multline}
  \ln(2 \text{Tr}[S^+(t) S^-]) 
  \approx 4 \Delta (\Delta - 1)^2  I_2(t)  - (\Delta - 1)^2 (\Delta/2) t^2\\
  + \frac{2 (\Delta - 1)^2 \ln[\cos(t\Delta/2)]}{\Delta^2}\epp
\end{multline}
This function does not have much value as an approximation of the 
considered infinite-temperature correlation function, but it contains
at least all known exact results,
\begin{equation}
  \text{Tr}[S^+(t) S^-] \to \begin{cases}
    \frac{1}{2}\exp(-t^2/4) & \text{for }\Delta=0\,,\\[2mm]
    \frac{1}{2} & \text{for } \Delta=1\,,\\[2mm]
    \frac{1}{4}(1 + \cos(\tilde t)) & \text{for } 
    \Delta\to\infty,\ \tilde t = t\Delta \text{ finite.}
  \end{cases}
\end{equation}
The last line corresponds to the Ising limit discussed in
Sec.~\ref{isinglimit}, and the Fourier transform of $1+\cos(\tilde t)$
gives the three $\delta$-peaks of $\Ph$, Eq.~\eqref{defphi}, where in
the limit $T\to\infty$ the side-bands have half the weight of the central peak.

\section{Physical part of the static correlation functions}%
\label{app:omphi}
The functions $\ph$, $\om$, and $\om'$ that determine all static
correlation functions of the Heisenberg-Ising chain are defined in
terms of solutions of non-linear and linear integral equations.
They were termed\cite{BoGo09} the physical part of the problem,
since the physical parameters like temperature or magnetic field
enter solely through these functions. We shall provide their definition
only for the massless case $- 1 \le \de < 0$.\cite{BDGKSW08} The
definitions for the massive case can be found in Ref.~\onlinecite{TGK10a}.

First of all let us define a basic pair of auxiliary functions as
the solution of the non-linear integral equations
\begin{widetext}
\begin{subequations}
\label{nlies}
\begin{align}
     \ln \fb (x) & = - \frac{\p h}{2 (\p - \g) T}
                     - \frac{2 \p J \sin (\g)}{T \g \ch (\p x/\g)}
                     + \int_{- \infty}^\infty \frac{\rd y}{2 \p}
                       F(x - y) \ln (1 + \fb (y))
                     - \int_{- \infty}^\infty \frac{\rd y}{2 \p}
                       F(x - y + \h^-) \ln (1 + \fbq (y)) \epc \\
     \ln \fbq (x) & = \frac{\p h}{2 (\p - \g) T}
                      - \frac{2 \p J \sin (\g)}{T \g \ch (\p x/\g)}
                      + \int_{- \infty}^\infty \frac{\rd y}{2 \p}
                       F(x - y) \ln (1 + \fbq (y))
                     - \int_{- \infty}^\infty \frac{\rd y}{2 \p}
                       F(x - y - \h^-) \ln (1 + \fb (y))
\end{align}
\end{subequations}
\end{widetext}
with kernel
\begin{equation}
     F(x) = \int_{- \infty}^\infty \rd k
            \frac{\sh \bigl( (\frac{\p}{2} - \g)k \bigr) \re^{\i kx}}
                 {2 \sh \bigl( (\p - \g)\frac{k}{2} \bigr)
                  \ch \bigl( \frac{\g k}{2} \bigr)} \epp
\end{equation}
Here we have introduced a parameter $\g$ which provides yet another
parameterization of the anisotropy, $\de = \cos(\g) - 1$. Eq.~\eqref{nlies} is valid
for $0 \le \g \le \p/2$ meaning that $- 1 \le \de < 0$. Below we shall also
use $\h = \i \g$. Note that the physical parameters temperature $T$, magnetic
field $h$, and coupling $J$ enter only through the driving terms of Eqs.~%
\eqref{nlies} into our formulae.

Except for the auxiliary functions $\fb$ and $\fbq$ we need two more pairs
of functions $g_\m^{(\pm)}$ and ${g'}_\m^{(\pm)}$ in order to define
$\ph$, $\om$, and $\om'$. Both pairs satisfy linear integral equations
involving $\fb$ and~$\fbq$,
\begin{widetext}
\begin{subequations}
\label{glies}
\begin{align}
     g_\m^{(+)} (x) & = \tst{\frac{\i \p}{\g}}
        \sech \Bigl( \tst{\frac{\p (x - \m)}{\g}} \Bigr)
        + \int_{- \infty}^\infty \frac{\rd y}{2 \p}
          \frac{F(x - y)}{1 + \fb^{-1} (y)} g_\m^{(+)} (y)
        - \int_{- \infty}^\infty \frac{\rd y}{2 \p}
          \frac{F(x - y + \h^-)}{1 + \fbq^{-1} (y)}
          g_\m^{(-)} (y) \epc \displaybreak[0] \\[2ex]
     g_\m^{(-)} (x) & = \tst{\frac{\i \p}{\g}}
        \sech \Bigl( \tst{\frac{\p (x - \m)}{\g}} \Bigr)
        + \int_{- \infty}^\infty \frac{\rd y}{2 \p}
          \frac{F(x - y)}{1 + \fbq^{-1} (y)}
          g_\m^{(-)} (y)
        - \int_{- \infty}^\infty \frac{\rd y}{2 \p}
          \frac{F(x - y - \h^-)}{1 + \fb^{-1} (y)} g_\m^{(+)} (y)
\end{align}
\end{subequations}
and
\begin{subequations}
\label{gplies}
\begin{align}
     {g'}_\m^{(+)} (x) & =
         \Bigl( \tst{\frac{\i \p}{\g} (x - \m) - \frac{\p}{2}} \Bigr)
         \sech \Bigl( \tst{\frac{\p (x - \m)}{\g}} \Bigr)
         + \g \int_{- \infty}^\infty \frac{\rd y}{2 \p}
           \frac{D(x - y)}{1 + \fb^{-1} (y)} g_\m^{(+)} (y)
         - \g \int_{- \infty}^\infty \frac{\rd y}{2 \p}
           \frac{D(x - y + \h^-)}{1 + \fbq^{-1} (y)} g_\m^{(-)} (y)
         \notag \\ & \mspace{145.mu}
         + \int_{- \infty}^\infty \frac{\rd y}{2 \p}
           \frac{F(x - y)}{1 + \fb^{-1} (y)} {g'}_\m^{(+)} (y)
         - \int_{- \infty}^\infty \frac{\rd y}{2 \p}
           \frac{F(x - y + \h^-)}{1 + \fbq^{-1} (y)} {g'}_\m^{(-)} (y)
           \epc \displaybreak[0] \\[2ex]
     {g'}_\m^{(-)} (x) & =
         \Bigl( \tst{\frac{\i \p}{\g} (x - \m) + \frac{\p}{2}} \Bigr)
         \sech \Bigl( \tst{\frac{\p (x - \m)}{\g}} \Bigr)
         + \g \int_{- \infty}^\infty \frac{\rd y}{2 \p}
           \frac{D(x - y)}{1 + \fbq^{-1} (y)} g_\m^{(-)} (y)
         - \g \int_{- \infty}^\infty \frac{\rd y}{2 \p}
           \frac{D(x - y - \h^-)}{1 + \fb^{-1} (y)} g_\m^{(+)} (y)
         \notag \\ & \mspace{145.mu}
         + \int_{- \infty}^\infty \frac{\rd y}{2 \p}
           \frac{F(x - y)}{1 + \fbq^{-1} (y)} {g'}_\m^{(-)} (y)
         - \int_{- \infty}^\infty \frac{\rd y}{2 \p}
           \frac{F(x - y - \h^-)}{1 + \fb^{-1} (y)} {g'}_\m^{(+)} (y)
           \epc
\end{align}
\end{subequations}
\end{widetext}
where
\begin{equation}
     D(x) = \int_{- \infty}^\infty \rd k
            \frac{\sin(kx) \sh \bigl( \frac{\p k}{2} \bigr)
                  \ch \bigl( (\frac{\p}{2} - \g)k \bigr)}
                 {4 \sh^2 \bigl( (\p - \g)\frac{k}{2} \bigr)
                  \ch^2 \bigl( \frac{\g k}{2} \bigr)} \epp
\end{equation}

The functions $\om (\m_1, \m_2)$, $\om' (\m_1, \m_2)$ and $\ph (\m)$
that determine the explicit form of the correlation functions of the
XXZ chain can be written as integrals involving $\fb$, $\fbq$,
$g_\m^{(\pm)}$ and ${g'}_\m^{(\pm)}$. The function
\begin{equation}
     \ph(\m) = \int_{- \infty}^\infty \frac{\rd x}{2 (\p - \g) \i}
               \biggl[ \frac{g_\m^{(+)} (x)}{1 + \fb^{-1} (x)}
                     - \frac{g_\m^{(-)} (x)}{1 + \fbq^{-1} (x)} \biggl]
\end{equation}
determines the magnetization $m(T,h) = - \tst{\2} \ph (0)$ which is the
only independent one-point function of the XXZ chain. The function
\begin{multline}
     \om(\m_1, \m_2) =
        - \tst{\2} K(\m_1 - \m_2) \\[1ex]
	- \int_{- \infty}^\infty \rd k \frac{\sh \bigl( (\p - \g)\frac{k}{2} \bigr)
                \cos(k(\m_1 - \m_2))}
               {\i \sh \bigl( \frac{\p k}{2} \bigr)
                   \ch \bigl( \frac{\g k}{2} \bigr)} \\
             - \int_{- \infty}^\infty
               \frac{\rd x}{\g \ch \bigl( \tst{\frac{\p (x - \m_2)}{\g}} \bigr)}
               \biggl[ \frac{g_{\m_1}^{(+)} (x)}{1 + \fb^{-1} (x)}
                     + \frac{g_{\m_1}^{(-)} (x)}{1 + \fbq^{-1} (x)}
               \biggl]
\end{multline}
with
\begin{equation}
     K(\m) = \cth(\m - \h) - \cth(\m + \h)
\end{equation}
also determines the energy per lattice site
\begin{equation}
     \< s_{j-1}^x s_j^x + s_{j-1}^y s_j^y + \D s_{j-1}^z s_j^z \>_T
            = \sh(\h) \om(0,0)/4
\end{equation}
of the XXZ chain. The function $\om' (\m_1, \m_2)$ is defined as
\begin{multline}
     \om'(\m_1, \m_2) =
        \tst{\frac{\h}{2}} K^{(+)} (\m_1 - \m_2)
        + \int_{- \infty}^\infty \rd k 
          \frac{\g \sin(k(\m_1 - \m_2))}
               {2 \i \tgh \bigl( \frac{\p k}{2} \bigr)
                   \ch^2 \bigl( \frac{\g k}{2} \bigr)} \displaybreak[0] \\
             - \int_{- \infty}^\infty
               \frac{\rd x}{\g \ch \bigl( \tst{\frac{\p (x - \m_2)}{\g}} \bigr)}
               \biggl[ \frac{f_{\m_1}^{(+)} (x)}{1 + \fb^{-1} (x)}
                     + \frac{f_{\m_1}^{(-)} (x)}{1 + \fbq^{-1} (x)}
               \biggl] \\
             + \int_{- \infty}^\infty
               \frac{\rd x \; (x - \m_2)}
	            {\g \ch \bigl( \tst{\frac{\p (x - \m_2)}{\g}} \bigr)}
               \biggl[ \frac{g_{\m_1}^{(+)} (x)}{1 + \fb^{-1} (x)}
                     + \frac{g_{\m_1}^{(-)} (x)}{1 + \fbq^{-1} (x)}
               \biggl] \epc
\end{multline}
where
\begin{subequations}
\begin{align}
     K^{(+)} (\m) & = \cth(\m - \h) + \cth(\m + \h) \epc \\
     f_\m^{(\pm)} (x) & = {g'}_\m^{(\pm)} (x)
                        \mp \tst{\frac{\i \g}{2}} g_\m^{(\pm)} (x) \epp
\end{align}
\end{subequations}

For the calculation of the moments in Sec.~\ref{sec:methmom} the
non-linear integral equations for $\fb$ and $\fbq$ as well as
their linear counterparts for $g_\m^{(\pm)}$ and ${g'}_\m^{(\pm)}$
were solved iteratively in Fourier space utilizing the fast
Fourier transformation algorithm. The derivatives of $g_\m^{(\pm)}$
and ${g'}_\m^{(\pm)}$ with respect to $\m$, needed in the computation
of the respective derivatives of $\ph$, $\om$, and $\om'$ satisfy
linear integral equations as well, which were obtained as derivatives
of the equations for $g_\m^{(\pm)}$ and ${g'}_\m^{(\pm)}$. Taking into
account derivatives is particularly simple in Fourier space.


\begin{thebibliography}{47}%
\makeatletter
\providecommand \@ifxundefined [1]{%
 \@ifx{#1\undefined}
}%
\providecommand \@ifnum [1]{%
 \ifnum #1\expandafter \@firstoftwo
 \else \expandafter \@secondoftwo
 \fi
}%
\providecommand \@ifx [1]{%
 \ifx #1\expandafter \@firstoftwo
 \else \expandafter \@secondoftwo
 \fi
}%
\providecommand \natexlab [1]{#1}%
\providecommand \enquote  [1]{``#1''}%
\providecommand \bibnamefont  [1]{#1}%
\providecommand \bibfnamefont [1]{#1}%
\providecommand \citenamefont [1]{#1}%
\providecommand \href@noop [0]{\@secondoftwo}%
\providecommand \href [0]{\begingroup \@sanitize@url \@href}%
\providecommand \@href[1]{\@@startlink{#1}\@@href}%
\providecommand \@@href[1]{\endgroup#1\@@endlink}%
\providecommand \@sanitize@url [0]{\catcode `\\12\catcode `\$12\catcode
  `\&12\catcode `\#12\catcode `\^12\catcode `\_12\catcode `\%12\relax}%
\providecommand \@@startlink[1]{}%
\providecommand \@@endlink[0]{}%
\providecommand \url  [0]{\begingroup\@sanitize@url \@url }%
\providecommand \@url [1]{\endgroup\@href {#1}{\urlprefix }}%
\providecommand \urlprefix  [0]{URL }%
\providecommand \Eprint [0]{\href }%
\@ifxundefined \urlstyle {%
  \providecommand \doi  [0]{\begingroup \@sanitize@url \@doi}%
  \providecommand \@doi [1]{\endgroup \@@startlink {\doibase
  #1}doi:\discretionary {}{}{}#1\@@endlink }%
}{%
  \providecommand \doi  [0]{doi:\discretionary{}{}{}\begingroup
  \urlstyle{rm}\Url }%
}%
\providecommand \doibase [0]{http://dx.doi.org/}%
\providecommand \Doi [0]{\begingroup \@sanitize@url \@Doi }%
\providecommand \@Doi  [1]{\endgroup\@@startlink{\doibase#1}\@@Doi}%
\providecommand \@@Doi [1]{#1\@@endlink}%
\providecommand \selectlanguage [0]{\@gobble}%
\providecommand \bibinfo  [0]{\@secondoftwo}%
\providecommand \bibfield  [0]{\@secondoftwo}%
\providecommand \translation [1]{[#1]}%
\providecommand \BibitemOpen [0]{}%
\providecommand \bibitemStop [0]{}%
\providecommand \bibitemNoStop [0]{.\EOS\space}%
\providecommand \EOS [0]{\spacefactor3000\relax}%
\providecommand \BibitemShut  [1]{\csname bibitem#1\endcsname}%
\bibitem [{\citenamefont {Kubo}\ and\ \citenamefont {Tomita}(1954)}]{KuTo54}%
  \BibitemOpen
  \bibfield  {author} {\bibinfo {author} {\bibfnamefont {R.}~\bibnamefont
  {Kubo}}\ and\ \bibinfo {author} {\bibfnamefont {K.}~\bibnamefont {Tomita}},\
  }\href@noop {} {\bibfield  {journal} {\bibinfo  {journal} {J. Phys. Soc.
  Jpn.},\ }\textbf {\bibinfo {volume} {9}},\ \bibinfo {pages} {888} (\bibinfo
  {year} {1954})}\BibitemShut {NoStop}%
\bibitem [{\citenamefont {Nagata}\ and\ \citenamefont {Tazuke}(1972)}]{NaTa72}%
  \BibitemOpen
  \bibfield  {author} {\bibinfo {author} {\bibfnamefont {K.}~\bibnamefont
  {Nagata}}\ and\ \bibinfo {author} {\bibfnamefont {Y.}~\bibnamefont
  {Tazuke}},\ }\href@noop {} {\bibfield  {journal} {\bibinfo  {journal} {J.
  Phys. Soc. Jpn.},\ }\textbf {\bibinfo {volume} {32}},\ \bibinfo {pages} {337}
  (\bibinfo {year} {1972})}\BibitemShut {NoStop}%
\bibitem [{\citenamefont {Ajiro}(2003)}]{Ajiro03}%
  \BibitemOpen
  \bibfield  {author} {\bibinfo {author} {\bibfnamefont {Y.}~\bibnamefont
  {Ajiro}},\ }\href@noop {} {\bibfield  {journal} {\bibinfo  {journal} {J.
  Phys. Soc. Jpn., Suppl. B},\ }\textbf {\bibinfo {volume} {72}},\ \bibinfo
  {pages} {12} (\bibinfo {year} {2003})}\BibitemShut {NoStop}%
\bibitem [{\citenamefont {Krug~von Nidda}\ \emph {et~al.}(2010)\citenamefont
  {Krug~von Nidda}, \citenamefont {B\"uttgen},\ and\ \citenamefont
  {Loidl}}]{KBL10}%
  \BibitemOpen
  \bibfield  {author} {\bibinfo {author} {\bibfnamefont {H.-A.}\ \bibnamefont
  {Krug~von Nidda}}, \bibinfo {author} {\bibfnamefont {N.}~\bibnamefont
  {B\"uttgen}}, \ and\ \bibinfo {author} {\bibfnamefont {A.}~\bibnamefont
  {Loidl}},\ }\href@noop {} {\bibfield  {journal} {\bibinfo  {journal} {Eur.
  Phys. J. Special Topics},\ }\textbf {\bibinfo {volume} {180}},\ \bibinfo
  {pages} {161} (\bibinfo {year} {2010})}\BibitemShut {NoStop}%
\bibitem [{\citenamefont {Oshikawa}\ and\ \citenamefont
  {Affleck}(1999)}]{OsAf99}%
  \BibitemOpen
  \bibfield  {author} {\bibinfo {author} {\bibfnamefont {M.}~\bibnamefont
  {Oshikawa}}\ and\ \bibinfo {author} {\bibfnamefont {I.}~\bibnamefont
  {Affleck}},\ }\href@noop {} {\bibfield  {journal} {\bibinfo  {journal} {Phys.
  Rev. Lett.},\ }\textbf {\bibinfo {volume} {82}},\ \bibinfo {pages} {5136}
  (\bibinfo {year} {1999})}\BibitemShut {NoStop}%
\bibitem [{\citenamefont {Oshikawa}\ and\ \citenamefont
  {Affleck}(2002)}]{OsAf02}%
  \BibitemOpen
  \bibfield  {author} {\bibinfo {author} {\bibfnamefont {M.}~\bibnamefont
  {Oshikawa}}\ and\ \bibinfo {author} {\bibfnamefont {I.}~\bibnamefont
  {Affleck}},\ }\href@noop {} {\bibfield  {journal} {\bibinfo  {journal} {Phys.
  Rev. B},\ }\textbf {\bibinfo {volume} {65}},\ \bibinfo {pages} {134410}
  (\bibinfo {year} {2002})}\BibitemShut {NoStop}%
\bibitem [{\citenamefont {Ogasahara}\ and\ \citenamefont
  {Miyashita}(2003)}]{OgMi03}%
  \BibitemOpen
  \bibfield  {author} {\bibinfo {author} {\bibfnamefont {A.}~\bibnamefont
  {Ogasahara}}\ and\ \bibinfo {author} {\bibfnamefont {S.}~\bibnamefont
  {Miyashita}},\ }\href@noop {} {\bibfield  {journal} {\bibinfo  {journal} {J.
  Phys. Soc. Jpn., Suppl. B},\ }\textbf {\bibinfo {volume} {72}},\ \bibinfo
  {pages} {44} (\bibinfo {year} {2003})}\BibitemShut {NoStop}%
\bibitem [{\citenamefont {Miyashita}\ \emph {et~al.}(1999)\citenamefont
  {Miyashita}, \citenamefont {Yoshino},\ and\ \citenamefont
  {Ogasahara}}]{MYO99}%
  \BibitemOpen
  \bibfield  {author} {\bibinfo {author} {\bibfnamefont {S.}~\bibnamefont
  {Miyashita}}, \bibinfo {author} {\bibfnamefont {T.}~\bibnamefont {Yoshino}},
  \ and\ \bibinfo {author} {\bibfnamefont {A.}~\bibnamefont {Ogasahara}},\
  }\href@noop {} {\bibfield  {journal} {\bibinfo  {journal} {J. Phys. Soc.
  Jpn.},\ }\textbf {\bibinfo {volume} {68}},\ \bibinfo {pages} {655} (\bibinfo
  {year} {1999})}\BibitemShut {NoStop}%
\bibitem [{\citenamefont {El~Shawish}\ \emph {et~al.}(2010)\citenamefont
  {El~Shawish}, \citenamefont {C\'epas},\ and\ \citenamefont
  {Miyashita}}]{ECM10}%
  \BibitemOpen
  \bibfield  {author} {\bibinfo {author} {\bibfnamefont {S.}~\bibnamefont
  {El~Shawish}}, \bibinfo {author} {\bibfnamefont {O.}~\bibnamefont {C\'epas}},
  \ and\ \bibinfo {author} {\bibfnamefont {S.}~\bibnamefont {Miyashita}},\
  }\href@noop {} {\bibfield  {journal} {\bibinfo  {journal} {Phys. Rev. B},\
  }\textbf {\bibinfo {volume} {81}},\ \bibinfo {pages} {224421} (\bibinfo
  {year} {2010})}\BibitemShut {NoStop}%
\bibitem [{\citenamefont {Maeda}\ \emph {et~al.}(2005)\citenamefont {Maeda},
  \citenamefont {Sakai},\ and\ \citenamefont {Oshikawa}}]{MSO05}%
  \BibitemOpen
  \bibfield  {author} {\bibinfo {author} {\bibfnamefont {Y.}~\bibnamefont
  {Maeda}}, \bibinfo {author} {\bibfnamefont {K.}~\bibnamefont {Sakai}}, \ and\
  \bibinfo {author} {\bibfnamefont {M.}~\bibnamefont {Oshikawa}},\ }\href@noop
  {} {\bibfield  {journal} {\bibinfo  {journal} {Phys. Rev. Lett.},\ }\textbf
  {\bibinfo {volume} {95}},\ \bibinfo {pages} {037602} (\bibinfo {year}
  {2005})}\BibitemShut {NoStop}%
\bibitem [{\citenamefont {van Vleck}(1948)}]{VanVleck48}%
  \BibitemOpen
  \bibfield  {author} {\bibinfo {author} {\bibfnamefont {J.~H.}\ \bibnamefont
  {van Vleck}},\ }\href@noop {} {\bibfield  {journal} {\bibinfo  {journal}
  {Phys. Rev.},\ }\textbf {\bibinfo {volume} {74}},\ \bibinfo {pages} {1168}
  (\bibinfo {year} {1948})}\BibitemShut {NoStop}%
\bibitem [{\citenamefont {Brockmann}\ \emph {et~al.}(2011)\citenamefont
  {Brockmann}, \citenamefont {G\"ohmann}, \citenamefont {Karbach},
  \citenamefont {Kl\"umper},\ and\ \citenamefont {Wei{\ss}e}}]{BGKKW11a}%
  \BibitemOpen
  \bibfield  {author} {\bibinfo {author} {\bibfnamefont {M.}~\bibnamefont
  {Brockmann}}, \bibinfo {author} {\bibfnamefont {F.}~\bibnamefont
  {G\"ohmann}}, \bibinfo {author} {\bibfnamefont {M.}~\bibnamefont {Karbach}},
  \bibinfo {author} {\bibfnamefont {A.}~\bibnamefont {Kl\"umper}}, \ and\
  \bibinfo {author} {\bibfnamefont {A.}~\bibnamefont {Wei{\ss}e}},\ }\href@noop
  {} {\bibfield  {journal} {\bibinfo  {journal} {Phys. Rev. Lett.},\ }\textbf
  {\bibinfo {volume} {107}},\ \bibinfo {pages} {017202} (\bibinfo {year}
  {2011})}\BibitemShut {NoStop}%
\bibitem [{\citenamefont {Jimbo}\ and\ \citenamefont {Miwa}(1995)}]{JiMi95}%
  \BibitemOpen
  \bibfield  {author} {\bibinfo {author} {\bibfnamefont {M.}~\bibnamefont
  {Jimbo}}\ and\ \bibinfo {author} {\bibfnamefont {T.}~\bibnamefont {Miwa}},\
  }\href@noop {} {\emph {\bibinfo {title} {Algebraic Analysis of Solvable
  Lattice Models}}}\ (\bibinfo  {publisher} {American Mathematical Society},\
  \bibinfo {year} {1995})\BibitemShut {NoStop}%
\bibitem [{\citenamefont {Bougourzi}\ \emph {et~al.}(1996)\citenamefont
  {Bougourzi}, \citenamefont {Couture},\ and\ \citenamefont {Kacir}}]{BCK96}%
  \BibitemOpen
  \bibfield  {author} {\bibinfo {author} {\bibfnamefont {A.~H.}\ \bibnamefont
  {Bougourzi}}, \bibinfo {author} {\bibfnamefont {M.}~\bibnamefont {Couture}},
  \ and\ \bibinfo {author} {\bibfnamefont {M.}~\bibnamefont {Kacir}},\
  }\href@noop {} {\bibfield  {journal} {\bibinfo  {journal} {Phys. Rev. B},\
  }\textbf {\bibinfo {volume} {54}},\ \bibinfo {pages} {R12669} (\bibinfo
  {year} {1996})}\BibitemShut {NoStop}%
\bibitem [{\citenamefont {Bougourzi}\ \emph {et~al.}(1998)\citenamefont
  {Bougourzi}, \citenamefont {Karbach},\ and\ \citenamefont
  {M\"uller}}]{BKM98}%
  \BibitemOpen
  \bibfield  {author} {\bibinfo {author} {\bibfnamefont {A.~H.}\ \bibnamefont
  {Bougourzi}}, \bibinfo {author} {\bibfnamefont {M.}~\bibnamefont {Karbach}},
  \ and\ \bibinfo {author} {\bibfnamefont {G.}~\bibnamefont {M\"uller}},\
  }\href@noop {} {\bibfield  {journal} {\bibinfo  {journal} {Phys. Rev. B},\
  }\textbf {\bibinfo {volume} {57}},\ \bibinfo {pages} {11429} (\bibinfo {year}
  {1998})}\BibitemShut {NoStop}%
\bibitem [{\citenamefont {Caux}\ \emph {et~al.}(2008)\citenamefont {Caux},
  \citenamefont {Mossel},\ and\ \citenamefont {Castillo}}]{CMP08}%
  \BibitemOpen
  \bibfield  {author} {\bibinfo {author} {\bibfnamefont {J.-S.}\ \bibnamefont
  {Caux}}, \bibinfo {author} {\bibfnamefont {J.}~\bibnamefont {Mossel}}, \ and\
  \bibinfo {author} {\bibfnamefont {I.~P.}\ \bibnamefont {Castillo}},\
  }\href@noop {} {\bibfield  {journal} {\bibinfo  {journal} {J. Stat. Mech.:
  Theor. Exp.},\ \bibinfo {pages} {P08006}} (\bibinfo {year}
  {2008})}\BibitemShut {NoStop}%
\bibitem [{\citenamefont {Boos}\ \emph {et~al.}(2008)\citenamefont {Boos},
  \citenamefont {Damerau}, \citenamefont {G\"ohmann}, \citenamefont
  {Kl\"umper}, \citenamefont {Suzuki},\ and\ \citenamefont
  {Wei{\ss}e}}]{BDGKSW08}%
  \BibitemOpen
  \bibfield  {author} {\bibinfo {author} {\bibfnamefont {H.}~\bibnamefont
  {Boos}}, \bibinfo {author} {\bibfnamefont {J.}~\bibnamefont {Damerau}},
  \bibinfo {author} {\bibfnamefont {F.}~\bibnamefont {G\"ohmann}}, \bibinfo
  {author} {\bibfnamefont {A.}~\bibnamefont {Kl\"umper}}, \bibinfo {author}
  {\bibfnamefont {J.}~\bibnamefont {Suzuki}}, \ and\ \bibinfo {author}
  {\bibfnamefont {A.}~\bibnamefont {Wei{\ss}e}},\ }\href@noop {} {\bibfield
  {journal} {\bibinfo  {journal} {J. Stat. Mech.: Theor. Exp.},\ \bibinfo
  {pages} {P08010}} (\bibinfo {year} {2008})}\BibitemShut {NoStop}%
\bibitem [{\citenamefont {Trippe}\ \emph {et~al.}(2010)\citenamefont {Trippe},
  \citenamefont {G\"ohmann},\ and\ \citenamefont {Kl\"umper}}]{TGK10a}%
  \BibitemOpen
  \bibfield  {author} {\bibinfo {author} {\bibfnamefont {C.}~\bibnamefont
  {Trippe}}, \bibinfo {author} {\bibfnamefont {F.}~\bibnamefont {G\"ohmann}}, \
  and\ \bibinfo {author} {\bibfnamefont {A.}~\bibnamefont {Kl\"umper}},\
  }\href@noop {} {\bibfield  {journal} {\bibinfo  {journal} {Eur. Phys. J. B},\
  }\textbf {\bibinfo {volume} {73}},\ \bibinfo {pages} {253} (\bibinfo {year}
  {2010})}\BibitemShut {NoStop}%
\bibitem [{\citenamefont {Jimbo}\ \emph {et~al.}(2009)\citenamefont {Jimbo},
  \citenamefont {Miwa},\ and\ \citenamefont {Smirnov}}]{JMS08}%
  \BibitemOpen
  \bibfield  {author} {\bibinfo {author} {\bibfnamefont {M.}~\bibnamefont
  {Jimbo}}, \bibinfo {author} {\bibfnamefont {T.}~\bibnamefont {Miwa}}, \ and\
  \bibinfo {author} {\bibfnamefont {F.}~\bibnamefont {Smirnov}},\ }\href@noop
  {} {\bibfield  {journal} {\bibinfo  {journal} {J. Phys. A},\ }\textbf
  {\bibinfo {volume} {42}},\ \bibinfo {pages} {304018} (\bibinfo {year}
  {2009})}\BibitemShut {NoStop}%
\bibitem [{\citenamefont {Boos}\ and\ \citenamefont
  {G\"ohmann}(2009)}]{BoGo09}%
  \BibitemOpen
  \bibfield  {author} {\bibinfo {author} {\bibfnamefont {H.}~\bibnamefont
  {Boos}}\ and\ \bibinfo {author} {\bibfnamefont {F.}~\bibnamefont
  {G\"ohmann}},\ }\href@noop {} {\bibfield  {journal} {\bibinfo  {journal} {J.
  Phys. A},\ }\textbf {\bibinfo {volume} {42}},\ \bibinfo {pages} {315001}
  (\bibinfo {year} {2009})}\BibitemShut {NoStop}%
\bibitem [{\citenamefont {Boos}\ \emph {et~al.}(2007)\citenamefont {Boos},
  \citenamefont {G\"ohmann}, \citenamefont {Kl\"umper},\ and\ \citenamefont
  {Suzuki}}]{BGKS07}%
  \BibitemOpen
  \bibfield  {author} {\bibinfo {author} {\bibfnamefont {H.}~\bibnamefont
  {Boos}}, \bibinfo {author} {\bibfnamefont {F.}~\bibnamefont {G\"ohmann}},
  \bibinfo {author} {\bibfnamefont {A.}~\bibnamefont {Kl\"umper}}, \ and\
  \bibinfo {author} {\bibfnamefont {J.}~\bibnamefont {Suzuki}},\ }\href@noop {}
  {\bibfield  {journal} {\bibinfo  {journal} {J. Phys. A},\ }\textbf {\bibinfo
  {volume} {40}},\ \bibinfo {pages} {10699} (\bibinfo {year}
  {2007})}\BibitemShut {NoStop}%
\bibitem [{\citenamefont {Sato}\ \emph {et~al.}(2011)\citenamefont {Sato},
  \citenamefont {Aufgebauer}, \citenamefont {Boos}, \citenamefont {G\"ohmann},
  \citenamefont {Kl\"umper}, \citenamefont {Takahashi},\ and\ \citenamefont
  {Trippe}}]{SABGKTT11}%
  \BibitemOpen
  \bibfield  {author} {\bibinfo {author} {\bibfnamefont {J.}~\bibnamefont
  {Sato}}, \bibinfo {author} {\bibfnamefont {B.}~\bibnamefont {Aufgebauer}},
  \bibinfo {author} {\bibfnamefont {H.}~\bibnamefont {Boos}}, \bibinfo {author}
  {\bibfnamefont {F.}~\bibnamefont {G\"ohmann}}, \bibinfo {author}
  {\bibfnamefont {A.}~\bibnamefont {Kl\"umper}}, \bibinfo {author}
  {\bibfnamefont {M.}~\bibnamefont {Takahashi}}, \ and\ \bibinfo {author}
  {\bibfnamefont {C.}~\bibnamefont {Trippe}},\ }\href@noop {} {\bibfield
  {journal} {\bibinfo  {journal} {Phys. Rev. Lett.},\ }\textbf {\bibinfo
  {volume} {106}},\ \bibinfo {pages} {257201} (\bibinfo {year}
  {2011})}\BibitemShut {NoStop}%
\bibitem [{\citenamefont {Typek}\ and\ \citenamefont {Guskos}(2006)}]{TyGu06}%
  \BibitemOpen
  \bibfield  {author} {\bibinfo {author} {\bibfnamefont {J.}~\bibnamefont
  {Typek}}\ and\ \bibinfo {author} {\bibfnamefont {N.}~\bibnamefont {Guskos}},\
  }\href@noop {} {\bibfield  {journal} {\bibinfo  {journal} {Rev. Adv. Mater.
  Sci.},\ }\textbf {\bibinfo {volume} {12}},\ \bibinfo {pages} {106} (\bibinfo
  {year} {2006})}\BibitemShut {NoStop}%
\bibitem [{\citenamefont {Kl\"umper}(1993)}]{Kluemper93}%
  \BibitemOpen
  \bibfield  {author} {\bibinfo {author} {\bibfnamefont {A.}~\bibnamefont
  {Kl\"umper}},\ }\href@noop {} {\bibfield  {journal} {\bibinfo  {journal} {Z.
  Phys. B},\ }\textbf {\bibinfo {volume} {91}},\ \bibinfo {pages} {507}
  (\bibinfo {year} {1993})}\BibitemShut {NoStop}%
\bibitem [{\citenamefont {Shiba}\ and\ \citenamefont {Adachi}(1981)}]{ShAd81}%
  \BibitemOpen
  \bibfield  {author} {\bibinfo {author} {\bibfnamefont {H.}~\bibnamefont
  {Shiba}}\ and\ \bibinfo {author} {\bibfnamefont {K.}~\bibnamefont {Adachi}},\
  }\href@noop {} {\bibfield  {journal} {\bibinfo  {journal} {J. Phys. Soc.
  Jpn.},\ }\textbf {\bibinfo {volume} {50}},\ \bibinfo {pages} {3278} (\bibinfo
  {year} {1981})}\BibitemShut {NoStop}%
\bibitem [{\citenamefont {Baxter}(1982)}]{Babook}%
  \BibitemOpen
  \bibfield  {author} {\bibinfo {author} {\bibfnamefont {R.~J.}\ \bibnamefont
  {Baxter}},\ }\href@noop {} {\emph {\bibinfo {title} {Exactly Solved Models in
  Statistical Mechanics}}}\ (\bibinfo  {publisher} {Academic Press, London},\
  \bibinfo {year} {1982})\BibitemShut {NoStop}%
\bibitem [{\citenamefont {Brandt}\ and\ \citenamefont {Jacoby}(1976)}]{BrJa76}%
  \BibitemOpen
  \bibfield  {author} {\bibinfo {author} {\bibfnamefont {U.}~\bibnamefont
  {Brandt}}\ and\ \bibinfo {author} {\bibfnamefont {K.}~\bibnamefont
  {Jacoby}},\ }\href@noop {} {\bibfield  {journal} {\bibinfo  {journal} {Z.
  Phys. B},\ }\textbf {\bibinfo {volume} {25}},\ \bibinfo {pages} {181}
  (\bibinfo {year} {1976})}\BibitemShut {NoStop}%
\bibitem [{\citenamefont {Vasil'ev}\ \emph {et~al.}(2001)\citenamefont
  {Vasil'ev}, \citenamefont {Ponomarenko}, \citenamefont {Manaka},
  \citenamefont {Yamada}, \citenamefont {Isobe},\ and\ \citenamefont
  {Ueda}}]{Exp_LiCuVO_01}%
  \BibitemOpen
  \bibfield  {author} {\bibinfo {author} {\bibfnamefont {A.~N.}\ \bibnamefont
  {Vasil'ev}}, \bibinfo {author} {\bibfnamefont {L.~A.}\ \bibnamefont
  {Ponomarenko}}, \bibinfo {author} {\bibfnamefont {H.}~\bibnamefont {Manaka}},
  \bibinfo {author} {\bibfnamefont {I.}~\bibnamefont {Yamada}}, \bibinfo
  {author} {\bibfnamefont {M.}~\bibnamefont {Isobe}}, \ and\ \bibinfo {author}
  {\bibfnamefont {Y.}~\bibnamefont {Ueda}},\ }\Doi {10.1103/PhysRevB.64.024419}
  {\bibfield  {journal} {\bibinfo  {journal} {Phys. Rev. B},\ }\textbf
  {\bibinfo {volume} {64}},\ \bibinfo {pages} {024419} (\bibinfo {year}
  {2001})}\BibitemShut {NoStop}%
\bibitem [{\citenamefont {Krug~von Nidda}\ \emph {et~al.}(2002)\citenamefont
  {Krug~von Nidda}, \citenamefont {Svistov}, \citenamefont {Eremin},
  \citenamefont {Eremina}, \citenamefont {Loidl}, \citenamefont {Kataev},
  \citenamefont {Validov}, \citenamefont {Prokofiev},\ and\ \citenamefont
  {A\ss{}mus}}]{Exp_LiCuVO_02}%
  \BibitemOpen
  \bibfield  {author} {\bibinfo {author} {\bibfnamefont {H.-A.}\ \bibnamefont
  {Krug~von Nidda}}, \bibinfo {author} {\bibfnamefont {L.~E.}\ \bibnamefont
  {Svistov}}, \bibinfo {author} {\bibfnamefont {M.~V.}\ \bibnamefont {Eremin}},
  \bibinfo {author} {\bibfnamefont {R.~M.}\ \bibnamefont {Eremina}}, \bibinfo
  {author} {\bibfnamefont {A.}~\bibnamefont {Loidl}}, \bibinfo {author}
  {\bibfnamefont {V.}~\bibnamefont {Kataev}}, \bibinfo {author} {\bibfnamefont
  {A.}~\bibnamefont {Validov}}, \bibinfo {author} {\bibfnamefont
  {A.}~\bibnamefont {Prokofiev}}, \ and\ \bibinfo {author} {\bibfnamefont
  {W.}~\bibnamefont {A\ss{}mus}},\ }\Doi {10.1103/PhysRevB.65.134445}
  {\bibfield  {journal} {\bibinfo  {journal} {Phys. Rev. B},\ }\textbf
  {\bibinfo {volume} {65}},\ \bibinfo {pages} {134445} (\bibinfo {year}
  {2002})}\BibitemShut {NoStop}%
\bibitem [{\citenamefont {Wei{\ss}e}(2004)}]{We04}%
  \BibitemOpen
  \bibfield  {author} {\bibinfo {author} {\bibfnamefont {A.}~\bibnamefont
  {Wei{\ss}e}},\ }\Doi {10.1140/epjb/e2004-00250-6} {\bibfield  {journal}
  {\bibinfo  {journal} {Eur. Phys. J. B},\ }\textbf {\bibinfo {volume} {40}},\
  \bibinfo {pages} {125} (\bibinfo {year} {2004})}\BibitemShut {NoStop}%
\bibitem [{\citenamefont {Wei{\ss}e}\ \emph {et~al.}(2006)\citenamefont
  {Wei{\ss}e}, \citenamefont {Wellein}, \citenamefont {Alvermann},\ and\
  \citenamefont {Fehske}}]{WWAF06}%
  \BibitemOpen
  \bibfield  {author} {\bibinfo {author} {\bibfnamefont {A.}~\bibnamefont
  {Wei{\ss}e}}, \bibinfo {author} {\bibfnamefont {G.}~\bibnamefont {Wellein}},
  \bibinfo {author} {\bibfnamefont {A.}~\bibnamefont {Alvermann}}, \ and\
  \bibinfo {author} {\bibfnamefont {H.}~\bibnamefont {Fehske}},\ }\Doi
  {10.1103/RevModPhys.78.275} {\bibfield  {journal} {\bibinfo  {journal} {Rev.
  Mod. Phys.},\ }\textbf {\bibinfo {volume} {78}},\ \bibinfo {pages} {275}
  (\bibinfo {year} {2006})}\BibitemShut {NoStop}%
\bibitem [{\citenamefont {Wei{\ss}e}\ and\ \citenamefont
  {Fehske}(2008)}]{WF08b}%
  \BibitemOpen
  \bibfield  {author} {\bibinfo {author} {\bibfnamefont {A.}~\bibnamefont
  {Wei{\ss}e}}\ and\ \bibinfo {author} {\bibfnamefont {H.}~\bibnamefont
  {Fehske}},\ }in\ \Doi {10.1007/978-3-540-74686-7_19} {\emph {\bibinfo
  {booktitle} {Computational Many-Particle Physics}}},\ \bibinfo {series}
  {Lecture Notes in Physics}, Vol.\ \bibinfo {volume} {739},\ \bibinfo {editor}
  {edited by\ \bibinfo {editor} {\bibfnamefont {H.}~\bibnamefont {Fehske}},
  \bibinfo {editor} {\bibfnamefont {R.}~\bibnamefont {Schneider}}, \ and\
  \bibinfo {editor} {\bibfnamefont {A.}~\bibnamefont {Wei{\ss}e}}}\ (\bibinfo
  {publisher} {Springer},\ \bibinfo {address} {Heidelberg},\ \bibinfo {year}
  {2008})\ pp.\ \bibinfo {pages} {545--577}\BibitemShut {NoStop}%
\bibitem [{\citenamefont {Lanczos}(1950)}]{La50}%
  \BibitemOpen
  \bibfield  {author} {\bibinfo {author} {\bibfnamefont {C.}~\bibnamefont
  {Lanczos}},\ }\href@noop {} {\bibfield  {journal} {\bibinfo  {journal} {J.
  Res. Nat. Bur. Stand.},\ }\textbf {\bibinfo {volume} {45}},\ \bibinfo {pages}
  {255} (\bibinfo {year} {1950})}\BibitemShut {NoStop}%
\bibitem [{\citenamefont {Arikawa}\ \emph {et~al.}(2006)\citenamefont
  {Arikawa}, \citenamefont {Karbach}, \citenamefont {M\"uller},\ and\
  \citenamefont {Wiele}}]{AKMW06}%
  \BibitemOpen
  \bibfield  {author} {\bibinfo {author} {\bibfnamefont {M.}~\bibnamefont
  {Arikawa}}, \bibinfo {author} {\bibfnamefont {M.}~\bibnamefont {Karbach}},
  \bibinfo {author} {\bibfnamefont {G.}~\bibnamefont {M\"uller}}, \ and\
  \bibinfo {author} {\bibfnamefont {K.}~\bibnamefont {Wiele}},\ }\href@noop {}
  {\bibfield  {journal} {\bibinfo  {journal} {J. Phys. A},\ }\textbf {\bibinfo
  {volume} {39}},\ \bibinfo {pages} {10623} (\bibinfo {year}
  {2006})}\BibitemShut {NoStop}%
\bibitem [{\citenamefont {Kitanine}\ \emph {et~al.}(2009)\citenamefont
  {Kitanine}, \citenamefont {Kozlowski}, \citenamefont {Maillet}, \citenamefont
  {Slavnov},\ and\ \citenamefont {Terras}}]{KKMST09b}%
  \BibitemOpen
  \bibfield  {author} {\bibinfo {author} {\bibfnamefont {N.}~\bibnamefont
  {Kitanine}}, \bibinfo {author} {\bibfnamefont {K.~K.}\ \bibnamefont
  {Kozlowski}}, \bibinfo {author} {\bibfnamefont {J.~M.}\ \bibnamefont
  {Maillet}}, \bibinfo {author} {\bibfnamefont {N.~A.}\ \bibnamefont
  {Slavnov}}, \ and\ \bibinfo {author} {\bibfnamefont {V.}~\bibnamefont
  {Terras}},\ }\href@noop {} {\bibfield  {journal} {\bibinfo  {journal} {J.
  Math. Phys.},\ }\textbf {\bibinfo {volume} {50}},\ \bibinfo {pages} {095209}
  (\bibinfo {year} {2009})}\BibitemShut {NoStop}%
\bibitem [{\citenamefont {Kitanine}\ \emph
  {et~al.}(2011){\natexlab{a}}\citenamefont {Kitanine}, \citenamefont
  {Kozlowski}, \citenamefont {Maillet}, \citenamefont {Slavnov},\ and\
  \citenamefont {Terras}}]{KKMST11a}%
  \BibitemOpen
  \bibfield  {author} {\bibinfo {author} {\bibfnamefont {N.}~\bibnamefont
  {Kitanine}}, \bibinfo {author} {\bibfnamefont {K.~K.}\ \bibnamefont
  {Kozlowski}}, \bibinfo {author} {\bibfnamefont {J.~M.}\ \bibnamefont
  {Maillet}}, \bibinfo {author} {\bibfnamefont {N.~A.}\ \bibnamefont
  {Slavnov}}, \ and\ \bibinfo {author} {\bibfnamefont {V.}~\bibnamefont
  {Terras}},\ }\href@noop {} {\bibfield  {journal} {\bibinfo  {journal} {J.
  Stat. Mech.: Theor. Exp.},\ }\textbf {\bibinfo {volume} {1105}},\ \bibinfo
  {pages} {P028} (\bibinfo {year} {2011}{\natexlab{a}})}\BibitemShut {NoStop}%
\bibitem [{\citenamefont {Biegel}\ \emph {et~al.}(2002)\citenamefont {Biegel},
  \citenamefont {Karbach},\ and\ \citenamefont {M\"uller}}]{BKM02a}%
  \BibitemOpen
  \bibfield  {author} {\bibinfo {author} {\bibfnamefont {D.}~\bibnamefont
  {Biegel}}, \bibinfo {author} {\bibfnamefont {M.}~\bibnamefont {Karbach}}, \
  and\ \bibinfo {author} {\bibfnamefont {G.}~\bibnamefont {M\"uller}},\
  }\href@noop {} {\bibfield  {journal} {\bibinfo  {journal} {Europhys.\
  Lett.},\ }\textbf {\bibinfo {volume} {59}},\ \bibinfo {pages} {882} (\bibinfo
  {year} {2002})}\BibitemShut {NoStop}%
\bibitem [{\citenamefont {Biegel}\ \emph {et~al.}(2003)\citenamefont {Biegel},
  \citenamefont {Karbach},\ and\ \citenamefont {M\"uller}}]{BKM03}%
  \BibitemOpen
  \bibfield  {author} {\bibinfo {author} {\bibfnamefont {D.}~\bibnamefont
  {Biegel}}, \bibinfo {author} {\bibfnamefont {M.}~\bibnamefont {Karbach}}, \
  and\ \bibinfo {author} {\bibfnamefont {G.}~\bibnamefont {M\"uller}},\
  }\href@noop {} {\bibfield  {journal} {\bibinfo  {journal} {J. Phys. A},\
  }\textbf {\bibinfo {volume} {36}},\ \bibinfo {pages} {5361} (\bibinfo {year}
  {2003})}\BibitemShut {NoStop}%
\bibitem [{\citenamefont {Sato}\ \emph {et~al.}(2004)\citenamefont {Sato},
  \citenamefont {Shiroishi},\ and\ \citenamefont {Takahashi}}]{SST04}%
  \BibitemOpen
  \bibfield  {author} {\bibinfo {author} {\bibfnamefont {J.}~\bibnamefont
  {Sato}}, \bibinfo {author} {\bibfnamefont {M.}~\bibnamefont {Shiroishi}}, \
  and\ \bibinfo {author} {\bibfnamefont {M.}~\bibnamefont {Takahashi}},\
  }\href@noop {} {\bibfield  {journal} {\bibinfo  {journal} {J. Phys. Soc.
  Jpn.},\ }\textbf {\bibinfo {volume} {73}},\ \bibinfo {pages} {3008} (\bibinfo
  {year} {2004})}\BibitemShut {NoStop}%
\bibitem [{\citenamefont {Caux}\ and\ \citenamefont {Maillet}(2005)}]{CaMa05}%
  \BibitemOpen
  \bibfield  {author} {\bibinfo {author} {\bibfnamefont {J.-S.}\ \bibnamefont
  {Caux}}\ and\ \bibinfo {author} {\bibfnamefont {J.~M.}\ \bibnamefont
  {Maillet}},\ }\href@noop {} {\bibfield  {journal} {\bibinfo  {journal} {Phys.
  Rev. Lett.},\ }\textbf {\bibinfo {volume} {95}},\ \bibinfo {pages} {077201}
  (\bibinfo {year} {2005})}\BibitemShut {NoStop}%
\bibitem [{\citenamefont {Kitanine}\ \emph
  {et~al.}(2011){\natexlab{b}}\citenamefont {Kitanine}, \citenamefont
  {Kozlowski}, \citenamefont {Maillet}, \citenamefont {Slavnov},\ and\
  \citenamefont {Terras}}]{KKMST11b}%
  \BibitemOpen
  \bibfield  {author} {\bibinfo {author} {\bibfnamefont {N.}~\bibnamefont
  {Kitanine}}, \bibinfo {author} {\bibfnamefont {K.~K.}\ \bibnamefont
  {Kozlowski}}, \bibinfo {author} {\bibfnamefont {J.~M.}\ \bibnamefont
  {Maillet}}, \bibinfo {author} {\bibfnamefont {N.~A.}\ \bibnamefont
  {Slavnov}}, \ and\ \bibinfo {author} {\bibfnamefont {V.}~\bibnamefont
  {Terras}},\ }\href@noop {} {\bibfield  {journal} {\bibinfo  {journal} {J.
  Stat. Mech.: Theor. Exp.},\ \bibinfo {pages} {P12010}} (\bibinfo {year}
  {2011}{\natexlab{b}})}\BibitemShut {NoStop}%
\bibitem [{\citenamefont {Johnson}\ \emph {et~al.}(1973)\citenamefont
  {Johnson}, \citenamefont {Krinsky},\ and\ \citenamefont {McCoy}}]{JKM73}%
  \BibitemOpen
  \bibfield  {author} {\bibinfo {author} {\bibfnamefont {J.~D.}\ \bibnamefont
  {Johnson}}, \bibinfo {author} {\bibfnamefont {S.}~\bibnamefont {Krinsky}}, \
  and\ \bibinfo {author} {\bibfnamefont {B.~M.}\ \bibnamefont {McCoy}},\
  }\href@noop {} {\bibfield  {journal} {\bibinfo  {journal} {Phys. Rev. A},\
  }\textbf {\bibinfo {volume} {8}},\ \bibinfo {pages} {2526} (\bibinfo {year}
  {1973})}\BibitemShut {NoStop}%
\bibitem [{\citenamefont {Jimbo}\ \emph {et~al.}(1992)\citenamefont {Jimbo},
  \citenamefont {Miki}, \citenamefont {Miwa},\ and\ \citenamefont
  {Nakayashiki}}]{JMMN92}%
  \BibitemOpen
  \bibfield  {author} {\bibinfo {author} {\bibfnamefont {M.}~\bibnamefont
  {Jimbo}}, \bibinfo {author} {\bibfnamefont {K.}~\bibnamefont {Miki}},
  \bibinfo {author} {\bibfnamefont {T.}~\bibnamefont {Miwa}}, \ and\ \bibinfo
  {author} {\bibfnamefont {A.}~\bibnamefont {Nakayashiki}},\ }\href@noop {}
  {\bibfield  {journal} {\bibinfo  {journal} {Phys. Lett. A},\ }\textbf
  {\bibinfo {volume} {168}},\ \bibinfo {pages} {256} (\bibinfo {year}
  {1992})}\BibitemShut {NoStop}%
\bibitem [{\citenamefont {Takahashi}\ \emph {et~al.}(2004)\citenamefont
  {Takahashi}, \citenamefont {Kato},\ and\ \citenamefont {Shiroishi}}]{TKS04}%
  \BibitemOpen
  \bibfield  {author} {\bibinfo {author} {\bibfnamefont {M.}~\bibnamefont
  {Takahashi}}, \bibinfo {author} {\bibfnamefont {G.}~\bibnamefont {Kato}}, \
  and\ \bibinfo {author} {\bibfnamefont {M.}~\bibnamefont {Shiroishi}},\
  }\href@noop {} {\bibfield  {journal} {\bibinfo  {journal} {J. Phys. Soc.
  Jpn.},\ }\textbf {\bibinfo {volume} {73}},\ \bibinfo {pages} {245} (\bibinfo
  {year} {2004})}\BibitemShut {NoStop}%
\bibitem [{\citenamefont {Kato}\ \emph {et~al.}(2004)\citenamefont {Kato},
  \citenamefont {Shiroishi}, \citenamefont {Takahashi},\ and\ \citenamefont
  {Sakai}}]{KSTS04}%
  \BibitemOpen
  \bibfield  {author} {\bibinfo {author} {\bibfnamefont {G.}~\bibnamefont
  {Kato}}, \bibinfo {author} {\bibfnamefont {M.}~\bibnamefont {Shiroishi}},
  \bibinfo {author} {\bibfnamefont {M.}~\bibnamefont {Takahashi}}, \ and\
  \bibinfo {author} {\bibfnamefont {K.}~\bibnamefont {Sakai}},\ }\href@noop {}
  {\bibfield  {journal} {\bibinfo  {journal} {J. Phys. A},\ }\textbf {\bibinfo
  {volume} {37}},\ \bibinfo {pages} {5097} (\bibinfo {year}
  {2004})}\BibitemShut {NoStop}%
\bibitem [{\citenamefont {B\"ohm}\ and\ \citenamefont
  {Leschke}(1992)}]{BoLe92}%
  \BibitemOpen
  \bibfield  {author} {\bibinfo {author} {\bibfnamefont {M.}~\bibnamefont
  {B\"ohm}}\ and\ \bibinfo {author} {\bibfnamefont {H.}~\bibnamefont
  {Leschke}},\ }\href@noop {} {\bibfield  {journal} {\bibinfo  {journal} {J.
  Phys. A},\ }\textbf {\bibinfo {volume} {25}},\ \bibinfo {pages} {1043}
  (\bibinfo {year} {1992})}\BibitemShut {NoStop}%
\bibitem [{\citenamefont {Vermaseren}(2000)}]{Ver00}%
  \BibitemOpen
  \bibfield  {author} {\bibinfo {author} {\bibfnamefont {J.~A.~M.}\
  \bibnamefont {Vermaseren}},\ }\href {http://arXiv.org/abs/math-ph/0010025}
  {\enquote {\bibinfo {title} {New features of form},}\ } (\bibinfo {year}
  {2000}),\ \bibinfo {note} {preprint}\BibitemShut {NoStop}%
\end{thebibliography}

%

\end{document}